\documentclass[10pt,%
  reprint,
  superscriptaddress,
 amsmath,amssymb,
 aps,
 pre,
]{revtex4-2}

\usepackage{graphicx}
\usepackage{dcolumn}
\usepackage{bm}
\usepackage[colorlinks=true,allcolors=blue]{hyperref}
\usepackage{amsmath}
\usepackage{SIunits}
\usepackage{color}
\usepackage[e]{esvect}

\renewcommand{\d}{\mathrm{d}}

\renewcommand{\vec}[1]{\bm{#1}}
\newcommand{\sgn}[1]{\mathrm{sgn}(#1)}
\newcommand{\w}{\mathrm{{w}}}

\newcommand{\vel}{{\mathrm{v}}}

\newcommand{\aref}[1]{{Appendix~\ref{#1}}}

\definecolor{mmcolor}{rgb}{0.8, 0.0, 0.2}

\definecolor{micolor}{rgb}{0.0, 0.7, 0.2}

\begin{document}

\title{Stabilization of active tissue deformation by a dynamic signaling gradient}

\author{Muhamet Ibrahimi}
\affiliation{Aix Marseille Univ, Université de Toulon, CNRS, CPT (UMR 7332), Turing Centre for Living Systems, Marseille, France.}
\affiliation{Laboratory of Artificial and Natural Evolution (LANE), Department of Genetics and Evolution, \\
University of Geneva, 1211 Geneva, Switzerland.}

\author{Matthias Merkel}\email{matthias.merkel@cnrs.fr}

\affiliation{Aix Marseille Univ, Université de Toulon, CNRS, CPT (UMR 7332), Turing Centre for Living Systems, Marseille, France.}%
\date{\today}

\begin{abstract}
	A key process during animal morphogenesis is oriented tissue deformation, which is often driven by internally generated active stresses.
	Yet, such active oriented materials are prone to well-known instabilities, raising the question of how oriented tissue deformation can be robust during morphogenesis.
	Here we study under which conditions active oriented deformation can be stabilized by the concentration pattern of a signaling molecule, which is secreted by a localized source region, diffuses across the tissue, and degrades.
	Consistent with earlier results, we find that oriented tissue deformation is always unstable in the gradient-contractile case, i.e.\ when active stresses act to contract the tissue along the direction of the signaling gradient, and we now show that this is true even in the limit of large diffusion.
	However, active deformation	can be stabilized in the gradient-extensile case, i.e.\ when active stresses act to extend the tissue along the direction of the signaling gradient. Specifically, we show that gradient-extensile systems can be stable when the tissue is already elongated in the direction of the gradient. 
	We moreover point out the existence of a formerly unknown, additional instability of the tissue shape change. This instability results from the interplay of active tissue shear and signal diffusion, and it indicates that some additional feedback mechanism may be required to control the target tissue shape.
	Taken together, our theoretical results provide quantitative criteria for robust active tissue deformation, and explain the lack of gradient-contractile systems in the biological literature, suggesting that the active matter instability acts as an evolutionary selection criterion.
\end{abstract}

\maketitle

\section{Introduction}
Oriented tissue deformation is a key process during animal morphogenesis \cite{Wolpert2015}, including for instance body axis elongation \cite{Zallen2004,Benazeraf2010,Shindo2018} and organ formation \cite{Johansen2003,Karner2009,Saxena2014,Etournay2015,Hopyan2017,Tao2019}.
In many cases, such tissue deformation is at least in part driven by internally generated active stresses \cite{Bertet2004,Bosveld2012,Collinet2015,Etournay2015,Tao2019}, and in recent years, deforming tissues have successfully been described as active materials \cite{Behrndt2012,Etournay2015,Streichan2018,Stokkermans2022,Gehrels2023,Dye2021,Serra2023,Gsell2023,Barrett2024}.
Moreover, to achieve persistent large-scale oriented deformation, tissues need some kind of tissue-wide orientational information, which can be represented for instance by a polar or a nematic field \cite{Etournay2015,Streichan2018,Claussen2023,Uba2023,Serra2023}. 
However, active polar or nematic materials are subject to the well-known Simha-Ramaswamy instability, where an initially globally ordered orientational field loses this order due to active flows \cite{Simha2002,Voituriez2005,Marchetti2013}.
This raises the question of how oriented tissue deformation can be stable during animal development.

In animals, oriented tissue deformation can be affected by tissue-scale concentration patterns of signaling molecules. 
Such patterns of signaling molecules are known to coordinate developmental processes on the tissue or even organism scale \cite{Wolpert2015}.
Signaling molecules can provide \emph{positional} information, in which case they are called morphogens \cite{VonderHardt2007,Benazeraf2010,Rogers2011a,Muller2013,Tkacik2014a,Mosby2024}.
Yet, in some cases, concentration gradients of signaling molecules also provide \emph{orientational} information that defines the axis of active stresses driving oriented tissue deformation \cite{Johansen2003,Ninomiya2004,Karner2009,Bosveld2012,Saxena2014}, typically by guiding cellular polarity and cell rearrangements \cite{Zallen2004,Karner2009,Bosveld2012,Shindo2018,Lavalou2021}.
Such alignment of orientational information with signaling gradients has recently also been discussed theoretically \cite{Wang2023}.

In a recent publication \cite{Ibrahimi2023}, we theoretically showed that oriented tissue deformation can be stabilized by signaling gradients.
We used a hydrodynamic model for an active polar material, and showed that the gradient of a scalar signaling field can stabilize the otherwise unstable material.
We showed that if the scalar field advects with the material flows \cite{Lefebvre2022,Plum2025b}, the stability of oriented material deformation depends on the coupling between scalar field gradient direction and deformation axis.
In the gradient-extensile case, i.e.\ when anisotropic active stresses act to extend the material along the gradient direction, the scalar field can stabilize the active material deformation.
However, in the gradient-contractile case, i.e.\ when anisotropic active stresses act to contract the material along the gradient direction, the deformation is unstable. 
Yet, one could always stabilize even a gradient-contractile system by sufficiently increasing the diffusion of the signaling molecule.

In our past publication \cite{Ibrahimi2023}, we have created a uniform scalar field gradient in a simple way through appropriate boundary conditions.
While providing advantages for analytical calculations, this is quite different from the real mechanism creating signaling gradients in developing biological tissues \cite{Wartlick2011,Rogers2011a,Muller2013,Wolpert2015,Mateus2020a,Romanova-Michaelides2022}.
Signaling molecules are typically produced by some source region, diffuse across the tissue, and degrade \cite{Wartlick2009,Romanova-Michaelides2022}.
As a consequence, depending on the source region, this can lead to approximately exponentially decaying \cite{Wartlick2011,Wang2016,Huang2017,Mateus2020a} or periodic \cite{Pare2014,Benton2016} concentration profiles.
However, it is not known how such signaling molecule dynamics would affect stability of active oriented tissue deformation.

\begin{figure}[h]
    \centering
    \includegraphics[width = 8.2 cm]{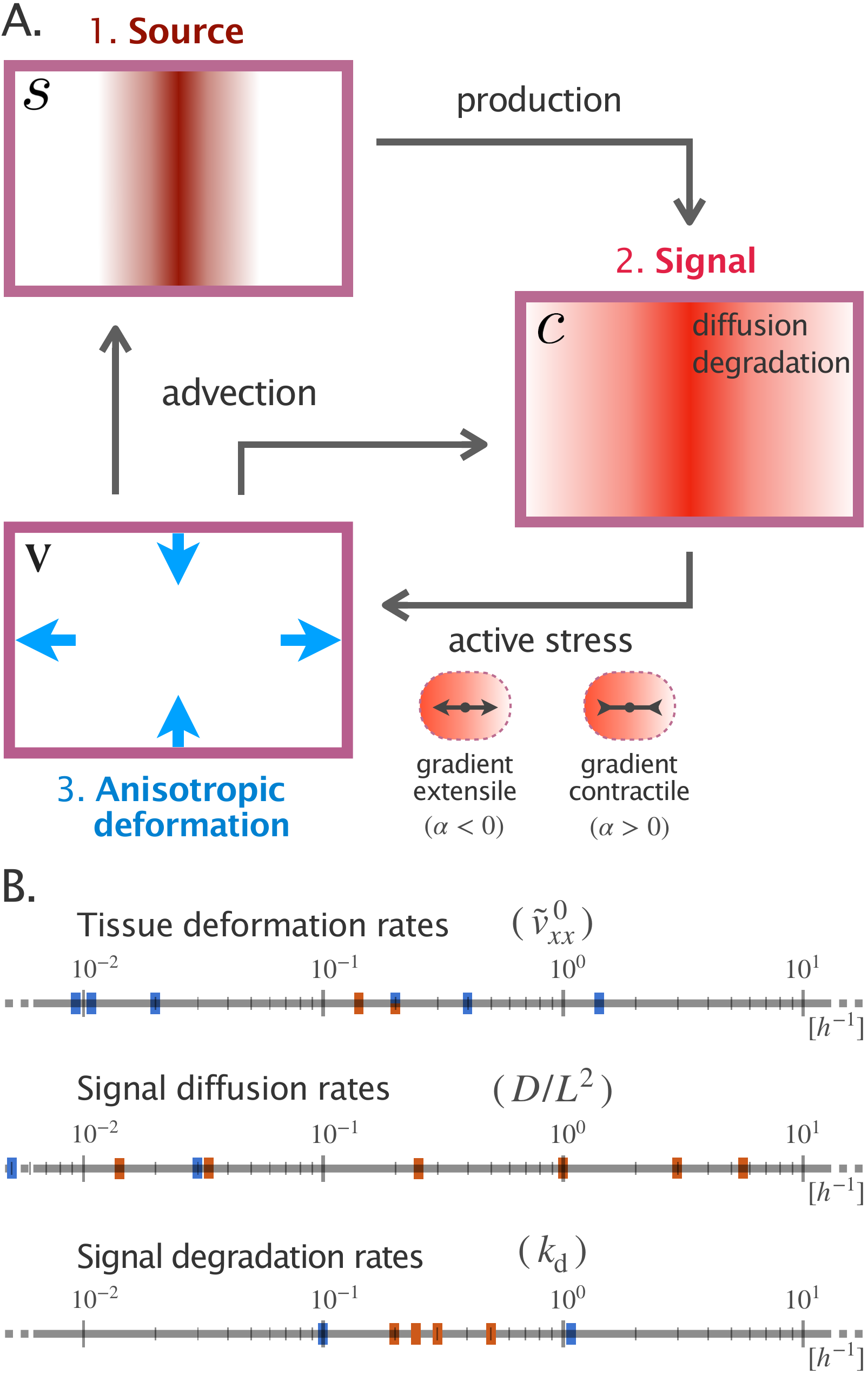}
    \caption{
    (A) Hydrodynamic model. A spatially varying source field $s(\vec{r},t)$ secretes a signaling molecule with concentration $c(\vec{r},t)$, which also diffuses and degrades. The gradient of $c$ creates extensile/contractile active anisotropic stresses, which drive viscous flows with velocity $\vec{\vel}(\vec{r},t)$. These flows affect both source and signaling fields through advection.
    (B) Typical orders of magnitude for tissue deformation rates $\tilde{v}^0_{xx}$ (top), signal diffusion rates $D/L^2$, where $D$ is the diffusion coefficient and $L$ is the linear dimension of the corresponding tissue (middle), and signal degradation rates $k_\mathrm{d}$ (bottom) from the biological literature. Blue marks are data from the fruit fly \textit{Drosophila melanogaster}, and orange marks are data from zebrafish, \textit{Danio rerio}. 
    Tissue deformation rates in the fruit fly were estimated from image time series of: egg chamber~\cite{Jia2016,Alegot2018}, dorsal thorax~\cite{Bosveld2012}, pupal wing~\cite{Etournay2015}, hindgut~\cite{Iwaki2001,Johansen2003}, renal tubules~\cite{Saxena2014}, and germ-band~\cite{Bertet2004}; and in zebrafish from mesodermal explants~\cite{Williams2020} and the tailbud~\cite{Steventon2016}. Signaling molecules: Decapentaplegic in the fruit fly~\cite{Kicheva2007,Wartlick2011,Romanova-Michaelides2022}; Cyclops, Squint, Lefty1 and Lefty2 in zebrafish~\cite{Muller2013,Wang2016}.} 
    \label{fig: model}
\end{figure}

In this article, we study the stability of active tissue deformation when accounting for more realistic signaling dynamics (\autoref{fig: model}A).
We consider an active viscous tissue with local flow velocity $\vec{\vel}(\vec{r}, t)$, where active stresses are defined by the gradient of a signaling molecule concentration field $c(\vec{r}, t)$.
The signaling field $c(\vec{r}, t)$ is in turn secreted by a source region that is defined by a field $s(\vec{r}, t)$, which is conserved and only advects with tissue flows. The signaling field $c(\vec{r}, t)$ not only advects with tissue flows, but also diffuses and degrades at constant rates everywhere.  

Using linear stability analysis, we find that  gradient-contractile tissues are always unstable, consistent with our earlier findings \cite{Ibrahimi2023}. Yet, we also show that for our more realistic signaling dynamics, a gradient-contractile system can never be stabilized by diffusion.
We further show that gradient-extensile systems can be stable, in particular if the tissue is longer in the direction of the gradient than transversal to it.
Finally, we find that an additional instability can appear for both gradient-extensile and gradient-contractile cases, arising from the interaction between overall tissue shear and diffusion. 
This instability may point to the necessity of an additional control mechanism for the final tissue dimensions.

In the following, we first introduce the model in \autoref{sec:model}. 
We then study the case of a uniform gradient in both $s(\vec{r}, t)$ and $c(\vec{r}, t)$ for fixed system size in \autoref{sec:uniform state}, for which the linear stability can be computed analytically. 
Next, in \autoref{sec: fixed localized profile}, still for fixed system size, we study the more realistic case of a localized source $s(\vec{r}, t)$, which gives rise to a non-uniform signaling gradient $c(\vec{r}, t)$.
We then discuss the stability of freely deforming systems in \autoref{sec:deforming system}.
Finally, in \autoref{sec:discussion}, we discuss our findings in the context of the biological literature. 

\section{Model}
\label{sec:model}
\subsection{Hydrodynamic fields and bulk dynamics}
We describe the tissue as a 2D active material, where we consider three interacting hydrodynamic fields: a scalar field $s(\vec{r},t)$, which describes the source, i.e.\ cells in the tissue which secrete the signaling molecule; a scalar field $c(\vec{r},t)$, which describes the signaling molecule concentration; and a vector field $\vec{\vel}(\vec{r},t)$, which describes the tissue flows (\autoref{fig: model}A).
We find that literature values for the respective rates of tissue deformation $\tilde{v}^0_{xx}$, signaling molecule diffusion $D/L^2$, where $D$ is the diffusion coefficient and $L$ denotes the linear dimension of the tissue, and signaling molecule degradation $k_\mathrm{d}$ are of roughly similar orders of magnitude (\autoref{fig: model}B). This suggests that none of them can be easily neglected, and we thus include all three effects in our model.

The source field $s(\vec{r},t)$, which describes signaling molecule secretion, only advects with material flows:
\begin{equation} \label{eq:source field dynamics}
  \frac{\partial s}{\partial t} + \partial_i(\vel_is)  = 0,
\end{equation}
where $\partial_i$ denotes the partial derivative with respect to the spatial dimension $i \in \{x,y\}$. Here and in the following, we label spatial dimensions by Latin indices $i,j,\dots$ and adopt the Einstein notation of summing over repeated indices. 

The signaling field $c(\vec{r},t)$ follows a generalized conservation equation:
\begin{equation}
    \frac{\partial c}{\partial t} + \partial_i(\vel_ic) = k_{\mathrm{p}}s -k_{\mathrm{d}} c + D\partial_i^2c. \label{eq: constitutive signaling field}
\end{equation}
Here, we have introduced production rate $k_\mathrm{p}$, degradation rate $k_{\mathrm{d}}$, and diffusion coefficient $D$. 
Note that there are several intra- and extra-cellular mechanisms of signaling molecule transport \cite{Muller2013} -- here we describe the emergent \emph{effective} transport on large length scales by advection and diffusion.

Tissue flows $\vec{\vel}(\vec{r},t)$ are governed by overdamped, incompressible, viscous dynamics:
\begin{align}
  \partial_i\sigma_{ij} &= 0, \label{eq:force balance}\\
  \partial_i\vel_i &= 0, \label{eq:incompressibility}
\end{align}
where the stress tensor is given by
\begin{equation}
    \sigma_{ij} = 2\eta \tilde{v}_{ij} - \Pi\delta_{ij} + \tilde\sigma^a_{ij}.\label{eq:stress}
\end{equation}
Here, $\eta$ is the shear viscosity, $\tilde{v}_{ij}:=(\partial_i\vel_j+\partial_j\vel_i-\partial_k\vel_k\delta_{ij})/2$ is the shear rate tensor, $\Pi$ is the hydrostatic pressure, and we use the following expression for the active stress $\tilde\sigma^a_{ij}$:
\begin{equation}
    \tilde\sigma^a_{ij} = \alpha \left[ (\partial_i c) (\partial_j c) -\frac{1}{2}(\partial_kc)^2\delta_{ij} \right]. \label{eq:active stress}
\end{equation}
This expression for the active anisotropic stress is the same as in Active Model H \cite{Tiribocchi2015a,Kirkpatrick2019a}. Its direction is determined by the sign of the coefficient $\alpha$: For $\alpha < 0$, the activity is gradient-extensile, i.e.\ it acts to extend the material along the local $c$ gradient direction and contract perpendicularly; while for $\alpha>0$, the activity is gradient-contractile, i.e.\ it acts to contract the material along the $c$ gradient direction and extend perpendicularly (\autoref{fig: model}A).

The expression for the active stress, Eq.~\eqref{eq:active stress}, is motivated by what is known about developing biological tissues.  Specifically, anisotropic internal stresses are usually generated by an anisotropic distribution of cytoskeletal elements within cells.  Such an anisotropic distribution is created by cell polarity, which in turn can be controlled by signaling gradients \cite{Bosveld2012,Gao2012,Merkel2014,Saxena2014,Lavalou2021}.
Rather than explicitly including cell polarity as polar field \cite{Ibrahimi2023}, we assume here for simplicity a direct coupling of active stresses to the signaling gradient.
This corresponds to the limit where the coupling of active stresses to the signaling gradient is fast compared to the other time scales of the system.
While this is clearly a simplification, we do not expect the fundamental results of the analysis to differ much in this limit, as suggested by our earlier work \cite{Ibrahimi2023}.
Note that Eq.~\eqref{eq:active stress} represents the lowest-order term for a coupling of an active, anisotropic stress to a scalar field \cite{Tiribocchi2015a}.

\subsection{Boundary Conditions}
We use periodic boundary conditions, where the periodic box has generally time-dependent dimensions $L_x(t)\times L_y(t)$ and constant area, $L_x(t)L_y(t)=\mathrm{const}$.
Throughout this article, we always use one of the following two boundary conditions:
\begin{enumerate}
    \item We fix the system dimensions $L_x$ and $L_y$, or
    \item we allow the system to freely deform by fixing the externally applied stress to zero, which results in a pure shear deformation of the periodic box.  Integration of Eq.~\eqref{eq:stress} implies that the instantaneous box shear rate is 
    \begin{equation}
        \frac{1}{L_x}\,\frac{\d L_x}{\d t} = -\frac{\alpha}{4\eta L_xL_y}\int{\big[ (\partial_xc)^2-(\partial_yc)^2 \big]\;\d A},\label{eq:shear rate}
    \end{equation}
    where the integral is over the whole periodic box.
\end{enumerate}
We apply fixed system dimensions in \autoref{sec:uniform state} and \autoref{sec: fixed localized profile}, and study a freely deforming system in \autoref{sec:deforming system}.

We investigate the stability of two stationary states: (i) a state where a linear source profile $s$ gives rise to a linear signaling profile $c$, and (ii) a state where a localized source $s$ generates a spatially decaying signaling field $c$. 
In both cases, the amplitude of the source field is defined by a constant $s_b$.
In case (i), which we study in \autoref{sec:uniform state}, the boundary conditions are periodic with the exception of the vertical boundaries for the scalar fields $s$ and $c$, where we set for all $y\in[0,L_y)$:
\begin{align}
    s(0,y) &= s(L_x,y) - s_b\label{eq:BC s}, \\
    c(0,y) &= c(L_x,y) - \frac{k_{\mathrm{p}}}{k_{\mathrm{d}}}s_b\label{eq:BC c},
\end{align}
with fixed $s_b$. These modified conditions remove the discontinuity that would otherwise appear for linear profiles in $s$ and $c$ at the boundary, and thus ensure that linear $s$ and $c$ profiles can be stationary \cite{Ibrahimi2023}. In case (ii), which we study in \autoref{sec: fixed localized profile} and \autoref{sec:deforming system}, we use standard periodic boundary conditions without offset.

\subsection{Dimensionless equations}
\label{sec:dimensionless units}
We nondimensionalize the set of equations Eqs.~\eqref{eq:source field dynamics}--\eqref{eq:BC c} in the following way:
\begin{itemize}
  \item We choose the unit of length such that the system area is one, $L_xL_y=1$.
  \item We choose the unit of the signaling field $c$ such that $k_\mathrm{p}=k_\mathrm{d}$.
  \item We choose the units of time and of the source field $s$, such that the free deformation rate of a squared system is 1. To this end, we set $\vert\alpha\vert/4\eta=1$, and in case (i) of a linear gradient, we also set $s_b=1$, while in case (ii) of a non-uniform gradient, we fix $s_b$ as described in \autoref{sec: fixed localized profile} (\aref{app:value of sb}).
\end{itemize}
The last condition allows to compare the linear stability of systems with different source profiles, because the reference time scale is always the respective free deformation rate.

Rescaled accordingly, the dimensionless equations read:
\begin{align}
  \frac{\partial s}{\partial t} +\vel_i\partial_i s &= 0, \label{eq:dimless s}\\
  \frac{\partial c}{\partial t} +\vel_i\partial_i c &= \left( D\partial_i^2 - k_{\mathrm{d}} \right) c + k_{\mathrm{d}}s, \label{eq:dimless c}\\
  0 &= \frac{1}{4}\partial_i^2\vel_j - \partial_j\Pi' + \sgn{\alpha}\,\partial_i \big[ (\partial_i c)(\partial_j c) \big], \label{eq:dimless v}\\
  \partial_i\vel_i &= 0. \label{eq:dimless incompressibility}
\end{align}
Here, we have introduced the sign function, $\sgn{\alpha}:=\alpha/ \vert\alpha\vert$, and have set $\Pi':=\Pi+\sgn{\alpha}(\partial_k c)^2/2$. The offset introduced at the boundary conditions in Eqs.~\eqref{eq:BC s} and \eqref{eq:BC c} is rescaled to $1$ for both scalar fields.

\section{Linear source profile} \label{sec:uniform state}
\begin{figure}
    \centering
    \includegraphics[width = \columnwidth]{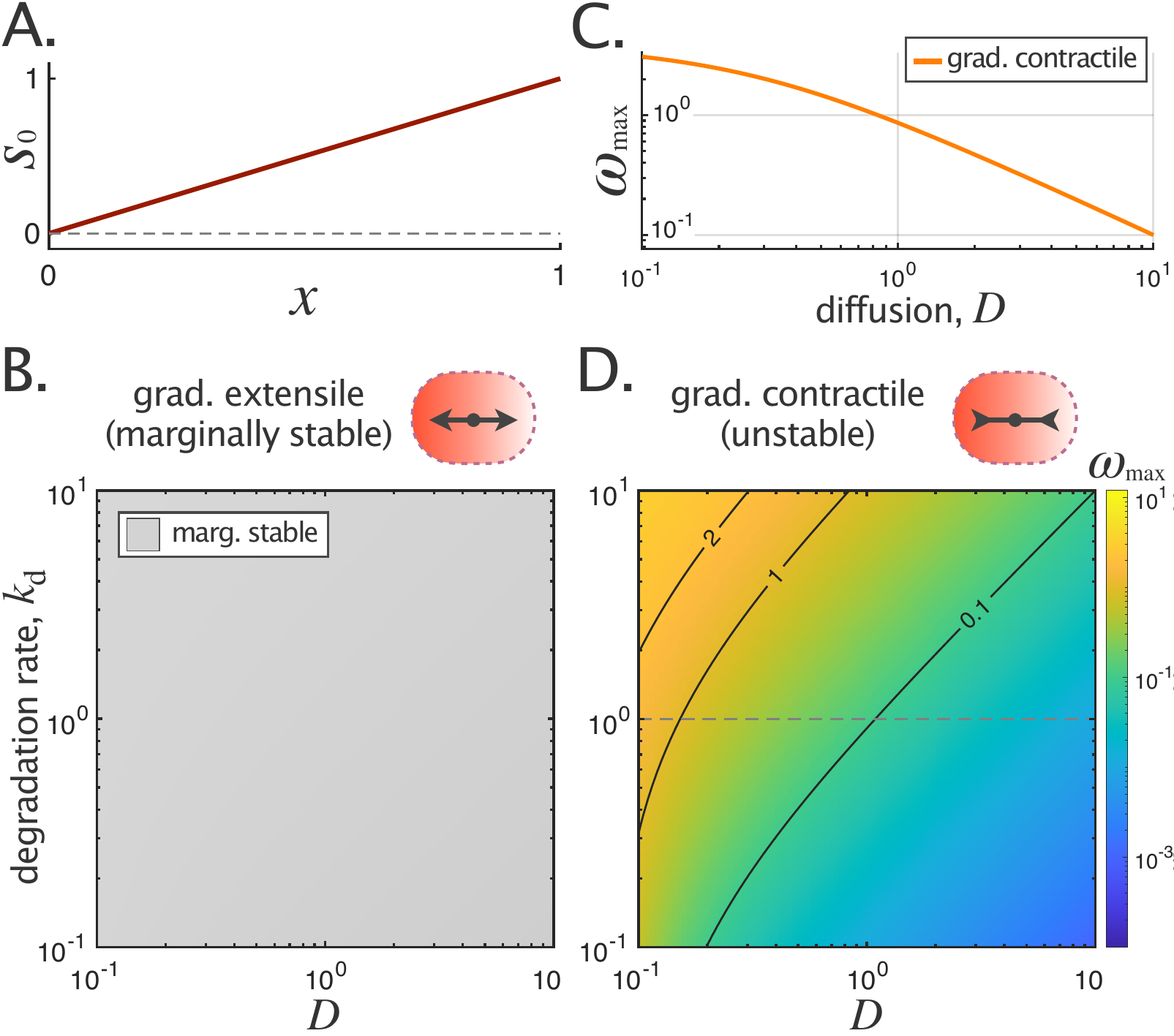}
    \caption{Stability of a fixed-size system with a linear source profile: gradient-extensile systems are marginally stable, whereas gradient-contractile systems are unstable.
    (A) The linear source profile, $s_0(x)=x$ in dimensionless units. In the stationary state, $c_0(x)=s_0(x)$.
    (B) Maximal perturbation growth rates $\omega_{\mathrm{max}}$ in the $D$--$k_\mathrm{d}$ parameter space show that gradient-extensile systems are marginally stable, i.e.\ $\omega_\mathrm{max}=0$, highlighted in light gray.
    (C) $\omega_{\mathrm{max}}$ as a function of signaling molecule diffusion $D$ for a gradient-contractile system with $k_\mathrm{d}=1$, plotted in log-log scaling. 
    (D) $\omega_{\mathrm{max}}$ in the $D$--$k_\mathrm{d}$ parameter space show that gradient-contractile systems are always unstable. The dashed gray line corresponds to the $k_\mathrm{d}=1$ curve plotted in panel C. Panels B-D were created from the result in \aref{app:linear source general},  where the perturbation wave vector $\bm{q}_\mathrm{min} = (0, 2\pi)$ in dimensionless units was used to maximize the perturbation growth rate.
    } 
    \label{fig: uniform}
\end{figure}

To build intuition, we study for fixed box dimensions $L_x=L_y=1$ the linear stability of a stationary state with linear spatial profiles in $s$ and $c$ (\autoref{fig: uniform}A), given by:
\begin{align}
    s_0 &= x, &
    c_0 &= x, &
    \vec{\vel}_0 &= \vec{0}. \label{eq: stat linear profile}
\end{align}
There are no flows in the stationary state, because the gradient of $c$ is homogeneous, and thus the active stress is homogeneous across the system of fixed dimensions.

In this case, linear stability can be computed analytically.
When we add a small perturbation around this state,
\begin{align}
    s &= x + \delta s, &
    c &= x + \delta c, &
    \vec{\vel} &= \vec{\delta\vel},
\end{align}
and linearize the dynamics, only constant prefactors occur. As a consequence the solutions in $(\delta s, \delta c, \vec{\delta\vel})$ are spatial Fourier modes with wave vectors $\vec{q} = q(\cos{\phi},\sin{\phi})$, whose amplitudes grow or shrink exponentially with time $t$ (\aref{app:linear source}).

We first discuss the limit without diffusion, $D=0$.
When we subtract the linearized versions of source and signaling dynamics, Eqs.~\eqref{eq:dimless s} and \eqref{eq:dimless c}, we obtain (\aref{app:linearized dynamics}):
\begin{equation}
    \frac{\partial}{\partial t}\big(\delta c - \delta s\big) = -k_\mathrm{d}\big(\delta c - \delta s\big).
\end{equation}
Thus, any difference between signaling field $c$ and source field $s$ decays at rate $k_\mathrm{d}$.  In addition, the combination of signaling dynamics and generated active flows, Eqs.~\eqref{eq:dimless c}--\eqref{eq:dimless incompressibility}, leads to the same results as for a single scalar field discussed in earlier publications \cite{Kirkpatrick2019a,Ibrahimi2023} (\aref{app:linear source, no diffusion}). 
Specifically, for the combined perturbation field $\delta c_\mathrm{eff}:=\delta c+k_\mathrm{d}/\omega_\mathrm{act}\delta s$, we obtain (Eq.~\eqref{eq:app:ceff} in \aref{app:linear source}):
\begin{equation}
	\frac{\partial\delta c_\mathrm{eff}}{\partial t} = \omega_\mathrm{act}\,\delta c_\mathrm{eff},\label{eq:ceff}
\end{equation}
where $\omega_\mathrm{act}:=4\,\sgn{\alpha}\sin^2{\phi}$ is the perturbation growth rate created by the active flows. It takes the same form without source field \cite{Ibrahimi2023}.
Thus, without diffusion, the system is marginally stable in the gradient-extensile case, i.e.\ the maximal perturbation growth rate is zero, and unstable in the gradient-contractile case.

In the presence of diffusion, $D>0$, gradient-extensile systems remain marginally stable (\autoref{fig: uniform}B). However, gradient-contractile systems can \emph{not} be stabilized by diffusion, even though the maximal perturbation growth rate decreases with diffusion (\autoref{fig: uniform}C,D).  This is in contrast to earlier findings for a system with a boundary-imposed signaling gradient, where we found that diffusion \emph{can} stabilize gradient-contractile systems \cite{Ibrahimi2023}.  To understand where the difference comes from, we consider the limit of very large $D$ (the general case is discussed in \aref{app:linear source}). The linearized version of the signaling dynamics, Eq.~\eqref{eq:dimless c}, in Fourier space reads:
\begin{equation}
    \frac{\partial \delta c}{\partial t} + \delta \vel_x = -\left( Dq^2 + k_{\mathrm{d}} \right) \delta c + k_{\mathrm{d}}\delta s. \label{eq:c linearized Fourier}
\end{equation}
The flow field $\vec{\delta \vel}$ enters on the left-hand side due to advection, where $\partial_xc_0=1$ in dimensionless units.  The flow $\vec{\delta \vel}$ is actively created by the signaling field itself through Eqs.~\eqref{eq:dimless v} and \eqref{eq:dimless incompressibility}, and its $x$ component in Fourier space reads: $\delta \vel_x = -\omega_\mathrm{act}\delta c$, with the active flow time scale $\omega_\mathrm{act}$ introduced before.
Denoting the minimal non-zero wave vector magnitude in our system by $q_\mathrm{min}$, we take the limit where $Dq_\mathrm{min}^2$ is much larger than all other relevant time scales in the system. In this case, according to Eq.~\eqref{eq:c linearized Fourier}, the Fourier amplitude $\delta c$ for each wave vector relaxes adiabatically fast towards
\begin{equation}
     \delta c \simeq \frac{k_\mathrm{d}}{Dq^2} \delta s.\label{eq:adiabatic limit - large diffusion}
\end{equation}
This effectively corresponds to a system where we again just have a single scalar field $s$, which advects with active tissue flows $\vec{\vel}$, but does \emph{not} diffuse. The active tissue flows are generated by the perturbations of $c$, and their $x$ component is given by $\delta \vel_x = -\omega_\mathrm{act}\delta c\simeq -(\omega_\mathrm{act}k_\mathrm{d}/Dq^2)\delta s$. Inserting this into the linearized version of the source dynamics in Fourier space, we obtain (compare to Eq.~\eqref{eq:ceff}):
\begin{equation}
    \frac{\partial \delta s}{\partial t} = -\delta\vel_x = \frac{\omega_\mathrm{act}k_\mathrm{d}}{Dq^2}\delta s.\label{eq:large-D-limit}
\end{equation}
Thus, the perturbation growth rate is given by the growth rate without diffusion, $\omega_\mathrm{act}$, multiplied by $k_\mathrm{d}/Dq^2$ (\autoref{fig: uniform}C). As a consequence, gradient-contractile systems are unstable even in the limit of a large diffusion constant.  The reason for this is that the source field itself advects with the active flows, but does not diffuse.
For the same reason, gradient-extensile systems remain only marginally stable even in the presence of signaling molecule diffusion.

\section{Localized source profile} 
\label{sec: fixed localized profile}
We next discuss how the linear stability changes when the source profile is not linear but is instead localized to some region of finite width $\w$. This is motivated by what we know about developing biological tissues \cite{Wolpert2015,Wartlick2009,Wartlick2011,Benton2016}. In this section, we discuss a fixed-size system $L_x=L_y=1$, while in the next one (\autoref{sec:deforming system}) we discuss a freely deforming system.

\begin{figure*}
    \centering
    \includegraphics[width = 0.9\textwidth]{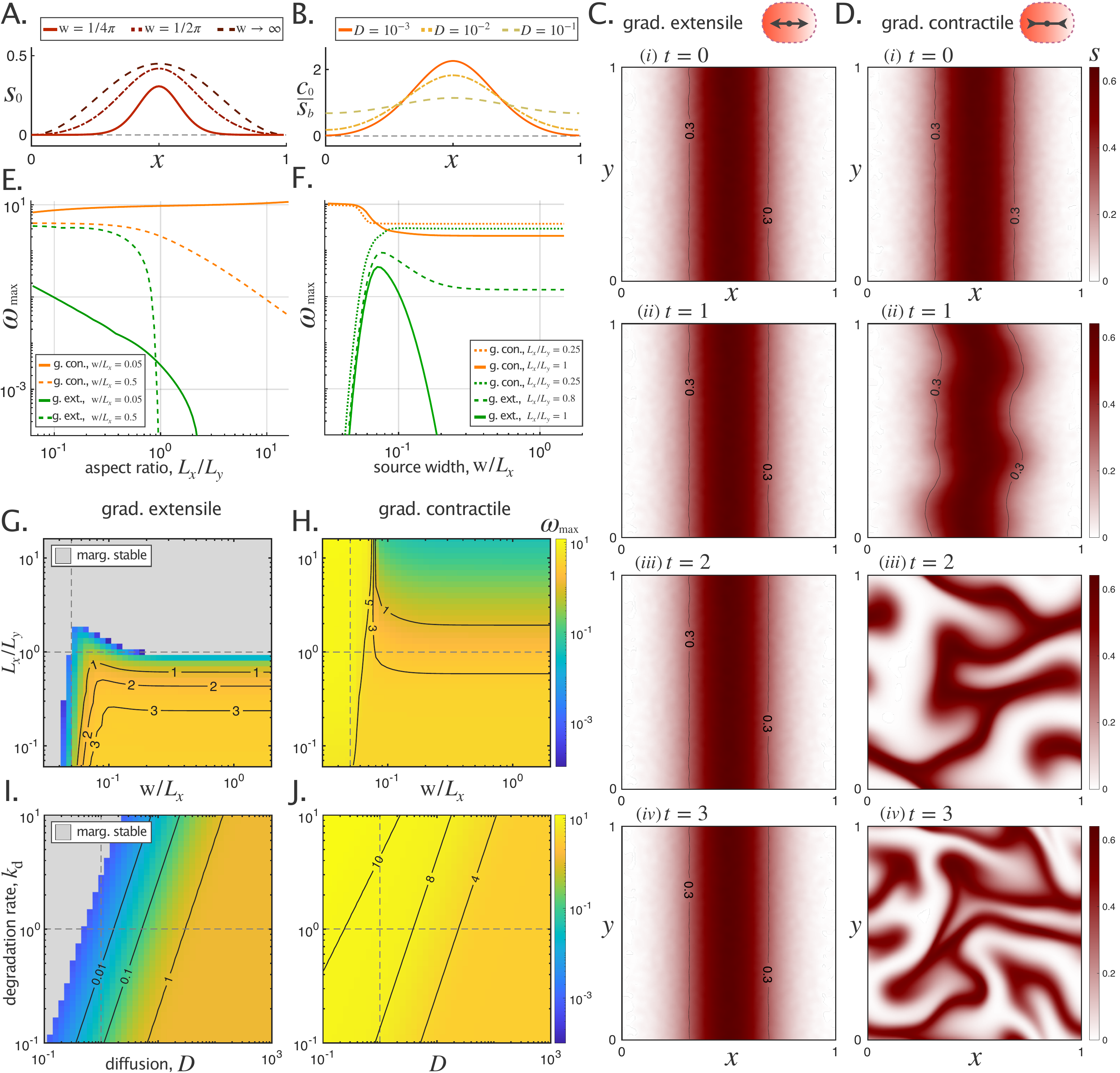}
    \caption{Stability of a fixed-size system with a localized source profile. 
    (A) The localized source profile $s_0(x)$ from Eq.~\eqref{eq:nonlinear source} for different values of the source width $\w$.
    (B) The localized signaling profile for different values of diffusion $D$. Here, $\w = 1/2\pi$ in all cases.
    (C,D) Snapshots of the source profile from numerical simulations of the perturbed stationary state (with $\w = 1/2\pi, D = 10^{-2}, k_{\mathrm{d}}=1$). While the source profile appears to be maintained in the (C) gradient-extensile case, it gets distorted in the (D) gradient-contractile case.
    (E) Effect of aspect ratio $L_x/L_y$ on maximal perturbation growth rates $\omega_\mathrm{max}$ for different source widths. Gradient-contractile systems (orange) are always unstable, while gradient-extensile systems (green) have in some regions an $\omega_\mathrm{max}$ that can numerically not be distinguished from zero. Parameter values: $k_\mathrm{d} = 1$ and $D = 1$.
    (F) Effect of the source width $\w/L_x$ on maximal perturbation growth rate for different  box aspect ratios.  Parameter values: $k_\mathrm{d} = 1$ and $D = 1$.
    (G,H) Color plots of maximal perturbation growth rates for gradient-extensile (G) and gradient-contractile (H) couplings with varying source widths (horizontal axis) and aspect ratios (vertical axis). Marginally stable regions ($\omega_\mathrm{max}=0$) are highlighted in light gray. Solid black curves represent contours, whereas dashed gray lines corresponds to solid curves in panels E,F. Parameter values: $k_\mathrm{d} = 1$ and $D = 1$.
    (I,J) Maximal perturbation growth rates in the $D$--$k_{\mathrm{d}}$ parameter space for gradient-extensile (I) and gradient-contractile (J) couplings. Numerically marginally stable regions are highlighted in light gray. Parameter values: $L_x/L_y = 1$ and $\w/L_x = 0.05$.
    } 
    \label{fig: fixed nonuniform}
\end{figure*}

As initial state, we use a source profile defined by a \emph{von Mises} distribution, which is a generalization of a Gaussian to periodic boundary conditions:
\begin{equation}
    s_0(\vec{r}) = s_b\Bigg[\exp{\bigg(\frac{\cos{(q_0x)}}{(q_0\w)^2}\bigg)}-\exp{\bigg(-\frac{1}{(q_0\w)^2}\bigg)}\Bigg] \label{eq:nonlinear source}
\end{equation}
with $q_0:=2\pi/L_x$.
Here, $\w$ is a parameter that adjusts the width of the source region, the second term in Eq.~\eqref{eq:nonlinear source} ensures that $s_0(\vec{r})$ is zero at its minimum, and $s_b$ is a prefactor whose value is fixed such that the box deformation rate is normalized to one (Eq.~\eqref{eq:s_b} in \aref{app:value of sb}).
As initial signaling field, we choose the $c$ field that is stationary in the absence of flows for the source given by Eq.~\eqref{eq:nonlinear source}. In Fourier space:
\begin{equation}
    c_0(\vec{q}) = \frac{k_{\mathrm{d}}}{k_{\mathrm{d}} + Dq^2}s_0(\vec{q}). \label{eq: localized stat c}
\end{equation}
Thus, for $D=0$, signaling and source profiles coincide, and examples for $s_0(x)=c_0(x)$ for different $\w$ values are shown in \autoref{fig: fixed nonuniform}A. Meanwhile, finite diffusion $D>0$ smoothens the signaling profile $c_0(x)$ as compared to the source profile $s_0(x)$ (\autoref{fig: fixed nonuniform}B). 

In this initial state, there are no flows in the system, $\vec{\vel}_0 = \vec{0}$. This is because the active stress in this case is parallel to the $x$ axis and homogeneous along the $y$ axis, because of incompressibility, and because of the fixed system size, which together imply that there are no Stokes flows (\aref{app:no non-affine flows}).

Numerical simulations of the full dynamics, Eqs.~\eqref{eq:dimless s}--\eqref{eq:dimless incompressibility}, are shown in \autoref{fig: fixed nonuniform}C,D and in Movies S1, S2 (shown is the source field) \footnote{See Supplemental Material at \url{https://arxiv.org/abs/2412.15774} for movies showing the simulations.}. We find that with chosen parameters ($\mathrm{w}=1/2\pi$, $k_\mathrm{d}=1$, and $D=10^{-2}$), the gradient-extensile system appears to be stable until simulation time $t=3$ (\autoref{fig: fixed nonuniform}C), yet an instability becomes apparent at much later times (\autoref{fig: extensile perturbation}A in \aref{app:fixed system nonuniform} and Movie S1 \cite{Note1}).
Meanwhile, the gradient-contractile system shown an instability already early on (\autoref{fig: fixed nonuniform}D and Movie S2 \cite{Note1}).

For a more systematic characterization of the system's stability, we perform again a linear stability analysis.
While the eigenmodes of the linearized dynamics still grow or shrink exponentially with time, they can only be represented by a superposition of several spatial Fourier modes (\aref{app:linearized dynamics, Fourier space}).
We thus compute the perturbation eigenmodes and their respective growth rates numerically (\aref{app:numerical solution linearized dynamics}).
Note that due to the numerical Fourier-space convolutions, we also find aliasing effect arising from a finite spatial discretization, which we treat using standard methods in the field (\aref{app:numerical solution linearized dynamics}).

We find again that gradient-contractile systems are always linearly unstable (\autoref{fig: fixed nonuniform}E,F,H,J). 
Specifically, systems are more unstable when the source width is below $\w\lesssim 0.06L_x$ (\autoref{fig: fixed nonuniform}F,H).
Moreover, increasing the diffusion coefficient $D$ decreases the perturbation growth rate, but it never reaches zero.  Instead, in contrast to the linear gradient (\autoref{sec:uniform state}), the maximum growth rate converges to a finite constant in the limit of large diffusion, $D\rightarrow\infty$ (\autoref{fig: large-diffusion limit}, \aref{app:localized source convergence perturbation growth rate}).
It converges to a finite constant, because we non-dimensionalize rates with respect to the free deformation rate, which also decreases as diffusion $D$ becomes larger due to an increased smoothing of $c$ (\aref{app:localized source convergence perturbation growth rate}). This is different from the linear-gradient scenario (\autoref{sec:uniform state}), where a constant signaling gradient, and thus a constant free deformation rate is imposed by the boundary conditions.

For gradient-extensile systems, the situation is more complex than for a linear gradient. Specifically, we find that the stability depends on the aspect ratio of the system. For aspect ratios $L_x/L_y\lesssim 2\dots3$, the system can be unstable, while it is numerically marginally stable for aspect ratios above (\autoref{fig: fixed nonuniform}E,G).
Here, by ``numerically marginally stable'' we mean that we numerically cannot distinguish between marginally stable and numerically small, but finite growth rates -- indeed, this distinction is also irrelevant in practice for a real biological system.
For aspect ratios between $\approx 1$ and $2\dots3$, the system is effectively unstable at intermediate source widths $\mathrm{w}/L_x$ between $0.04$ and $0.2$ and otherwise numerically marginally stable (light gray regions in \autoref{fig: fixed nonuniform}G). 
For even lower aspect ratios, we only observe numerical marginal stability for very narrow sources with $\mathrm{w}/L_x\lesssim0.04$ (\autoref{fig: fixed nonuniform}F,G).
The type of perturbation creating the most unstable mode appears to be qualitatively always the same: recirculating active flows create spatial variations of the source region width, which create corresponding modulations in the signaling field, leading to the recirculating active flows (\autoref{fig: extensile perturbation}A,B).
Finally, for given aspect ratio and source width, the system is generally more stable for small signaling length scales $\sqrt{D/k_\mathrm{d}}$ (\autoref{fig: fixed nonuniform}I).
In the large-diffusion limit, the maximum perturbation growth rate converges to a constant for the same reason as in the gradient-contractile case (\autoref{fig: large-diffusion limit}, \aref{app:localized source convergence perturbation growth rate}).

Note that even if gradient-extensile systems are formally unstable in a substantial part of the parameter space, for aspect ratios close to one the maximal perturbation growth rates can be several orders of magnitude smaller as compared to the gradient-contractile case. For instance, the maximal growth rate for the example simulation in \autoref{fig: fixed nonuniform}C, \autoref{fig: extensile perturbation}A, and Movie S1 \cite{Note1} is positive, but very small, $\omega_\mathrm{max}=1.9\times 10^{-3}$ (as compared to $\omega_\mathrm{max}=4.2$ in the gradient-contractile example in \autoref{fig: fixed nonuniform}D).

Taken together, the gradient-extensile system is effectively stable when it is more elongated in the direction of the gradient, or when the signaling length scale $\sqrt{D/k_\mathrm{d}}$ is small.


\section{Freely deforming system}
\label{sec:deforming system}
Here we study the linear stability of a system with a localized source $s_0$, Eq.~\eqref{eq:nonlinear source}, under free deformation of the periodic box. 

\begin{figure}
	\centering
	\includegraphics[width = 0.93\linewidth]{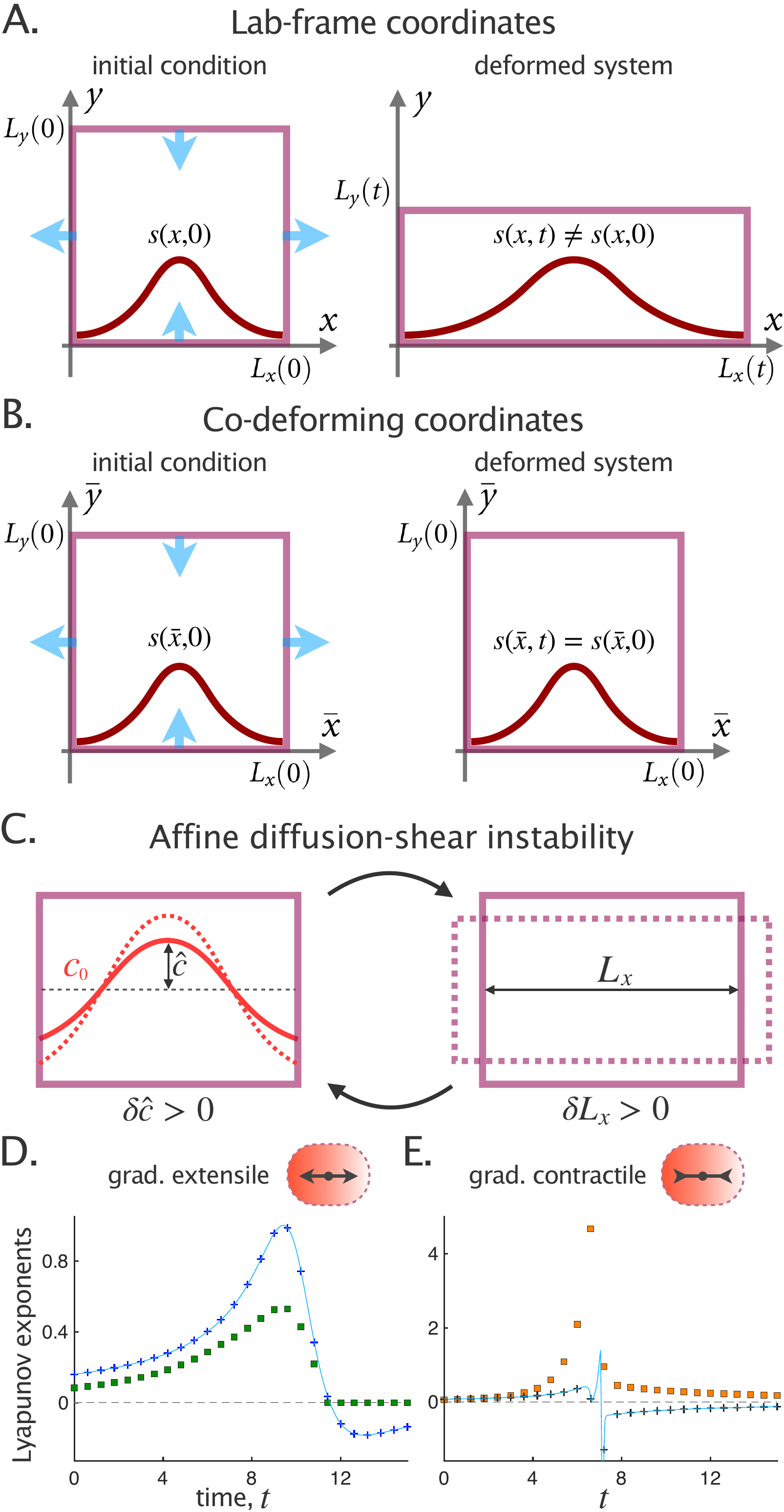}
	\caption{
		Stability of a freely deforming system with a diffusive signaling molecule, studied in co-deforming coordinates.
		(A) In lab-frame coordinates, a localized source $s(\vec{r}, t)$ is not stationary when the system deforms affinely, since its width changes proportionally with the system dimension $L_x$ due to advection.
		(B) However, in co-deforming coordinates, Eqs.~\eqref{eq: codef x} and \eqref{eq: codef y}, the localized source $s(\vec{\bar{r}}, t)$ remains stationary under affine deformation.
		(C) Illustration of the affine diffusion-shear instability:  diffusion reduces the amplitude $\hat{c}$ of the signaling field, but a larger $L_x$ decreases the effective diffusion coefficient observed in co-deforming coordinates. Thus, a larger $L_x$ leads to a larger $\hat{c}$, which in turn increases active shear and thus $L_x$.
		(D) Local Lyapunov exponents (LLE) for the gradient-extensile system over time. 
		Blue crosses indicate the numerically computed LLE of the affine diffusion-shear perturbation with the largest growth rate, and the light blue curve displays the analytical prediction according to Eq.~\eqref{eq:affine ds instability, growth rate} in \aref{app:localized source, affine diffusion-shear instability, cosine}.
		Green squares denote the maximal LLE of all other perturbations. 
		(E) The same results as in panel D, but for the gradient-contractile system. Specifically, black crosses indicate the LLE of the affine diffusion-shear perturbation with the largest growth rate, and orange squares denote the maximal LLE of all other perturbations.
	}
	\label{fig: deforming nonuniform}
\end{figure}

Even for homogeneous shear deformation, the source profile $s_0$ is not stationary, but becomes distorted due to advection with the box deformation (\autoref{fig: deforming nonuniform}A).
To address this, following our earlier work \cite{Ibrahimi2023}, we define the co-deforming coordinates, $\vec{\bar{r}}=(\bar{x},\bar{y})$, based on the lab-frame coordinates $\vec{r}=(x,y)$ by rescaling both axes:
\begin{align}
\bar{x} &:= \frac{L_x(0)}{L_x(t)}x = l_x^{-1}(t)x \label{eq: codef x} \\ 
\bar{y} &:= \frac{L_y(0)}{L_y(t)}y = l_x(t)y, \label{eq: codef y}
\end{align}
where $l_x(t):=L_x(t)/L_x(0)$ is the box shear.
Moreover, we also introduce a flow field in co-deforming coordinates, $\vec{\bar{\vel}}(\vec{\bar{r}})$, which corresponds to non-affine flows that may occur in addition to the pure shear deformation of the periodic box (\aref{app:co-deforming velocity}). Thus, for $\vec{\bar{\vel}}=\vec{0}$ the system deforms affinely, and the source field $s_0(\vec{\bar{r}})$ is stationary with respect to the co-deforming coordinates (\autoref{fig: deforming nonuniform}B). 

Yet, in the general case of a finite diffusion constant, the signaling field $c$ does not reach a stationary state -- not even in co-deforming coordinates.
This is because diffusion as observed in co-deforming coordinates depends on box shear $l_x$ through the transformation rules in Eqs.~\eqref{eq: codef x} and \eqref{eq: codef y}. 
Thus, as a consequence of the permanently changing $l_x$, there is no stationary state in the freely deforming case.
Hence, no standard linear stability analysis is possible -- we instead analyze the local Lyapunov exponents of the system.

To compute the local Lyapunov exponents, we consider the system state by combining source and signaling profiles in co-deforming coordinates with box shear, $\vv{z}(t)=[s(\vec{\bar{r}},t),c(\vec{\bar{r}},t),l_x(t)]$. This state constantly evolves through Eqs.~\eqref{eq:shear rate} and \eqref{eq:dimless s}--\eqref{eq:dimless incompressibility}, which we summarize here as $\d \vv{z}/\d t = \vv{F}(\vv{z})$.
As initial condition, we use the source profile $s_0$ from Eq.~\eqref{eq:nonlinear source}, and the corresponding signaling profile $c_0$ that \emph{would} be stationary in a fixed system, Eq.~\eqref{eq: localized stat c}.
To simplify our discussion here, we will focus on the limit $\w\rightarrow\infty$, where the source $s_0(\bm{r})$ given by Eq.~\eqref{eq:nonlinear source} corresponds to a cosine profile, $s_0(\bm{r})\sim\cos{(2\pi x/L_x)}+1$. The corresponding signaling profile is then also a cosine profile. 
We discuss the unperturbed box dimension dynamics, $l_x(t)$, in \aref{app:localized source, system size dynamics}, and focus here only on the perturbation dynamics.


Local Lyapunov exponents characterize whether and how quickly two slightly different solutions, $\vv{z}$ and $\vv{z}'$, converge or diverge.
The difference between both, $\delta \vv{z} :=\vv{z}'-\vv{z}$, follows the dynamics $\d \delta \vv{z}/\d t = J(\vv{z})\cdot\delta \vv{z}$, where $J(\vv{z})$ is the Jacobian of $\vv{F}(\vv{z})$. The local Lyapunov exponents at $\vv{z}$ are then given by the eigenvalue spectrum of $J(\vv{z})$.
When applied to a stationary state, this approach coincides with linear stability analysis.

In the limit of no diffusion, $D=0$ (previous section, \autoref{sec: fixed localized profile}), the linearized dynamics of source and signaling, Eqs.~\eqref{eq:gen lin source} and \eqref{eq:gen lin signaling} (\aref{app:linearized dynamics, real space}), do not depend on the perturbation of the box shear, $\delta l_x$.  Thus, the perturbation dynamics of $s$ and $c$ decouples from that of $l_x$, and most of the Lyapunov spectrum corresponds to the eigenvalue spectrum of the fixed system (\autoref{sec: fixed localized profile}).  The remaining eigenvalue of the local Lyapunov spectrum is given by the system size perturbation dynamics:
\begin{equation}
	\frac{\d \delta l_x}{\d t} = \frac{\sgn{\alpha}}{l_x^2(t)L_x^2(0)}\delta l_x.
\end{equation}
The prefactor in this equation is positive for gradient-contractile systems and negative for gradient-extensile systems.
Hence, for $D=0$, the gradient-contractile system has positive local Lyapunov exponents, i.e.\ the system dynamics strongly depends on the initial conditions. Meanwhile, for gradient-extensile systems, the maximal local Lyapunov exponent is given by the maximal eigenvalue for the fixed-size system (\autoref{sec: fixed localized profile}).

\subsection{Affine diffusion-shear instability}

For finite diffusion, $D>0$, the equations defining the linearized dynamics, Eqs.~\eqref{eq:gen lin source}--\eqref{eq:gen lin size}, are all coupled to each other and the full Jacobian needs to be diagonalized in order to obtain the local Lyapunov exponents. At each time point of the trajectory $\vv{z}(t)$, we plot the maximal Lyapunov exponent in \autoref{fig: deforming nonuniform}D,E.

We first discuss the gradient-extensile case, where we find that initially, perturbation growth rates can be positive (blue crosses and green squares in \autoref{fig: deforming nonuniform}D). However, at later times, as the system shears and $l_x$ grows sufficiently, all perturbation growth rates become maximally zero (\autoref{fig: deforming nonuniform}D).
Initially, i.e.\ at low $l_x$, we found that the fastest-growing perturbation mode corresponds to a new kind of instability that we call ``affine diffusion-shear instability'' (blue crosses in \autoref{fig: deforming nonuniform}D). The nature of this instability is quite different from the active scalar matter instability \cite{Kirkpatrick2019a,Ibrahimi2023} -- specifically, it affects neither the non-affine part of the flow field, $\bm{\bar{\vel}}$ nor the source field $s$.
Moreover, the signaling profile keeps its cosine shape, and only its amplitude $\hat{c}$ and the box shear $l_x$ are affected by the perturbation. The instability arises because diffusion diminishes the amplitude of the signaling cosine profile, $\hat{c}$. Thus, increasing $l_x$ by some positive amount $\delta l_x$ decreases the effect of diffusion, which acts to increase $\hat{c}$, which in turn increases the active stresses making $l_x$ grow further (\autoref{fig: deforming nonuniform}C). 
In \aref{app:localized source, affine diffusion-shear instability, cosine}, Eq.~\eqref{eq:affine ds instability, growth rate}, we obtain an analytical prediction for the growth rate of this instability (light blue curve in \autoref{fig: deforming nonuniform}D), which matches our numerical results (blue crosses). 
From our analytical prediction, we find that this instability appears as long as the diffusion length $\lambda:=\sqrt{D/k_\mathrm{d}}$ is larger than $L_x$, or more precisely $\lambda>L_x/(2\pi\sqrt{3})$ (\aref{app:localized source, affine diffusion-shear instability, cosine}).
Remarkably, this instability does not disappear even when introducing an arbitrary non-linearity into the active stress (\aref{app:non-linear active stress}).

For the gradient-contractile case, we find several modes with a positive growth rate (orange squares in \autoref{fig: deforming nonuniform}E). Moreover, we again have an eigenmode of the Jacobian that is an affine diffusion-shear perturbation (black crosses in \autoref{fig: deforming nonuniform}E, compare to the predicted growth rate marked by the light blue curve). 

Finally, in \aref{app:localized source, affine diffusion-shear instability, general} we study the affine diffusion-shear instability for the general case of a \emph{finite} source width $\w$. We find a positive growth rate whenever the order of magnitude of the diffusion length becomes larger than the source width, $\lambda\gtrsim\w$ (\aref{app:localized source, affine diffusion-shear instability, general}).

\section{Discussion} \label{sec:discussion}
In developing animals, biological tissues can undergo active anisotropic deformation \cite{Wolpert2015}.  However, classical active matter results predict that such deformations are unstable under certain conditions \cite{Simha2002,Voituriez2005,Kirkpatrick2019a}.
In several systems, the local direction of active tissue deformation is controlled by the gradient of a signaling molecule concentration field \cite{Bosveld2012,Lavalou2021,Saxena2014}.
Such signaling molecules are typically produced by the cells in a localized source region of the tissue, and they can form gradients through diffusion and degrade \cite{Wolpert2015}.
However, both signaling pattern and source region are also advected by the flows generated by the active deformation \cite{Lefebvre2022}.
Here, we study the stability of a system which includes source field, signaling field, and signaling-gradient-controlled active tissue deformation.
We find that systems where the active stresses are
gradient-contractile are always unstable, while gradient-extensile systems can be marginally stable under certain conditions.
Our findings are consistent with, and potentially explain, an observed abundance of gradient-extensile systems in the biological literature \cite{Johansen2003,Ninomiya2004,Gao2012,Benazeraf2010,Alegot2018,Saxena2014,Pare2014,Lienkamp2012, Ossipova2015,Williams2020} and the virtual absence of gradient-contractile systems (also discussed in \cite{Ibrahimi2022}), suggesting that the instability discussed here may act as an evolutionary selection criterion.

In our previous work, we studied a simplified signaling dynamics, where the overall gradient was imposed by the boundary conditions \cite{Ibrahimi2023}.
Here, we show that more realistic signaling dynamics, including secretion, diffusion, and degradation, lead to a number of important differences.
First, in the gradient-contractile case, neither degradation nor diffusion can stabilize the system.
While diffusion can decrease the growth rates of perturbations, we find that it never suffices to fully stabilize the system. This is because both signal and source fields are advected with the active flows, while only the signal diffuses.
In the case of a linear source profile, the maximal perturbation growth rate could become arbitrarily small for a sufficiently large diffusion coefficient. However, in the case of a localized source, the maximal perturbation growth rate converges to a finite multiple of the active shear rate in the limit of a large diffusion coefficient (\aref{app:localized source convergence perturbation growth rate}).

Second, we show that for a localized source region, gradient-extensile systems can become unstable whenever the tissue is too short in the direction of the gradient as compared to the transversal direction. Conversely, we numerically obtain marginal stability whenever the tissue is more elongated in the direction of the gradient or when the signaling lengths scale, $\sqrt{D/k_\mathrm{d}}$, is small. 
However, we also find that even in the unstable regime, the maximum growth rate of gradient-extensile systems are often orders of magnitude smaller than the active shear rate, in particular for aspect ratios close to one (Figs.~\ref{fig: fixed nonuniform}E-J). It is not impossible that during development, biological tissues transiently accept an instability with a small growth rate until entering a stable regime once they become more elongated.

Finally, in deforming systems, an ``affine diffusion-shear instability'' can appear. Different from the other instabilities discussed here, which are all related to the Simha-Ramaswamy instability, the affine diffusion-shear instability does not require non-affine tissue flows. Instead, it is only reflected in an interplay between the amplitude of the signaling molecule and the overall tissue shear rate.
This instability appears in tissues with a localized source whose width is smaller than the signaling diffusion length $\sqrt{D/k_\mathrm{d}}$. It occurs even when adding any non-linearity in the dependence of active stress on the signaling gradient (\aref{app:non-linear active stress}).
In many developing tissues, the signaling profile has a larger extent than the source region  \cite{Wolpert2015,Wartlick2011,Johansen2003}, and our work suggests that this instability should occur.
Although this instability does not perturb the \emph{patterns} of source, signaling molecule, or flow, it perturbs their amplitude and, most importantly, leads to a lack of control of the final tissue dimensions. This points to an additional feedback mechanism not included in our model that may be required to control the final tissue shape.

Our study further motivates experiments, and a key question is whether there are any gradient-contractile biological systems at all.
There are many systems where signaling molecules are known to be necessary for active anisotropic tissue deformation.
However, in most systems, it is not yet known whether the signal controls the directionality or only the magnitude of the active stresses.
For instance, there are still open debates about what controls the directionality of active stresses in what is probably the best-studied anisotropic tissue deformation process, which is germ band extension (GBE) in the fruit fly \emph{Drosophila melanogaster}.
Recent work on GBE proposed that a dorso-ventral signaling gradient modulates the \emph{magnitude} of junctional myosin, which plays the role of active stresses \cite{Lefebvre2024}. Yet, different mechanisms have been proposed for the control of the myosin anisotropy and its \emph{directionality}. Classically, these include anterior-posterior-oriented gradients \cite{Pare2014,Collinet2015}, which would correspond to a gradient-extensile coupling. But more recently, a group has proposed that a dorso-ventrally aligned tissue tension could provide a global mechanical orienting signal \cite{Lefebvre2022,Lefebvre2024}.
Thus, in many cases, more experiments are needed to cleanly disentangle the signals that merely modulate the amplitude of active stresses from those that control their direction. For instance, to modify the direction of the signaling pattern while keeping a similar overall magnitude, one could combine null-mutants with a graded over-expression \cite{Lavalou2021} or the external supply of a gradient, e.g.\ through signaling-molecule-soaked beads \cite{VonderHardt2007,Wolpert2015}.
We believe that systematically combining insights from active matter theory with experiments in this way will help decode the biological mechanisms underlying oriented tissue deformation and its robustness.

\begin{acknowledgments}
We thank the Centre Interdisciplinaire de Nanoscience de Marseille (CINaM) for providing office space.
The project leading to this publication has received funding from France 2030, the French Government program managed by the French National Research Agency (ANR-16-CONV-0001), and from the Excellence Initiative of Aix-Marseille University - A*MIDEX.
This project was also supported by the grant RobustTissue attributed to M.M.\ by the French National Research Agency (ANR-22-CE30-0039).
\end{acknowledgments}

\appendix
\section{Co-deforming coordinates}
\label{app: codef coord}

\subsection{Definition}
Following Ref.~\cite{Ibrahimi2023}, we introduce the co-deforming coordinate system, $(\bm{\bar r},\bar t)=(\bar x, \bar y, \bar t)$, to analytically solve the linearized dynamics of systems under pure shear deformation.

The co-deforming coordinates map to the lab coordinates $(\bm{r}, t)=(x, y, t)$ in the following way:
\begin{align}
  r_i &= s_{ij}(\bar{t})\bar{r}_j \label{eq:co_deforming r}\\
  t &= \bar{t}, \label{eq:co_deforming t}
\end{align}
where $\bm{s}(\bar{t})$ is a time-dependent shear tensor, given by
\begin{equation}
  \bm{s}(\bar{t}) = \begin{pmatrix}
                l_x(\bar{t}) & 0 \\ 0 & l_x^{-1}(\bar{t})
              \end{pmatrix}. \label{eq:shear tensor}
\end{equation}
Thus, while at some time $t$, lab coordinates range from $0\leq x<L_x(t)$ and $0\leq y<L_y(t)$, co-deforming coordinates map these affinely to the box dimensions at time zero, with $0\leq \bar{x}<L_x(0)$ and $0\leq \bar{y}<L_y(0)$.

Note that for fixed system dimensions, where $L_x(t)=L_x(0)$, and thus $l_x(t)=1$, both, co-deforming and lab coordinates coincide, $\bm{\bar{r}}\equiv\bm{r}$.

\subsection{Partial derivatives}
As a direct consequence of Eqs.~\eqref{eq:co_deforming r} and \eqref{eq:co_deforming t}, partial derivatives of some quantity $f$ transform as:
\begin{align}
  \bar\partial_jf &:= \frac{\partial f(\bm{\bar{r}}, \bar{t})}{\partial \bar{r}_j} = (\partial_if)s_{ij} \label{eq:trafo partial derivative r}\\
  \bar\partial_tf &:= \frac{\partial f(\bm{\bar{r}}, \bar{t})}{\partial \bar{t}} = \partial_tf + (\partial_if)\dot{s}_{ij}\bar{r}_j, \label{eq:trafo partial derivative t}
\end{align}
where $\partial_if := \partial f(\bm{r}, t)/\partial r_i$, $\partial_tf := \partial f(\bm{r}, t)/\partial t$, and $\dot{s}_{ij} := \d s_{ij}/\d t= \d s_{ij}/\d\bar{t}$. Thus, the partial time derivative in co-deforming coordinates, $\bar\partial_{{t}}f$, i.e.\ for fixed $\bm{\bar{r}}$, includes a term related to the box shear rate as compared to the partial time derivative with respect to lab coordinates.

\subsection{Velocity and velocity gradient}
\label{app:co-deforming velocity}
To obtain the mapping for the velocity field, we consider a tracer particle that is perfectly advected with the flows.  The velocity of that tracer particle corresponds to a total time derivative $\vel_i=\d r_i/\d t$, for which we obtain by insertion of Eq.~\eqref{eq:co_deforming r}:
\begin{equation}
	\vel_i = \dot{s}_{ij}\bar{r}_j + s_{ij}\bar{\vel}_j, \label{eq:co-deforming v}
\end{equation}
where $\bar\vel_i:=\d \bar{r}_i/\d t=\d \bar{r}_i/\d \bar{t}$ is the co-deforming velocity, with $\bar{r}(\bar{t})$ being the tracer trajectory in co-deforming coordinates.

The first term in Eq.~\eqref{eq:co-deforming v} corresponds to a motion due to the affine transformation according to box coordinates. Thus, $\bar\vel_i$ can be interpreted as the non-affine component of the flow field.
This can also be seen more explicitly by computing the velocity gradient from Eq.~\eqref{eq:co-deforming v}:
\begin{equation}
	\partial_i\vel_j = v_{ij}^0 + s^{-1}_{li}s_{jk}\bar\partial_l\bar{\vel}_k,
	\label{eq:co-deforming grad v}
\end{equation}
where $v^0_{ij} := s^{-1}_{ki}\dot{s}_{jk}$ is the average box deformation rate tensor. For $s_{ij}$ as defined in Eq.~\eqref{eq:shear tensor}, the box deformation rate tensor is:
\begin{equation}
	v^0_{ij}
	= \begin{pmatrix}
		\dot{l}_x/l_x & 0 \\ 0 & -\dot{l}_x/l_x
	\end{pmatrix}
	= \begin{pmatrix}
	\dot{L}_x/L_x & 0 \\ 0 & -\dot{L}_x/L_x
	\end{pmatrix}.
	\label{eq:box shear rate}
\end{equation}

\subsection{Total derivative}
To obtain a transformation formula for the convective derivative, we consider again our tracer and the presence of some spatio-temporal field $f$.  The convective derivative corresponds to the total derivative of the value of $f$ that the tracer locally sees. Thus, we expect analogous expressions for the convective derivative in both lab and co-deforming systems, $\dot{f} := \d f/\d t= \d f/\d \bar t$. Indeed, using Eqs.~\eqref{eq:trafo partial derivative r}, \eqref{eq:trafo partial derivative t}, and \eqref{eq:co-deforming v}, we obtain:
\begin{equation}
	\dot{f} = \partial_tf + \vel_i(\partial_i f) = \bar\partial_tf + \bar\vel_i(\bar\partial_i f).
\end{equation}

\subsection{Fourier transform}
Moreover, we define the co-deforming Fourier transformation of a quantity $f$ such that
\begin{equation}
  f(\vec{\bar r}) = \sum_{\vec{\bar{q}}}f(\vec{\bar q})\,e^{i\vec{\bar q}\cdot\vec{\bar r}}, \label{eq:codef fourier}
\end{equation}
where the sum is over all co-deforming wave vectors $\vec{\bar{q}}=(\bar{q}_x, \bar{q}_y)$ with $\bar q_i=2\pi n_i/L_i(0)$, where $n_i\in\mathbb{Z}$ and $i\in\lbrace x, y\rbrace$.
A co-deforming Fourier mode with wave vector $\vec{\bar q}$ is distorted over time by the overall system deformation:
\begin{align}
    q_x(\vec{\bar q}, t) &= l_x^{-1}(t)\bar{q}_x \label{eq:kx_kbarx}\\
    q_y(\vec{\bar q}, t) &= l_x(t)\bar{q}_y.\label{eq:ky_kbary}
\end{align}
As in Eq.~\eqref{eq:codef fourier}, we have the usual derivation rule, where the Fourier transform of $\bar\partial_j f(\bm{\bar r}, \bar t)$ is $i\bar{k}_jf(\bm{\bar q}, \bar t)$.
From Eqs.~\eqref{eq:codef fourier} and \eqref{eq:co_deforming r}, it also follows that a given co-deforming Fourier mode with wave vector $\bm{\bar{q}}$ corresponds to a lab-frame Fourier mode with wave vector $\bm{q}$ with components
\begin{equation}
  q_i = \bar{q}_js^{-1}_{ji}, \label{eq:trafo codef k}
\end{equation}
because then we have $\bm{\bar{q}}\cdot\bm{\bar{r}} = \bm{q}\cdot\bm{r}$.

\section{Dynamics in co-deforming coordinates}
Using \aref{app: codef coord}, we can express the dimensionless dynamic equations, Eqs.~\eqref{eq:dimless s}--\eqref{eq:dimless incompressibility}, in co-deforming coordinates: 
\begin{align}
  \bar\partial_t s + \bar{\vel}_i(\bar\partial_i s) &= 0 \label{eq:codef s}\\
  \bar\partial_t c + \bar{\vel}_i(\bar\partial_i c) &= \big(Ds^{-2}_{ij}\bar\partial_i\bar\partial_j - k_{\mathrm{d}}\big)c + k_{\mathrm{d}}s \label{eq:codef c} \\
  0 &= \frac{1}{4}s_{jl}\partial_i^2\bar\vel_l - \partial_j\Pi + \partial_i\tilde\sigma^a_{ij} \label{eq:codef v} \\
  s_{il}\partial_i\bar\vel_l &= 0, \label{eq:codef incompressibility}
\end{align}
where $\tilde\sigma^a_{ij}$ is given by Eq.~\eqref{eq:active stress}.
Eqs.~\eqref{eq:codef v} and \eqref{eq:codef incompressibility} are still expressed using lab-frame derivatives, because this will allow for a more convenient solution of the non-affine flows in the next section.

Finally, the system size dynamics for the freely deforming system, Eq.~\eqref{eq:shear rate}, becomes:
\begin{equation}
	\frac{\d l_x}{\d t} = -\sgn{\alpha}l_x\iint{\Big[ l_x^{-2}(\bar\partial_xc)^2-l_x^2(\bar\partial_yc)^2 \Big]\; \d\bar{x}\,\d\bar{y}},\label{eq:codef shear}
\end{equation}
where the integral is over the box at time point 0, i.e.\ the integration variables are the co-deforming coordinates $\bar{x}$ and $\bar{y}$.

\subsection{Non-affine flows}
For deforming systems, we will generally consider the dynamics of source and signaling fields in co-deforming coordinates, Eqs.~\eqref{eq:codef s} and \eqref{eq:codef c}.
To obtain the non-affine flows, we solve Eqs.~\eqref{eq:codef v} and \eqref{eq:codef incompressibility} by taking the lab-frame Fourier transform, which yields after some transformations:
\begin{equation}
	\bar\vel_k = \frac{4is^{-1}_{ki}q_l}{q^2} \left( \delta_{ij} - \frac{q_i q_j}{q^2} \right)\tilde\sigma^a_{lj}
	\label{eq:solution non-affine velocity}
\end{equation}
with $\tilde\sigma^a_{ij}$ given by Eq.~\eqref{eq:active stress}.

Note that Eq.~\eqref{eq:solution non-affine velocity} is undefined for $\bm{q}=0$. Indeed, any constant homogeneous velocity can always be added to the flow field. Here and in the following, we will ignore this, because it does not affect the physics in any way.

\subsection{Absence of non-affine flows\texorpdfstring{ for $\partial_yc=0$}{}}
\label{app:no non-affine flows}
If the signaling field $c$ is homogeneous in $y$ direction, i.e.\ $\partial_yc=\partial_{\bar{y}}c=0$, then there are no non-affine flows, i.e.\ there are no flows in co-deforming coordinates, $\vec{\bar\vel}=0$. For fixed system dimensions, this corresponds to a complete absence of flows, $\vec{\vel}=0$.

Formally, the absence of non-affine flows follows from Eq.~\eqref{eq:solution non-affine velocity}, which links $\vec{\bar\vel}$ in Fourier space to the stress tensor in Fourier space. From Eq.~\eqref{eq:active stress} we obtain for $\partial_{y}c=0$, that $\tilde{\sigma}^a_{xy}=0$ everywhere, and $\tilde{\sigma}^a_{xx}$ is independent of $y$.  Thus, in Fourier space, $\tilde{\sigma}^a_{xx}$ is nonzero only for wave vectors $\vec{q}$ that have a vanishing $y$ component, $q_y=0$.  Insertion in Eq.~\eqref{eq:solution non-affine velocity} yields indeed $\vec{\bar\vel}=0$.

\section{Linearized dynamics}
\label{app:linearized dynamics}
\subsection{In co-deforming real space coordinates}
\label{app:linearized dynamics, real space}
We perturb the system around the dynamic state $\vv{z_0}=(s_0(\vec{\bar{r}}), c_0(\vec{\bar{r}}), l_{x0})$ to linear order, where $s_0$ and $c_0$ only depend on $\bar{x}$:
\begin{align}
    s(\vec{\bar{r}}) &= s_0(\bar{x}) + \delta s(\vec{\bar{r}}), \label{eq: app s} \\
    c(\vec{\bar{r}})  &= c_0(\bar{x}) + \delta c(\vec{\bar{r}}), \label{eq: app c}\\
    l_x &= l_{x0} + \delta l_x. \label{eq: app lx} 
\end{align}
Because $c_0$ only depends on $\bar{x}$, we also have in the state $\vv{z_0}$ that $\tilde{\sigma}^a_{xy}=0$ and there are no non-affine flows, $\vec{\bar\vel} = 0$ (\aref{app:no non-affine flows}). 

Combining Eqs.~\eqref{eq: app s}--\eqref{eq: app lx} with Eqs.~\eqref{eq:codef s}, \eqref{eq:codef c}, and \eqref{eq:codef shear} yields the following linearized dynamics:
\begin{align}
    \bar\partial_t \delta s &= -s_0'\delta\bar\vel_x, \label{eq:gen lin source} \\
    \bar\partial_t \delta c &= -c_0'\delta\bar\vel_x
    + k_{\mathrm{d}}(\delta s - \delta c) \nonumber\\
    &\qquad\qquad + Ds^{-2}_{ij}\bar\partial_i\bar\partial_j\delta c -2D(\bar\partial_x^2c_0)\frac{\delta l_x}{l_{x0}^3}, \label{eq:gen lin signaling} \\
    \frac{\d \delta l_x(t)}{\d t} &= \frac{2\sgn{\alpha}}{l_{x0}} \iint{(\bar\partial_x^2c_0)\delta c\,\d\bar x \d\bar y} - \frac{\d l_{x0}}{\d t}\,\frac{\delta l_x}{l_{x0}}, \label{eq:gen lin size} 
\end{align}
where we have introduced the notation $s_0':=\bar\partial_x s_0$ and $c_0':=\bar\partial_x c_0$, and we used a partial integration in the linearized $l_x$ dynamics.

\subsection{In co-deforming Fourier space coordinates}
\label{app:linearized dynamics, Fourier space}
In co-deforming Fourier space, the linearized dynamics read:
\begin{align}
    \bar\partial_t \delta s &= -s_0' \ast \delta\bar\vel_x  \label{eq:linearized s dynamics, Fourier}\\
    \bar\partial_t \delta c &= -c_0' \ast \delta\bar\vel_x
    + k_{\mathrm{d}}(\delta s - \delta c) \nonumber\\
    &\qquad\qquad - Ds^{-2}_{ij}\bar{q}_i\bar{q}_j\delta c +2D\bar{q}_x^2c_0\frac{\delta l_x}{l_{x0}^3}, \label{eq:linearized c dynamics, Fourier} \\
    \frac{\d \delta l_x(t)}{\d t} &= -\frac{2\sgn{\alpha}}{l_{x0}} \sum_{\vec{\bar{q}}}{\bar{q}_x^2c_0^\dag\delta c} -\frac{\d l_{x0}}{\d t}\,\frac{\delta l_x}{l_{x0}}. \label{eq:linearized lx dynamics, Fourier}
\end{align}
Here $s_0'$ and $c_0'$ represent the co-deforming Fourier transforms of $\bar\partial_x s_0$ and $\bar\partial_x c_0$, respectively. Furthermore, to rewrite the $\delta l_x$ dynamics, we have used Parseval's identity, where the dagger, $\cdot^\dag$, denotes the complex conjugate. We have moreover used the following definition of convolution operator, $\ast$, in co-deforming Fourier space:
\begin{equation}
    f(\vec{\bar{q}})\ast g(\vec{\bar{q}}) := 
    \sum_{\vec{\bar{q}'}} f(\vec{\bar{q}'})g(\vec{\bar{q}}-\vec{\bar{q}'}), 
	\label{eq:codef convolution}
\end{equation}
where the sum is over all co-deforming wave vectors $\vec{\bar{q}'}=(\bar{q}'_x, \bar{q}'_y)$ with $\bar q_i'=2\pi n'_i/L_i(0)$, where $n'_i\in\mathbb{Z}$ and $i\in\lbrace x, y\rbrace$.
For a fixed system size, we ignore the $\delta l_x$ dynamics and the $\delta l_x$ term in the $\delta c$ dynamics.

To close Eqs.~\eqref{eq:linearized s dynamics, Fourier}--\eqref{eq:linearized lx dynamics, Fourier}, we still need to insert the $x$-component of non-affine flow field perturbation, $\delta \bar{\vel}_x$. 
For convenience, we will use \emph{lab-frame} Fourier modes to compute $\delta \bar{\vel}_x$.
Specifically, we use Eq.~\eqref{eq:solution non-affine velocity}, which expresses $\vec{\bar\vel}$ in terms of the active stress tensor $\tilde{\sigma}^a_{ij}$. We obtain to linear order:
\begin{equation}
	\delta\bar\vel_x(q,\phi) = \frac{4i\sin{\phi}}{ql_{x0}} \Big[\sin{2\phi}\; \delta\tilde\sigma^a_{xx} - \cos{2\phi}\; \delta\tilde\sigma^a_{xy}\Big], \label{eq:stokes delta vx}
\end{equation} 
where $q$ and $\phi$ denote the amplitude and orientation, respectively, of the lab-frame wave vector $\vec{q} = q(\cos{\phi},\sin{\phi})$. 

We further express $\delta\tilde{\sigma}^a_{ij}$ in terms of $\delta c$ using Eq.~\eqref{eq:active stress}. To linear order, in lab-frame Fourier space:
\begin{equation}
	\delta\tilde\sigma^a_{xj} = \frac{i\sgn{\alpha}}{l_{x0}} c_0'\ast(q_j\delta c) \quad\text{for $j\in\lbrace x, y\rbrace$}. \label{eq:delta active stress}
\end{equation}
The other two components of the symmetric traceless tensor $\delta\tilde{\sigma}^a_{ij}$ are obtained through the relations $\delta\tilde\sigma^a_{yy} = -\delta\tilde\sigma^a_{xx}$ and $\delta\tilde\sigma^a_{yx} = \delta\tilde\sigma^a_{xy}$.
The convolution operator in lab-frame Fourier coordinates in Eq.~\eqref{eq:delta active stress} is defined analogously to Eq.~\eqref{eq:codef convolution} as:
\begin{equation}
	f(\vec{q})\ast g(\vec{q}) := 
	\sum_{\vec{q'}} f(\vec{q'})g(\vec{q}-\vec{q'}), 
	\label{eq:lab convolution}
\end{equation}
where the sum is over all lab-frame wave vectors $\vec{q'}=(q'_x, q'_y)$ with $q_i'=2\pi n'_i/L_i(t)$, where $n'_i\in\mathbb{Z}$ and $i\in\lbrace x, y\rbrace$.

Combining \eqref{eq:stokes delta vx} and \eqref{eq:delta active stress}, we obtain:
\begin{equation}
	\begin{split}
		\delta\bar\vel_x = &-\frac{4\sgn{\alpha}\sin{\phi}}{ql_{x0}^2} \bigg\lbrace \sin{2\phi}\Big[c_0'\ast(q_x\delta c)\Big] \\
		&\qquad\qquad\qquad\qquad
		-\cos{2\phi}\Big[c_0'\ast(q_y\delta c)\Big]\bigg\rbrace.
	\end{split}
	\label{eq:delta vx from delta c}  
\end{equation}
The system of equations, Eqs.~\eqref{eq:linearized s dynamics, Fourier}--\eqref{eq:linearized lx dynamics, Fourier}, and \eqref{eq:delta vx from delta c}, is now closed.

\subsection{Numerical solution}
\label{app:numerical solution linearized dynamics}
In the special case of a linear source profile, we can solve the linearized dynamics analytically, because the gradients of $s_0$ and $c_0$ are constant, and thus the convolutions in Eqs.~\eqref{eq:linearized s dynamics, Fourier}, \eqref{eq:linearized c dynamics, Fourier}, and \eqref{eq:delta vx from delta c} collapse to a simple scaling factor.
However, in general this is not the case, and we need to solve the linearized dynamics numerically.

To numerically compute the convolutions in Eqs.~\eqref{eq:linearized s dynamics, Fourier}, \eqref{eq:linearized c dynamics, Fourier}, and \eqref{eq:delta vx from delta c}, we need to take into account that the numerical real-space representations of our fields are spatially discretised.
To discuss the main ideas, we will discuss here the case of a single spatial dimension, $\bar{x}$, in co-deforming coordinates.  The generalisation to lab frame coordinates and/or several dimensions will be straightforward.
Specifically, we consider the periodic interval $\bar{x}\in [-L_x(0)/2, +L_x(0)/2)$ to be discretised by $N_x$ equal-length steps $\Delta\bar x=L_x(0)/N_x$, where $N_x$ is an even integer.  Any function $f(\bar{x})$ is represented by a set of numbers $f_k$ with $k=-N_x/2, -N_x/2+1, \dots, 0, \dots, N_x/2-1$, such that $f_k=f(k\Delta\bar x)$.
As a consequence, the Fourier series are not infinite any more as in Eq.~\eqref{eq:codef fourier}, but instead cut off at some finite wave vectors.
In particular, the numerical co-deforming Fourier transform of $f$ is represented by a set of complex numbers $\tilde{f}_j$ with $j=-N_x/2, -N_x/2+1, \dots, 0, \dots, N_x/2-1$, where $\tilde{f}_j=f(\bar{q}_x=j\bar{q}_0)$ with $\bar{q}_0:=2\pi/L_x(0)$.
Eq.~\eqref{eq:codef fourier} then becomes:
\begin{equation}
	f_k = \sum_{j=-N_x/2}^{N_x/2-1}{\tilde{f}_je^{2\pi i (jk/N_x)}},
\end{equation}
where we used that $\bar{q}_0\Delta\bar x = 2\pi/N_x$.

We define the convolution of two Fourier transforms, $\tilde{f}_j$ and $\tilde{g}_j$ as follows:
\begin{equation}
	(\tilde{f}\ast\tilde{g})_j = \sum_{j'=-N_x/2}^{N_x/2-1}{\tilde{f}_{j'}\tilde{g}_{j-j'}} \label{eq:circular convolution}
\end{equation}
for $j=-N_x/2, -N_x/2+1, \dots, 0, \dots, N_x/2-1$.  Yet, we see that the index $j-j'$ of $\tilde{g}$ can get outside of the range $-N_x/2, -N_x/2+1, \dots, N_x/2-1$. 
To prevent aliasing effects, we set $\tilde{g}_j\equiv 0$ whenever $j<-N_x/2$ or $j\geq N_x/2$.
Note that this technique for de-aliasing is akin to the zero-padding method \cite{Roberts2011,Bowman2011}, frequently used for spectral solvers. 

To numerically solve the linearized dynamics, Eqs.~\eqref{eq:linearized s dynamics, Fourier}--\eqref{eq:linearized lx dynamics, Fourier}, we use an exponential ansatz for the time dependence of the perturbations, 
\begin{align}
	\delta s(\vec{\bar{q}}, t) &= \delta\hat{s}(\vec{\bar{q}})e^{\omega t} \\
	\delta c(\vec{\bar{q}}, t) &= \delta\hat{c}(\vec{\bar{q}})e^{\omega t} \\
	\delta l_x(t) &= \delta\hat{l}_xe^{\omega t}.
\end{align}
Inserting this in Eqs.~\eqref{eq:linearized s dynamics, Fourier}--\eqref{eq:linearized lx dynamics, Fourier} yields an eigenvalue problem:
\begin{equation}
	\omega \delta\hat{\vv{z}}=M\cdot\delta\hat{\vv{z}}, \label{eq:ev problem}
\end{equation}
where the components of $\delta\hat{\vv{z}}$ are $\delta\hat{l}_x$ and the values of $\delta\hat{s}(\vec{\bar{q}})$ and $\delta\hat{c}(\vec{\bar{q}})$ for all $N_xN_y$ discrete values of the co-deforming wave vector $\vec{\bar{q}}$.  Thus, $\delta\hat{\vv{z}}$ has $2N_xN_y+1$ components.
The matrix $M$ usually couples not only $\delta\hat{s}$ and $\delta\hat{c}$ components for the same wave vector $\vec{\bar{q}}$ to each other, but due to the convolutions in Eqs.~\eqref{eq:linearized s dynamics, Fourier}, \eqref{eq:linearized c dynamics, Fourier}, and \eqref{eq:delta vx from delta c}, there is also mixing occurring across different wave vectors.

\section{Fixed-size system with linear source}
\label{app:linear source}
For linear source and signaling fields, the active flows, Eq.~\eqref{eq:delta vx from delta c}, simplify to:
\begin{equation}
	\delta \vel_x = -\omega_{\mathrm{act}}\delta c \qquad
\end{equation}
with
\begin{equation}
	\omega_{\mathrm{act}} = 4\sgn{\alpha}\sin^2{\phi}, \label{eq:omega active}
\end{equation}
where $\omega_{\mathrm{act}}$ is the perturbation growth rate for the case of a boundary-provided gradient without source field \cite{Ibrahimi2023}.
With this, the linearized dynamics, Eqs.~\eqref{eq:linearized s dynamics, Fourier}--\eqref{eq:linearized c dynamics, Fourier} become:
\begin{align}
	\partial_t \delta s  &= \omega_{\mathrm{act}}\delta c, \label{eq:fixed, linear, lin s dynamics}\\
	\partial_t \delta c  &= \left[\omega_{\mathrm{act}} - Dq^2 - k_{\mathrm{d}} \right] \delta c + k_{\mathrm{d}}\delta s. \label{eq:fixed, linear, lin c dynamics}
\end{align}
Thus, the convolutions in Eqs.~\eqref{eq:linearized s dynamics, Fourier}, \eqref{eq:linearized c dynamics, Fourier}, and \eqref{eq:delta vx from delta c} collapse to a constant prefactor.
As a consequence, the solutions of the system are Fourier modes.
In the following, we first discuss the limiting case of no diffusion before presenting the general solution.

\subsection{No diffusion\texorpdfstring{, $D = 0$}{}}
\label{app:linear source, no diffusion}
To solve the eigenvalue problem in this case, we define
\begin{align}
	\delta\Delta &= \delta c - \delta s \\
	\delta c_\mathrm{eff} &= \delta c + \frac{k_\mathrm{d}}{\omega_\mathrm{act}}\delta s.
\end{align}
Inserting this into Eqs.~\eqref{eq:fixed, linear, lin s dynamics} and \eqref{eq:fixed, linear, lin c dynamics}, we obtain:
\begin{align}
	\partial_t \delta\Delta &= -k_\mathrm{d}\delta\Delta \\
	\partial_t \delta c_\mathrm{eff} &= \omega_\mathrm{act} \delta c_\mathrm{eff}.\label{eq:app:ceff}
\end{align}
Hence, the difference between both signaling and source, $\delta\Delta$, decays at rate $k_\mathrm{d}$, while a superposition of both, $\delta c_\mathrm{eff}$, grows with the same growth rate as the system without source \cite{Ibrahimi2023}.

\subsection{General case}
\label{app:linear source general}
With an exponential Ansatz, where $\delta s\sim\delta c\sim e^{\omega t}$, we obtain an eigenvalue problem, whose solutions in $\omega$ are:
\begin{equation}
	\omega_{\pm} = \frac{1}{2}\left[ \mathcal{T} \pm \sqrt{\mathcal{T}^2 + 4k_{\mathrm{d}}\;\omega_{\mathrm{act}}} \right],
\end{equation}
where $\mathcal{T} = \omega_{\mathrm{act}} -Dq^2 - k_{\mathrm{d}}$. The corresponding eigenvectors in $(\delta s,\delta c)$ are
\begin{equation}
	\bm{v}_+  = 
	\begin{pmatrix}
		\omega_{+}\\
		\omega_{\mathrm{act}}
	\end{pmatrix} 
	\quad \mathrm{and} \quad \bm{v}_-  = 
	\begin{pmatrix}
		\omega_-\\ \omega_{\mathrm{act}}
	\end{pmatrix},
\end{equation}
respectively.

\section{Fixed-size system with localized source}
\label{app:fixed system nonuniform}
\begin{figure}
    \centering
    \includegraphics[width = \columnwidth]{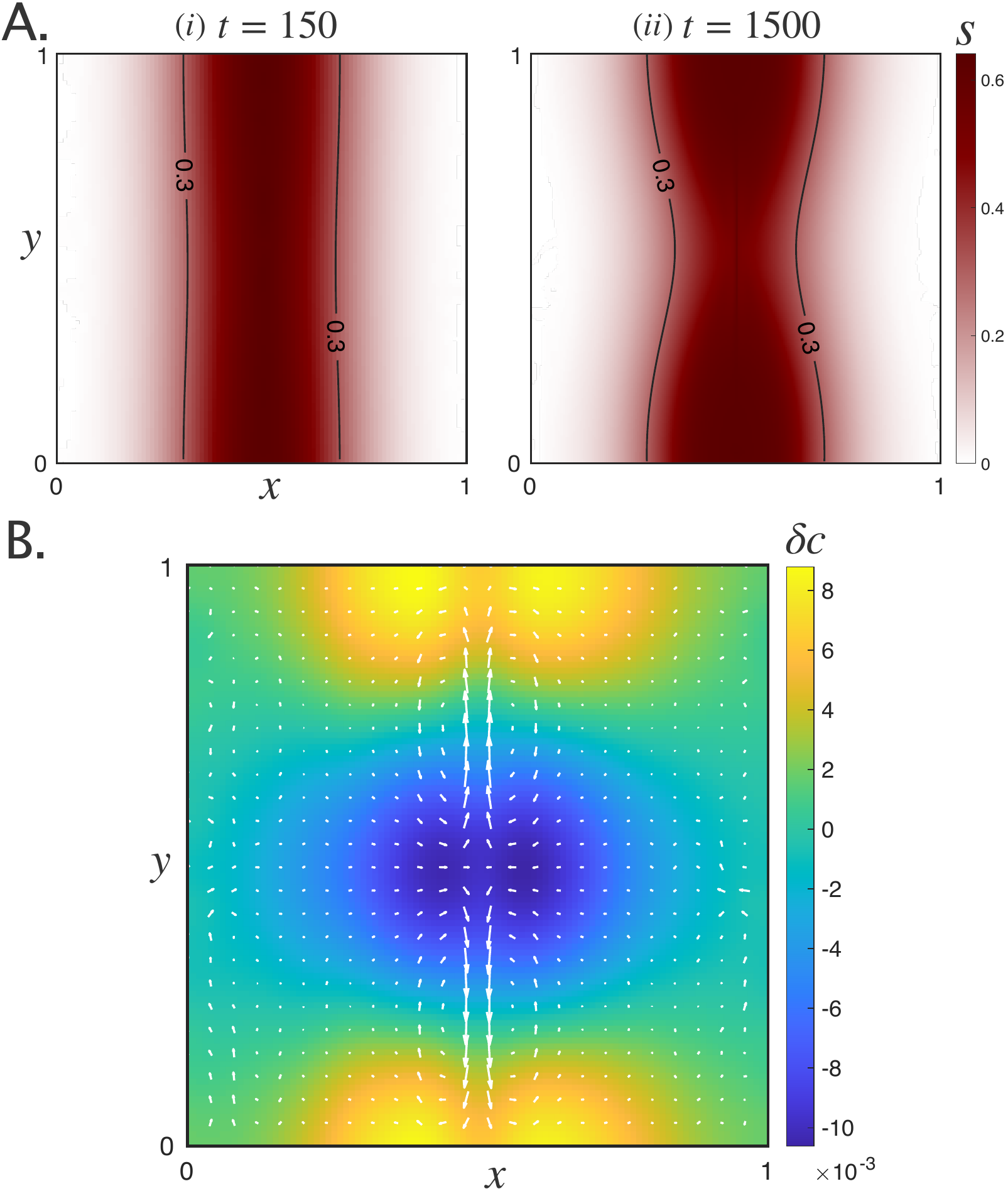}
    \caption{Instability in a gradient-extensile system with fixed size and a localized source profile. 
   	(A) Snapshots of the late-time dynamics of the system in \autoref{fig: fixed nonuniform}C, showing the source profile at $t = 300$ (left) and $t = 1500$ (right). 
    (B) Perturbation mode with maximum growth rate for system from \autoref{fig: fixed nonuniform}C and panel A. The color plot shows the signaling molecule perturbation, $\delta c$, and white arrows show the flow field perturbation, $\delta\vec{\vel}$.
    } 
    \label{fig: extensile perturbation}
\end{figure}

\subsection{Value of \texorpdfstring{$s_b$}{sb} in dimensionless units}
\label{app:value of sb}

As we discussed in \autoref{sec:dimensionless units}, we set the value of $s_b$ such that the magnitude of the deformation rate of a system with $L_x=L_y=1$ is one.
This means that we set
\begin{equation}
	s_b = \left[\sum_{q_x}{\frac{k_{\mathrm{d}}^2\vert m(q_x)\vert^2}{(k_{\mathrm{d}} + Dq_x^2)^2}}\right]^{-1/2} \label{eq:s_b}
\end{equation}
where $m(q_x)$ is the Fourier transform of the function
\begin{equation}
	m(x) = \exp{\bigg(\frac{\cos{(q_0x)}}{(q_0\w)^2}\bigg)}
\end{equation}
which is defined on the interval $x\in [-1/2,1/2)$, and where $q_0=2\pi/L_x$.
Indeed, using Eqs.~\eqref{eq:shear rate} and \eqref{eq: localized stat c}, one obtains that with the value of $s_b$ from Eq.~\eqref{eq:s_b}, a von Mises source profile, Eq.~\eqref{eq:nonlinear source}, and $L_x=L_y=1$, the absolute value of the shear rate is one.

\subsection{Convergence to a finite perturbation growth rate for \texorpdfstring{$D\rightarrow\infty$}{large diffusion}}
\label{app:localized source convergence perturbation growth rate}

\begin{figure}
    \centering
    \includegraphics[width = \columnwidth]{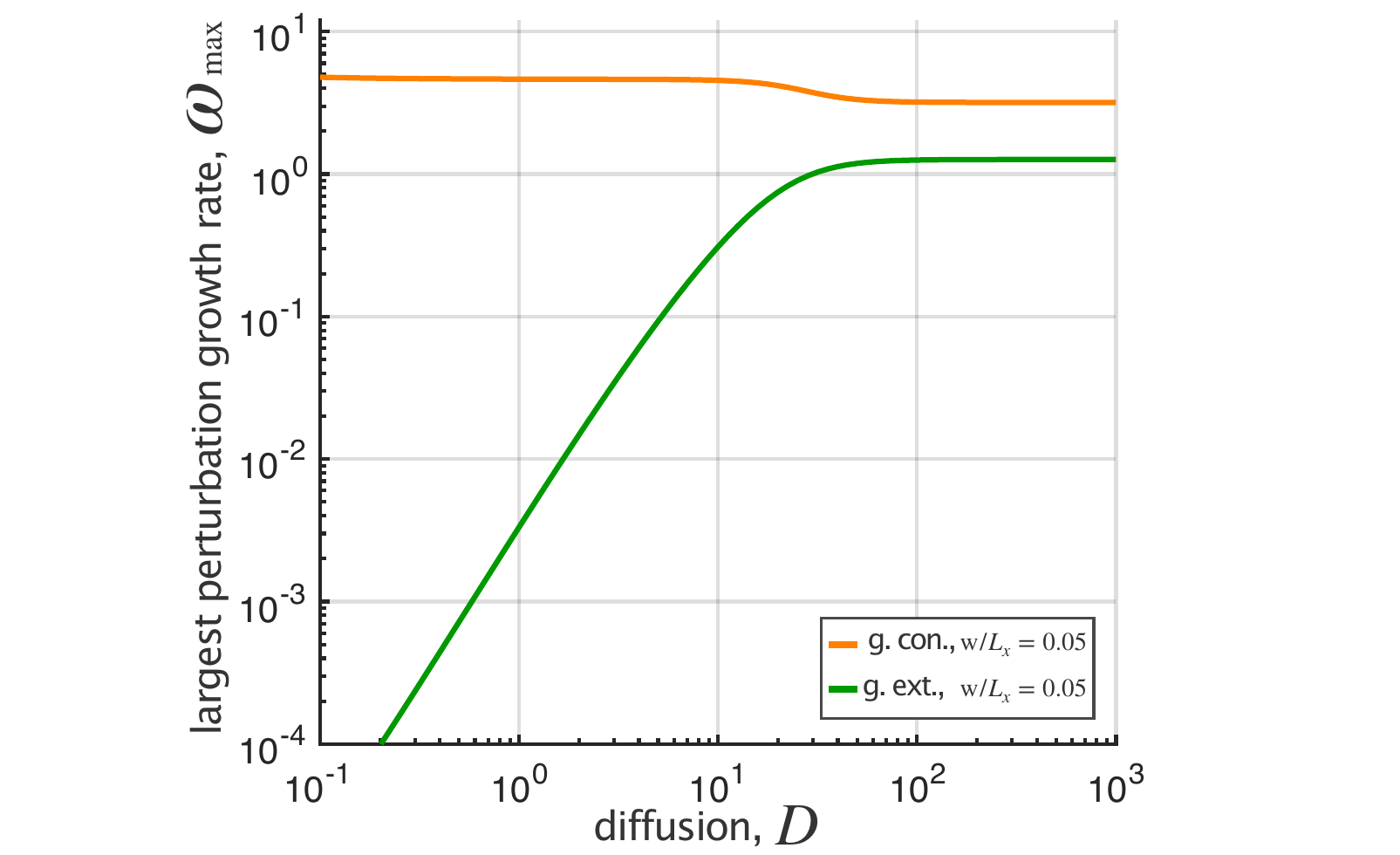}
    \caption{Largest perturbation growth rates $\omega_{\mathrm{max}}$ of a fixed-size system as a function of the signaling molecule diffusion. Parameter values, $k_\mathrm{d} = 1$ and $L_x = L_y = 1$. 
    } 
    \label{fig: large-diffusion limit}
\end{figure}

The large-diffusion limit can be understood similarly to \autoref{sec:uniform state}, where we assume that $Dq_\mathrm{min}^2$ is faster than any other rate to eliminate the signaling perturbation dynamics. Specifically, only the source perturbation dynamics, Eq.~\eqref{eq:linearized s dynamics, Fourier}, remains, where the signaling perturbation $\delta c$ depends directly on the source perturbation through Eq.~\eqref{eq:adiabatic limit - large diffusion}, which implies $\delta c\sim D^{-1}\delta s$.
Moreover, for large $D$, Eq.~\eqref{eq: localized stat c} implies that $c_0\sim D^{-1} s_0$, and thus $c_0'\sim D^{-1} s_0'$.

In dimensionless units, we set $s_b$ such that the free deformation rate is one, which implies $s_0\sim s_b\sim D$ (compare Eq.~\eqref{eq:s_b}). Thus, we have $c_0'\sim D^{0}$. As a consequence, the perturbation of active stresses and active flows scale as $\delta\bar\vel_x\sim\delta\tilde\sigma^a_{ij}\sim c_0'\delta c\sim D^{-1}$ (compare Eqs.~\eqref{eq:delta active stress} and \eqref{eq:delta vx from delta c}). The active flows enter in the advective term of the source field, Eq.~\eqref{eq:linearized s dynamics, Fourier}, where they are multiplied with the stationary source field, which scales as $s_0'\sim s_b\sim D$. Hence, in the large-$D$ limit, the source dynamics does not depend on $D$ any more. As a consequence, the largest perturbation growth rate is constant.

In dimensionful units, i.e.\ if we imposed a constant source magnitude $s_b\sim D^{0}$, we would have $c_0'\sim D^{-1}$. Hence, the perturbation growth rate would scale as $c_0'\delta c\sim D^{-2}$, and the free deformation rate would scale as $\sim\tilde\sigma^a_{xx}\sim(c_0')^2\sim D^{-2}$.
Hence, the perturbation growth rate decreases with $D$ in the same way as the free deformation rate.

\section{Deforming system with localized source}

\subsection{System dimension dynamics}
\label{app:localized source, system size dynamics}

The dynamics of system dimension $L_x$, and thus box shear $l_x$, is given by Eq.~\eqref{eq:shear rate}.

\begin{figure} 
	\centering
	\includegraphics[width = \columnwidth]{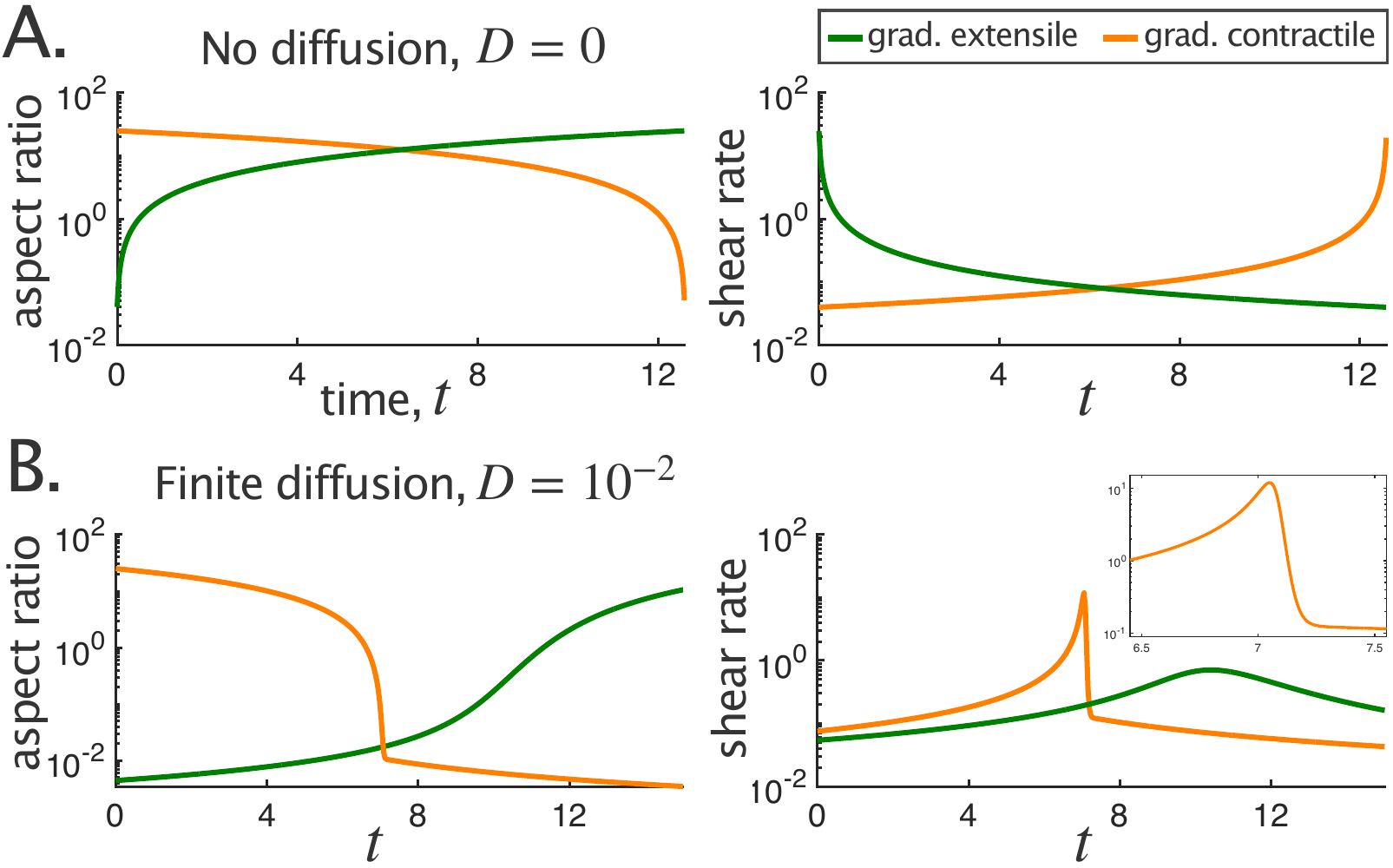}
	\caption{ System size dynamics of a freely deforming system in the limit $\w\rightarrow\infty$, where the source has a cosine shape. Here, $k_\mathrm{d} = 1$.
	(A) Aspect ratios (left) and shear rates (right) over time for the gradient-extensile (green) and contractile (orange) systems, starting at initial aspect ratios of $L_x(0)/L_y(0) = 1/225$ and $L_x(0)/L_y(0) = 25$, respectively. Here the signaling molecule is non-diffusive, $D = 0$.
	(B) The same data as panel A shown for a system with diffusive signaling molecule with $D = 10^{-2}$. Here, the gradient-extensile (green) and gradient-contractile (orange) systems  start from $L_x(0)/L_y(0) = 0.04$ and $L_x(0)/L_y(0) = 25$, respectively. Inset: zoomed around the sharp drop in the gradient-contractile case.
    } 
	\label{fig: system size dynamics}
\end{figure}

We first discuss the case without diffusion, $D=0$, where source and signaling fields, $s(\bar{x})$ and $c(\bar{x})$, are stationary in co-deforming coordinates.
In this case, the shear rate is essentially given by the integral over $(\partial_x c_0)^2$. Because the signaling profile, $c_0$, is stationary in co-deforming coordinates, we have that for varying system size, $(\partial_x c_0)^2\sim1/l_x^2(\bar\partial_x c_0)^2\sim1/l_x^2$.
So, using dimensionless units (\autoref{sec:dimensionless units}), we can write
\begin{equation}
	\frac{1}{l_x}\,\frac{\d l_x}{\d t} = -\frac{\sgn{\alpha}}{L_x^2(0)l_x^2}.
\end{equation}
The solution of this is:
\begin{equation}
	l_x(t) = \sqrt{1 -\frac{2\sgn{\alpha}t}{L_{x}(0)^2} }.\label{eq:lx(t)}
\end{equation}
Thus, gradient-extensile systems ($\alpha<0$) deform more and more slowly over time due to the widening of the signaling profile (green curves in \autoref{fig: system size dynamics}A). Meanwhile, gradient-contractile systems ($\alpha>0$) deform faster and faster over time due to the compression of the signaling profile, until the deformation rate diverges at some critical time point $t_\mathrm{crit}=L_x^2(0)/2$ (orange curves in \autoref{fig: system size dynamics}A).

This box dimension dynamics is changed when allowing for finite diffusion, $D>0$ (\autoref{fig: system size dynamics}B).
Specifically, the shear rate of gradient-extensile systems first increases, before it decreases like in the case without diffusion (compare green curves in \autoref{fig: system size dynamics}A,B, right panels).
In gradient contractile systems, the shear rate first increases, like in the case without diffusion before it rather abruptly drops (compare orange curves in \autoref{fig: system size dynamics}A,B right panels and \autoref{fig: system size dynamics}B inset).

To better understand these effects of diffusion, we first note that without any non-affine flows (i.e.\ $\bm{\bar{\vel}}=\vec{0}$), both signaling and source profiles keep their cosine profiles, $s(\bm{r},t)=\hat{s}[\cos{(2\pi x/L_x)}+1]$ and $c(\bm{r},t)=\hat{c}(t)\cos{(2\pi x/L_x)} + c_\mathrm{offset}(t)$, while only the signaling amplitude $\hat{c}(t)$ and offset $c_\mathrm{offset}(t)$ vary over time. We can thus simplify the dynamics to:
\begin{align}
  \frac{\d\hat{c}}{\d t} &= -k_\mathrm{d}(\hat{c}-\hat{s}) - 4\pi^2D\frac{\hat{c}}{L_x^2(0)l_x^2} \label{eq:cos-simplified-c}\\
  \frac{1}{l_x}\,\frac{\d l_x}{\d t} &= -\frac{2\pi^2\sgn{\alpha}\hat{c}^2}{L_x^2(0)l_x^2}, \label{eq:cos-simplified-Lx}
\end{align}
where $\hat{s}$ is constant since $\bm{\bar{\vel}}=\vec{0}$, and the offset $c_\mathrm{offset}(t)$ does not matter for the box dimension dynamics.
Eq.~\eqref{eq:cos-simplified-Lx} implies that the shear rate scales as $\sim\hat{c}^2/l_x^2$. Further, the diffusion term in  Eq.~\eqref{eq:cos-simplified-c} can be neglected as long as $l_x\gtrsim \sqrt{D/k_\mathrm{d}}/L_x(0)$, and consequentially $\hat{c}\simeq\hat{s}$.  In other words, the box shear rate for large $l_x$ is the same as in the case without diffusion, where the shear rate scales as $\sim l_x^{-2}$, for both gradient-extensile and gradient-contractile systems, consistent with \autoref{fig: system size dynamics}A,B.
Conversely, taking Eq.~\eqref{eq:cos-simplified-c} to the adiabatic limit of large diffusion $D$ and/or small box shear $l_x$, we have $\hat{c}\sim(k_\mathrm{d}/D)L_x^2(0)l_x^2\hat{s}\sim l_x^2$. Thus, in the limit of small box shear $l_x$, the shear rate scales as $\sim\hat{c}^2/l_x^2\sim l_x^2$, i.e.\ it increases with $l_x$, again consistent with \autoref{fig: system size dynamics}A,B.

\subsection{Affine diffusion-shear instability}
\label{app:localized source, affine diffusion-shear instability}
To discuss the affine diffusion-shear instability, we start from the linearized dynamics in Fourier space, Eqs.~\eqref{eq:linearized s dynamics, Fourier}--\eqref{eq:linearized lx dynamics, Fourier}, where we set non-affine flows to zero, $\vec{\delta\bar\vel}=0$, we ignore any variations of the source field, $\delta s=0$, and we ignore any $y$-dependence of the perturbations. We obtain from Eqs.~\eqref{eq:linearized c dynamics, Fourier} and \eqref{eq:linearized lx dynamics, Fourier} with an exponential Ansatz for the time dependency, $\delta c(\bar q_x, t)=\delta\tilde{c}(\bar q_x)e^{\omega t}$ and $\delta l_x(t)=\delta\tilde{l}_xe^{\omega t}$:
\begin{align}
    \omega\delta\tilde{c}(\bar q_x) &= -k_\mathrm{d}\big(1 + \lambda^2q_x^2\big)\delta\tilde{c}(\bar q_x) + 2k_\mathrm{d}\lambda^2q_x^2c_0(\bar q_x)\frac{\delta\tilde{l}_x}{l_x} \label{eq:affine diffusion-shear c perturbation}\\
    \omega\frac{\delta\tilde{l}_x}{l_x} &= - 2\sgn{\alpha}\sum_{\bar q_x}{q_x^2c_0^\dag(\bar q_x)\delta\tilde{c}(\bar q_x)} \nonumber\\
    &\qquad\qquad +\sgn{\alpha}\frac{\delta\tilde{l}_x}{l_x}\sum_{\bar q_x}{q_x^2\vert c_0(\bar q_x)\vert^2}. \label{eq:affine diffusion-shear lx perturbation}
\end{align}
Here, we also substituted the diffusion length scale $\lambda=\sqrt{D/k_\mathrm{d}}$ and the lab-frame wave vector $q_x=\bar q_x/l_x$.
If and only if this linear system of equations has a positive eigenvalue $\omega$, then there is an instability.

\subsubsection{Cosine source profile}
\label{app:localized source, affine diffusion-shear instability, cosine}
For the case of a cosine profile, the only nonzero values of $c_0(\bar q_x)$ are $c_0\equiv c_0(\pm\bar q_0)$ with $\bar q_0 = 2\pi/L_x(0)$. 
Then, Eqs.~\eqref{eq:affine diffusion-shear c perturbation} and \eqref{eq:affine diffusion-shear lx perturbation} become:
\begin{align}
    \omega\delta\tilde{c} &= -k_\mathrm{d}\big(1 + \lambda^2q_0^2\big)\delta\tilde{c} + 2k_\mathrm{d}\lambda^2q_0^2c_0\frac{\delta\tilde{l}_x}{l_x} \label{eq:affine diffusion-shear c perturbation2}\\
    \omega\frac{\delta\tilde{l}_x}{l_x} &= -4\sgn{\alpha}q_0^2c_0^\dag\delta\tilde{c} +2\sgn{\alpha}q_0^2\vert c_0\vert^2\frac{\delta\tilde{l}_x}{l_x}, \label{eq:affine diffusion-shear lx perturbation2}
\end{align}
where $q_0(t):=\bar q_0/l_x(t)=2\pi/L_x(t)$.
There are two eigenvalues, $\omega_1$ and $\omega_2$ with $\omega_1\leq\omega_2$, and the largest is given by:
\begin{equation}
	\omega_2 = \frac{1}{2}\left[ \mathcal{T} + \sqrt{\mathcal{T}^2 - 4\mathcal{D}} \right],\label{eq:affine ds instability, growth rate}
\end{equation}
where $\mathcal{T} := -k_\mathrm{d}(1+ \lambda^2q_0^2) +2\sgn{\alpha}q_0^2\vert c_0\vert^2$ and $\mathcal{D} := -2\sgn{\alpha}k_\mathrm{d}q_0^2\vert c_0\vert^2(1 - 3\lambda^2q_0^2)$ are trace and determinant, respectively, of the matrix on the right-hand side of Eqs.~\eqref{eq:affine diffusion-shear c perturbation2} and \eqref{eq:affine diffusion-shear lx perturbation2}.

In the gradient-extensile case, $\sgn{\alpha}=-1$, we obtain that $\omega_1+\omega_2 = \mathcal{T}<0$. Thus, at least one eigenvalue is negative. In addition, the sign of the product, $\omega_1\omega_2=\mathcal{D}$, will indicate whether both eigenvalues are negative or at least one is positive.
The product is given by $\omega_1\omega_2=\mathcal{D} = 2k_\mathrm{d}q_0^2\vert c_0\vert^2(1 - 3\lambda^2q_0^2)$.
Hence, there is a positive eigenvalue if and only if $\omega_1\omega_2<0$, which is equivalent to $3\lambda^2q_0^2>1$, and thus to $2\pi\lambda\sqrt{3}>L_x$.

\subsubsection{Arbitrary source profile}
\label{app:localized source, affine diffusion-shear instability, general}
For the case of an arbitrary source profile $s_0$, we focus on the gradient-extensile case, $\sgn{\alpha}=-1$. We discuss the existence of positive solutions $\omega$ by first expressing $\delta\tilde{c}(\bar q_x)$ in terms of $\delta\tilde{l}_x$ and $\omega$ using Eq.~\eqref{eq:affine diffusion-shear c perturbation}:
\begin{equation}
    \delta\tilde{c}(\bar q_x) = \frac{2\lambda^2q_x^2c_0(\bar q_x)}{1 + \omega/k_\mathrm{d} + \lambda^2q_x^2}\,\frac{\delta\tilde{l}_x}{l_x}.
\end{equation}
Insertion into Eq.~\eqref{eq:affine diffusion-shear lx perturbation} yields:
\begin{equation}
    \begin{split}
        \omega &= \sum_{\bar q_x}{q_x^2\vert c_0(\bar q_x)\vert^2\frac{4\lambda^2q_x^2}{1 + \omega/k_\mathrm{d} + \lambda^2q_x^2}}
        -\sum_{\bar q_x}{q_x^2\vert c_0(\bar q_x)\vert^2}\\
            &= \sum_{\bar q_x}{q_x^2\vert c_0(\bar q_x)\vert^2\frac{3\lambda^2q_x^2-1-\omega/k_\mathrm{d}}{1 + \omega/k_\mathrm{d} + \lambda^2q_x^2}}.
    \end{split}\label{eq:affine diffusion-shear implicit}
\end{equation}
To solve this self-consistent equation for $\omega$, we look for the zeros of the function
\begin{equation}
    f_\lambda(\omega) :=
        - \omega + \sum_{\bar q_x}{q_x^2\vert c_0(\bar q_x)\vert^2\frac{3\lambda^2q_x^2-1-\omega/k_\mathrm{d}}{1 + \omega/k_\mathrm{d} + \lambda^2q_x^2}}.
\end{equation}
Specifically, we now show that iff $f_\lambda(0)>0$, there exists a positive solution in $\omega$, i.e.\ an $\omega>0$ for which $f_\lambda(\omega)=0$.  This means that there is an affine diffusion-shear instability iff $f_\lambda(0)>0$.
To show this, we first assume that there is an $\omega>0$ for which $f_\lambda(\omega)=0$. Then, we have:
\begin{equation}
    \begin{split}
        0 
        &< \omega \\
        &\quad= \sum_{\bar q_x}{q_x^2\vert c_0(\bar q_x)\vert^2\frac{3\lambda^2q_x^2-1-\omega/k_\mathrm{d}}{1 + \omega/k_\mathrm{d} + \lambda^2q_x^2}} \\
        &\quad< \sum_{\bar q_x}{q_x^2\vert c_0(\bar q_x)\vert^2\frac{3\lambda^2q_x^2-1}{1 + \lambda^2q_x^2}} \\
        &\qquad= f_\lambda(0).
    \end{split}
\end{equation}
In these transformations, we have subsequently used $\omega>0$, $f_\lambda(\omega)=0$, $\omega>0$, and the definition of $f_\lambda(\omega)$. Thus, if there is an $\omega>0$ for which $f_\lambda(\omega)=0$, then $f_\lambda(0)>0$.

Second, we assume that $f_\lambda(0)>0$. To show that there is a positive zero of $f_\lambda$, we need to show that there is a positive $\hat\omega>0$ for which $f_\lambda(\hat\omega)<0$. Then, the Intermediate Value Theorem implies that there is an $\omega$ with $0<\omega<\hat\omega$ such that $f_\lambda(\omega)=0$.  Indeed, for $\hat\omega=3\sum_{\bar q_x}{q_x^2\vert c_0(\bar q_x)\vert^2}$, we find that:
\begin{equation}
    \begin{split}
        f_\lambda(\hat\omega) 
        &= -3\sum_{\bar q_x}{q_x^2\vert c_0(\bar q_x)\vert^2} \\
        &\qquad\qquad\qquad+ \sum_{\bar q_x}{q_x^2\vert c_0(\bar q_x)\vert^2\frac{3\lambda^2q_x^2-1-\hat\omega/k_\mathrm{d}}{1 + \hat\omega/k_\mathrm{d} + \lambda^2q_x^2}} \\
        &= \sum_{\bar q_x}{q_x^2\vert c_0(\bar q_x)\vert^2\frac{-2(1+\hat\omega/k_\mathrm{d})}{1 + \hat\omega/k_\mathrm{d} + \lambda^2q_x^2}} \\
        &<0.
    \end{split}
\end{equation}
Thus, if $f_\lambda(0)>0$ then $f_\lambda$ has a positive zero and the affine diffusion-shear instability exists.
Taken together, the sign of $f_\lambda(0)$ directly indicates the existence of an instability.

Finally, we discuss the sign of $f_\lambda(0)$.
Using Eq.~\eqref{eq: localized stat c}, we can express $f_\lambda(0)$ in terms of the source profile:
\begin{equation}
    f_\lambda(0) = \frac{1}{\lambda^2}\sum_{\bar q_x}{\vert s_0(\bar q_x)\vert^2\frac{\lambda^2q_x^2(3\lambda^2q_x^2-1)}{(1 + \lambda^2q_x^2)^3}}.
\end{equation}
Choosing for $s_0$ the von Mises profile in Eq.~\eqref{eq:nonlinear source} with finite width $\w$, we numerically obtain $f_\lambda(0)>0$ iff $\lambda > g\w$ with $g\approx 0.639$ in the limit where $\w\ll L_x$ (i.e.\ where the von Mises profile becomes a Gaussian with standard deviation $\w$). However, the precise value of $g$ depends on the shape of the profile $s_0$. For a rectangular profile of $s_0$, we numerically even find that the instability always occurs.

\section{Non-linearity in the active stress}
\label{app:non-linear active stress}
We consider the system of equations, Eqs.~\eqref{eq:source field dynamics}--\eqref{eq:active stress}, where we introduce a non-linearity into the active stress, Eq.~\eqref{eq:active stress}:
\begin{equation}
    \tilde\sigma^a_{ij} = \alpha\, h\big([\partial_k c]^2\big)\left[ (\partial_i c) (\partial_j c) -\frac{1}{2}(\partial_kc)^2\delta_{ij} \right],
    \label{eq:active stress nonlinear}
\end{equation}
where $z\mapsto h(z)$ is some real function expressing the nonlinearity.  In the following, we assume that $h(z)>0$ for any real $z$, and that the function $z\mapsto zh(z)$ is monotonously increasing, i.e.\ the magnitude of the active stress increases monotonously with the magnitude of scalar field gradient.

Then, the dimensionless, linearized dynamics become:
\begin{align}
	\bar\partial_t \delta s &= -s_0'\delta\bar\vel_x, \label{eq:gen lin source - nl stress} \\
	\bar\partial_t \delta c &= -c_0'\delta\bar\vel_x
	+ k_{\mathrm{d}}(\delta s - \delta c) \nonumber\\
	&\qquad\qquad + Ds^{-2}_{ij}\bar\partial_i\bar\partial_j\delta c -2D(\bar\partial_x^2c_0)\frac{\delta l_x}{l_{x0}^3}, \label{eq:gen lin signaling - nl stress} \\
	\frac{\d \delta l_x(t)}{\d t} &= -2\sgn{\alpha} \iint{\Big[h(z)+zh'(z)\Big]\frac{c_0'}{l_{x0}}\,\delta c'\,\d\bar x \d\bar y} \nonumber\\
	&\qquad + \sgn{\alpha} \big(I_0+2\Delta I\big)\,\delta l_x, \label{eq:gen lin size - nl stress} 
\end{align}
where $\delta c':=\bar\partial_x\delta c$, and
\begin{align}
	I_0 &:= \iint{zh(z)\,\d\bar x \d\bar y} \\
	\Delta I &:= \iint{z^2h'(z)\,\d\bar x \d\bar y}
\end{align}
with $z(\bm{\bar{r}}):=(c_0'(\bm{\bar{x}})/l_{x0})^2$.

We further focus on the gradient-extensile case, $\sgn{\alpha}=-1$, without non-affine flows, $\bm{\bar\vel}=\bm{0}$, and, for simplicity, we consider the limit $\w\rightarrow\infty$, such that $s_0(\bar{x})\sim\cos{(\bar{q}_0\bar{x})}$, and thus also $ c_0(\bar{x})=\hat{c}_0\cos{(\bar{q}_0\bar{x})}$ with $\bar{q}_0=2\pi/L_x(0)$ and some constant prefactor $\hat{c}_0$.
As a consequence, based on Eq.~\eqref{eq:gen lin signaling - nl stress}, all Fourier modes of $\delta c$ decay except for $\delta c=\delta\hat c\cos{(\bar{q}_0\bar{x})}$. Hence, $\delta c(\bar{x}) = (\delta\hat c/\hat c_0)c_0(\bar{x})$.
As a consequence, Eqs.~\eqref{eq:gen lin signaling - nl stress} and \eqref{eq:gen lin size - nl stress} simplify to:
\begin{align}
	\frac{\d\delta\hat c}{\d t} &= -(k_{\mathrm{d}} + Dq_0^2)\delta\hat c +2Dq_0^2\hat c_0\frac{\delta l_x}{l_{x0}}, \label{eq:Fourier gen lin signaling - nl stress} \\
	\frac{\d \delta l_x(t)}{\d t} &= 2 \big(I_0+\Delta I\big)l_{x0}\frac{\delta\hat c}{\hat c_0}- \big(I_0+2\Delta I\big)\,\delta l_x. \label{eq:Fourier gen lin size - nl stress} 
\end{align}
This is a linear system of differential equations with constant coefficients. It is linearly stable if both eigenvalues, $\omega_1$ and $\omega_2$, of the coefficient matrix are negative. This implies that the trace of the matrix needs to be negative: $0>\omega_1+\omega_2=-k_{\mathrm{d}} - Dq_0^2 -I_0-2\Delta I$, and that the determinant of the matrix needs to be positive: $0<\omega_1\omega_2=(k_{\mathrm{d}} + Dq_0^2)(I_0+2\Delta I) - 4Dq_0^2(I_0+\Delta I)=k_{\mathrm{d}}(I_0+2\Delta I) - Dq_0^2(3I_0+2\Delta I)$. For the system to be stable also for large diffusion, i.e.\ for $k_{\mathrm{d}}\ll Dq_0^2$, the latter condition implies $-3I_0-2\Delta I>0$. However, we have assumed that $h(z)>0$, implying $I_0>0$, and that $z\mapsto zh(z)$ increases monotonically, implying that $h(z)+zh'(z)\geq0$ and thus $I_0+\Delta I\geq 0$. This contradicts determinant condition for a stable system. Taken together, this means that the diffusion-shear instability arises for large diffusion for any non-linearity $h$ in the active stress.
We expect this result to generalize also to finite source widths $\w$; higher-order Fourier modes in $c_0$ would increase prefactor of the first term in Eq.~\eqref{eq:Fourier gen lin size - nl stress}, and thus make the system more unstable.

\bibliography{source_morphogen,refs_mm}

\begin{thebibliography}{62}%
\makeatletter
\providecommand \@ifxundefined [1]{%
 \@ifx{#1\undefined}
}%
\providecommand \@ifnum [1]{%
 \ifnum #1\expandafter \@firstoftwo
 \else \expandafter \@secondoftwo
 \fi
}%
\providecommand \@ifx [1]{%
 \ifx #1\expandafter \@firstoftwo
 \else \expandafter \@secondoftwo
 \fi
}%
\providecommand \natexlab [1]{#1}%
\providecommand \enquote  [1]{``#1''}%
\providecommand \bibnamefont  [1]{#1}%
\providecommand \bibfnamefont [1]{#1}%
\providecommand \citenamefont [1]{#1}%
\providecommand \href@noop [0]{\@secondoftwo}%
\providecommand \href [0]{\begingroup \@sanitize@url \@href}%
\providecommand \@href[1]{\@@startlink{#1}\@@href}%
\providecommand \@@href[1]{\endgroup#1\@@endlink}%
\providecommand \@sanitize@url [0]{\catcode `\\12\catcode `\$12\catcode
  `\&12\catcode `\#12\catcode `\^12\catcode `\_12\catcode `\%12\relax}%
\providecommand \@@startlink[1]{}%
\providecommand \@@endlink[0]{}%
\providecommand \url  [0]{\begingroup\@sanitize@url \@url }%
\providecommand \@url [1]{\endgroup\@href {#1}{\urlprefix }}%
\providecommand \urlprefix  [0]{URL }%
\providecommand \Eprint [0]{\href }%
\providecommand \doibase [0]{https://doi.org/}%
\providecommand \selectlanguage [0]{\@gobble}%
\providecommand \bibinfo  [0]{\@secondoftwo}%
\providecommand \bibfield  [0]{\@secondoftwo}%
\providecommand \translation [1]{[#1]}%
\providecommand \BibitemOpen [0]{}%
\providecommand \bibitemStop [0]{}%
\providecommand \bibitemNoStop [0]{.\EOS\space}%
\providecommand \EOS [0]{\spacefactor3000\relax}%
\providecommand \BibitemShut  [1]{\csname bibitem#1\endcsname}%
\let\auto@bib@innerbib\@empty
\bibitem [{\citenamefont {Wolpert}\ \emph {et~al.}(2015)\citenamefont
  {Wolpert}, \citenamefont {Tickle},\ and\ \citenamefont
  {Arias}}]{Wolpert2015}%
  \BibitemOpen
  \bibfield  {author} {\bibinfo {author} {\bibfnamefont {L.}~\bibnamefont
  {Wolpert}}, \bibinfo {author} {\bibfnamefont {C.}~\bibnamefont {Tickle}},\
  and\ \bibinfo {author} {\bibfnamefont {A.~M.}\ \bibnamefont {Arias}},\
  }\href@noop {} {\emph {\bibinfo {title} {Principles of Development}}}\
  (\bibinfo  {publisher} {Oxford University Press},\ \bibinfo {year}
  {2015})\BibitemShut {NoStop}%
\bibitem [{\citenamefont {Zallen}\ and\ \citenamefont
  {Wieschaus}(2004)}]{Zallen2004}%
  \BibitemOpen
  \bibfield  {author} {\bibinfo {author} {\bibfnamefont {J.~A.}\ \bibnamefont
  {Zallen}}\ and\ \bibinfo {author} {\bibfnamefont {E.}~\bibnamefont
  {Wieschaus}},\ }\bibfield  {title} {\bibinfo {title} {{Patterned Gene
  Expression Directs Bipolar Planar Polarity in Drosophila}},\ }\href
  {https://doi.org/10.1016/S1534-5807(04)00060-7} {\bibfield  {journal}
  {\bibinfo  {journal} {Dev. Cell}\ }\textbf {\bibinfo {volume} {6}},\ \bibinfo
  {pages} {343} (\bibinfo {year} {2004})}\BibitemShut {NoStop}%
\bibitem [{\citenamefont {B{\'{e}}naz{\'{e}}raf}\ \emph
  {et~al.}(2010)\citenamefont {B{\'{e}}naz{\'{e}}raf}, \citenamefont
  {Francois}, \citenamefont {Baker}, \citenamefont {Denans}, \citenamefont
  {Little},\ and\ \citenamefont {Pourqui{\'{e}}}}]{Benazeraf2010}%
  \BibitemOpen
  \bibfield  {author} {\bibinfo {author} {\bibfnamefont {B.}~\bibnamefont
  {B{\'{e}}naz{\'{e}}raf}}, \bibinfo {author} {\bibfnamefont {P.}~\bibnamefont
  {Francois}}, \bibinfo {author} {\bibfnamefont {R.~E.}\ \bibnamefont {Baker}},
  \bibinfo {author} {\bibfnamefont {N.}~\bibnamefont {Denans}}, \bibinfo
  {author} {\bibfnamefont {C.~D.}\ \bibnamefont {Little}},\ and\ \bibinfo
  {author} {\bibfnamefont {O.}~\bibnamefont {Pourqui{\'{e}}}},\ }\bibfield
  {title} {\bibinfo {title} {{A random cell motility gradient downstream of FGF
  controls elongation of an amniote embryo}},\ }\href
  {https://doi.org/10.1038/nature09151} {\bibfield  {journal} {\bibinfo
  {journal} {Nature}\ }\textbf {\bibinfo {volume} {466}},\ \bibinfo {pages}
  {248} (\bibinfo {year} {2010})}\BibitemShut {NoStop}%
\bibitem [{\citenamefont {Shindo}(2018)}]{Shindo2018}%
  \BibitemOpen
  \bibfield  {author} {\bibinfo {author} {\bibfnamefont {A.}~\bibnamefont
  {Shindo}},\ }\bibfield  {title} {\bibinfo {title} {{Models of convergent
  extension during morphogenesis}},\ }\bibfield  {journal} {\bibinfo  {journal}
  {Wiley Interdisciplinary Reviews: Developmental Biology}\ }\textbf {\bibinfo
  {volume} {7}},\ \href {https://doi.org/10.1002/wdev.293} {10.1002/wdev.293}
  (\bibinfo {year} {2018})\BibitemShut {NoStop}%
\bibitem [{\citenamefont {Johansen}\ \emph {et~al.}(2003)\citenamefont
  {Johansen}, \citenamefont {Iwaki},\ and\ \citenamefont
  {Lengyel}}]{Johansen2003}%
  \BibitemOpen
  \bibfield  {author} {\bibinfo {author} {\bibfnamefont {K.~A.}\ \bibnamefont
  {Johansen}}, \bibinfo {author} {\bibfnamefont {D.~D.}\ \bibnamefont
  {Iwaki}},\ and\ \bibinfo {author} {\bibfnamefont {J.~A.}\ \bibnamefont
  {Lengyel}},\ }\bibfield  {title} {\bibinfo {title} {{Localized JAK/STAT
  signaling is required for oriented cell rearrangement in a tubular
  epithelium}},\ }\href {https://doi.org/10.1242/dev.00202} {\bibfield
  {journal} {\bibinfo  {journal} {Development}\ }\textbf {\bibinfo {volume}
  {130}},\ \bibinfo {pages} {135} (\bibinfo {year} {2003})}\BibitemShut
  {NoStop}%
\bibitem [{\citenamefont {Karner}\ \emph {et~al.}(2009)\citenamefont {Karner},
  \citenamefont {Chirumamilla}, \citenamefont {Aoki}, \citenamefont {Igarashi},
  \citenamefont {Wallingford},\ and\ \citenamefont {Carroll}}]{Karner2009}%
  \BibitemOpen
  \bibfield  {author} {\bibinfo {author} {\bibfnamefont {C.~M.}\ \bibnamefont
  {Karner}}, \bibinfo {author} {\bibfnamefont {R.}~\bibnamefont
  {Chirumamilla}}, \bibinfo {author} {\bibfnamefont {S.}~\bibnamefont {Aoki}},
  \bibinfo {author} {\bibfnamefont {P.}~\bibnamefont {Igarashi}}, \bibinfo
  {author} {\bibfnamefont {J.~B.}\ \bibnamefont {Wallingford}},\ and\ \bibinfo
  {author} {\bibfnamefont {T.~J.}\ \bibnamefont {Carroll}},\ }\bibfield
  {title} {\bibinfo {title} {{Wnt9b signaling regulates planar cell polarity
  and kidney tubule morphogenesis}},\ }\href {https://doi.org/10.1038/ng.400}
  {\bibfield  {journal} {\bibinfo  {journal} {Nature Genetics}\ }\textbf
  {\bibinfo {volume} {41}},\ \bibinfo {pages} {793} (\bibinfo {year}
  {2009})}\BibitemShut {NoStop}%
\bibitem [{\citenamefont {Saxena}\ \emph {et~al.}(2014)\citenamefont {Saxena},
  \citenamefont {Denholm}, \citenamefont {Bunt}, \citenamefont {Bischoff},
  \citenamefont {VijayRaghavan},\ and\ \citenamefont {Skaer}}]{Saxena2014}%
  \BibitemOpen
  \bibfield  {author} {\bibinfo {author} {\bibfnamefont {A.}~\bibnamefont
  {Saxena}}, \bibinfo {author} {\bibfnamefont {B.}~\bibnamefont {Denholm}},
  \bibinfo {author} {\bibfnamefont {S.}~\bibnamefont {Bunt}}, \bibinfo {author}
  {\bibfnamefont {M.}~\bibnamefont {Bischoff}}, \bibinfo {author}
  {\bibfnamefont {K.}~\bibnamefont {VijayRaghavan}},\ and\ \bibinfo {author}
  {\bibfnamefont {H.}~\bibnamefont {Skaer}},\ }\bibfield  {title} {\bibinfo
  {title} {{Epidermal Growth Factor Signalling Controls Myosin II Planar
  Polarity to Orchestrate Convergent Extension Movements during Drosophila
  Tubulogenesis}},\ }\href {https://doi.org/10.1371/journal.pbio.1002013}
  {\bibfield  {journal} {\bibinfo  {journal} {PLoS Biology}\ }\textbf {\bibinfo
  {volume} {12}},\ \bibinfo {pages} {1002013} (\bibinfo {year}
  {2014})}\BibitemShut {NoStop}%
\bibitem [{\citenamefont {Etournay}\ \emph {et~al.}(2015)\citenamefont
  {Etournay}, \citenamefont {Popovi{\'{c}}}, \citenamefont {Merkel},
  \citenamefont {Nandi}, \citenamefont {Blasse}, \citenamefont {Aigouy},
  \citenamefont {Brandl}, \citenamefont {Myers}, \citenamefont {Salbreux},
  \citenamefont {J{\"{u}}licher},\ and\ \citenamefont {Eaton}}]{Etournay2015}%
  \BibitemOpen
  \bibfield  {author} {\bibinfo {author} {\bibfnamefont {R.}~\bibnamefont
  {Etournay}}, \bibinfo {author} {\bibfnamefont {M.}~\bibnamefont
  {Popovi{\'{c}}}}, \bibinfo {author} {\bibfnamefont {M.}~\bibnamefont
  {Merkel}}, \bibinfo {author} {\bibfnamefont {A.}~\bibnamefont {Nandi}},
  \bibinfo {author} {\bibfnamefont {C.}~\bibnamefont {Blasse}}, \bibinfo
  {author} {\bibfnamefont {B.}~\bibnamefont {Aigouy}}, \bibinfo {author}
  {\bibfnamefont {H.}~\bibnamefont {Brandl}}, \bibinfo {author} {\bibfnamefont
  {G.}~\bibnamefont {Myers}}, \bibinfo {author} {\bibfnamefont
  {G.}~\bibnamefont {Salbreux}}, \bibinfo {author} {\bibfnamefont
  {F.}~\bibnamefont {J{\"{u}}licher}},\ and\ \bibinfo {author} {\bibfnamefont
  {S.}~\bibnamefont {Eaton}},\ }\bibfield  {title} {\bibinfo {title}
  {{Interplay of cell dynamics and epithelial tension during morphogenesis of
  the Drosophila pupal wing}},\ }\href {https://doi.org/10.7554/eLife.07090}
  {\bibfield  {journal} {\bibinfo  {journal} {Elife}\ }\textbf {\bibinfo
  {volume} {4}},\ \bibinfo {pages} {e07090} (\bibinfo {year}
  {2015})}\BibitemShut {NoStop}%
\bibitem [{\citenamefont {Hopyan}(2017)}]{Hopyan2017}%
  \BibitemOpen
  \bibfield  {author} {\bibinfo {author} {\bibfnamefont {S.}~\bibnamefont
  {Hopyan}},\ }\bibfield  {title} {\bibinfo {title} {Biophysical regulation of
  early limb bud morphogenesis},\ }\href
  {https://doi.org/10.1016/j.ydbio.2017.06.034} {\bibfield  {journal} {\bibinfo
   {journal} {Developmental Biology}\ }\textbf {\bibinfo {volume} {429}},\
  \bibinfo {pages} {429} (\bibinfo {year} {2017})}\BibitemShut {NoStop}%
\bibitem [{\citenamefont {Tao}\ \emph {et~al.}(2019)\citenamefont {Tao},
  \citenamefont {Zhu}, \citenamefont {Lau}, \citenamefont {Whitley},
  \citenamefont {Samani}, \citenamefont {Xiao}, \citenamefont {Chen},
  \citenamefont {Hahn}, \citenamefont {Liu}, \citenamefont {Valencia},
  \citenamefont {Wu}, \citenamefont {Wang}, \citenamefont {Fenelon},
  \citenamefont {Pasiliao}, \citenamefont {Hu}, \citenamefont {Wu},
  \citenamefont {Spring}, \citenamefont {Ferguson}, \citenamefont {Karuna},
  \citenamefont {Henkelman}, \citenamefont {Dunn}, \citenamefont {Huang},
  \citenamefont {Ho}, \citenamefont {Atit}, \citenamefont {Goyal},
  \citenamefont {Sun},\ and\ \citenamefont {Hopyan}}]{Tao2019}%
  \BibitemOpen
  \bibfield  {author} {\bibinfo {author} {\bibfnamefont {H.}~\bibnamefont
  {Tao}}, \bibinfo {author} {\bibfnamefont {M.}~\bibnamefont {Zhu}}, \bibinfo
  {author} {\bibfnamefont {K.}~\bibnamefont {Lau}}, \bibinfo {author}
  {\bibfnamefont {O.~K.~W.}\ \bibnamefont {Whitley}}, \bibinfo {author}
  {\bibfnamefont {M.}~\bibnamefont {Samani}}, \bibinfo {author} {\bibfnamefont
  {X.}~\bibnamefont {Xiao}}, \bibinfo {author} {\bibfnamefont {X.~X.}\
  \bibnamefont {Chen}}, \bibinfo {author} {\bibfnamefont {N.~A.}\ \bibnamefont
  {Hahn}}, \bibinfo {author} {\bibfnamefont {W.}~\bibnamefont {Liu}}, \bibinfo
  {author} {\bibfnamefont {M.}~\bibnamefont {Valencia}}, \bibinfo {author}
  {\bibfnamefont {M.}~\bibnamefont {Wu}}, \bibinfo {author} {\bibfnamefont
  {X.}~\bibnamefont {Wang}}, \bibinfo {author} {\bibfnamefont {K.~D.}\
  \bibnamefont {Fenelon}}, \bibinfo {author} {\bibfnamefont {C.~C.}\
  \bibnamefont {Pasiliao}}, \bibinfo {author} {\bibfnamefont {D.}~\bibnamefont
  {Hu}}, \bibinfo {author} {\bibfnamefont {J.}~\bibnamefont {Wu}}, \bibinfo
  {author} {\bibfnamefont {S.}~\bibnamefont {Spring}}, \bibinfo {author}
  {\bibfnamefont {J.}~\bibnamefont {Ferguson}}, \bibinfo {author}
  {\bibfnamefont {E.~P.}\ \bibnamefont {Karuna}}, \bibinfo {author}
  {\bibfnamefont {R.~M.}\ \bibnamefont {Henkelman}}, \bibinfo {author}
  {\bibfnamefont {A.}~\bibnamefont {Dunn}}, \bibinfo {author} {\bibfnamefont
  {H.}~\bibnamefont {Huang}}, \bibinfo {author} {\bibfnamefont {H.-Y.~H.}\
  \bibnamefont {Ho}}, \bibinfo {author} {\bibfnamefont {R.}~\bibnamefont
  {Atit}}, \bibinfo {author} {\bibfnamefont {S.}~\bibnamefont {Goyal}},
  \bibinfo {author} {\bibfnamefont {Y.}~\bibnamefont {Sun}},\ and\ \bibinfo
  {author} {\bibfnamefont {S.}~\bibnamefont {Hopyan}},\ }\bibfield  {title}
  {\bibinfo {title} {Oscillatory cortical forces promote three dimensional cell
  intercalations that shape the murine mandibular arch},\ }\href
  {https://doi.org/10.1038/s41467-019-09540-z} {\bibfield  {journal} {\bibinfo
  {journal} {Nature Communications}\ }\textbf {\bibinfo {volume} {10}},\
  \bibinfo {pages} {1703} (\bibinfo {year} {2019})}\BibitemShut {NoStop}%
\bibitem [{\citenamefont {Bertet}\ \emph {et~al.}(2004)\citenamefont {Bertet},
  \citenamefont {Sulak},\ and\ \citenamefont {Lecuit}}]{Bertet2004}%
  \BibitemOpen
  \bibfield  {author} {\bibinfo {author} {\bibfnamefont {C.}~\bibnamefont
  {Bertet}}, \bibinfo {author} {\bibfnamefont {L.}~\bibnamefont {Sulak}},\ and\
  \bibinfo {author} {\bibfnamefont {T.}~\bibnamefont {Lecuit}},\ }\bibfield
  {title} {\bibinfo {title} {{Myosin-dependent junction remodelling controls
  planar cell intercalation and axis elongation}},\ }\href
  {https://doi.org/10.1038/nature02590} {\bibfield  {journal} {\bibinfo
  {journal} {Nature}\ }\textbf {\bibinfo {volume} {429}},\ \bibinfo {pages}
  {667} (\bibinfo {year} {2004})}\BibitemShut {NoStop}%
\bibitem [{\citenamefont {Bosveld}\ \emph {et~al.}(2012)\citenamefont
  {Bosveld}, \citenamefont {Bonnet}, \citenamefont {Guirao}, \citenamefont
  {Tlili}, \citenamefont {Wang}, \citenamefont {Petitalot}, \citenamefont
  {Marchand}, \citenamefont {Bardet}, \citenamefont {Marcq}, \citenamefont
  {Graner},\ and\ \citenamefont {Bellaiche}}]{Bosveld2012}%
  \BibitemOpen
  \bibfield  {author} {\bibinfo {author} {\bibfnamefont {F.}~\bibnamefont
  {Bosveld}}, \bibinfo {author} {\bibfnamefont {I.}~\bibnamefont {Bonnet}},
  \bibinfo {author} {\bibfnamefont {B.}~\bibnamefont {Guirao}}, \bibinfo
  {author} {\bibfnamefont {S.}~\bibnamefont {Tlili}}, \bibinfo {author}
  {\bibfnamefont {Z.}~\bibnamefont {Wang}}, \bibinfo {author} {\bibfnamefont
  {A.}~\bibnamefont {Petitalot}}, \bibinfo {author} {\bibfnamefont
  {R.}~\bibnamefont {Marchand}}, \bibinfo {author} {\bibfnamefont {P.-L.}\
  \bibnamefont {Bardet}}, \bibinfo {author} {\bibfnamefont {P.}~\bibnamefont
  {Marcq}}, \bibinfo {author} {\bibfnamefont {F.}~\bibnamefont {Graner}},\ and\
  \bibinfo {author} {\bibfnamefont {Y.}~\bibnamefont {Bellaiche}},\ }\bibfield
  {title} {\bibinfo {title} {{Mechanical Control of Morphogenesis by
  Fat/Dachsous/Four-Jointed Planar Cell Polarity Pathway}},\ }\href
  {https://doi.org/10.1126/science.1221071} {\bibfield  {journal} {\bibinfo
  {journal} {Science (80-. ).}\ }\textbf {\bibinfo {volume} {336}},\ \bibinfo
  {pages} {724} (\bibinfo {year} {2012})}\BibitemShut {NoStop}%
\bibitem [{\citenamefont {Collinet}\ \emph {et~al.}(2015)\citenamefont
  {Collinet}, \citenamefont {Rauzi}, \citenamefont {Lenne},\ and\ \citenamefont
  {Lecuit}}]{Collinet2015}%
  \BibitemOpen
  \bibfield  {author} {\bibinfo {author} {\bibfnamefont {C.}~\bibnamefont
  {Collinet}}, \bibinfo {author} {\bibfnamefont {M.}~\bibnamefont {Rauzi}},
  \bibinfo {author} {\bibfnamefont {P.~F.}\ \bibnamefont {Lenne}},\ and\
  \bibinfo {author} {\bibfnamefont {T.}~\bibnamefont {Lecuit}},\ }\bibfield
  {title} {\bibinfo {title} {Local and tissue-scale forces drive oriented
  junction growth during tissue extension},\ }\href
  {https://doi.org/10.1038/ncb3226} {\bibfield  {journal} {\bibinfo  {journal}
  {Nature Cell Biology}\ }\textbf {\bibinfo {volume} {17}},\ \bibinfo {pages}
  {1247} (\bibinfo {year} {2015})}\BibitemShut {NoStop}%
\bibitem [{\citenamefont {Behrndt}\ \emph {et~al.}(2012)\citenamefont
  {Behrndt}, \citenamefont {Salbreux}, \citenamefont {Campinho}, \citenamefont
  {Hauschild}, \citenamefont {Oswald}, \citenamefont {Roensch}, \citenamefont
  {Grill},\ and\ \citenamefont {Heisenberg}}]{Behrndt2012}%
  \BibitemOpen
  \bibfield  {author} {\bibinfo {author} {\bibfnamefont {M.}~\bibnamefont
  {Behrndt}}, \bibinfo {author} {\bibfnamefont {G.}~\bibnamefont {Salbreux}},
  \bibinfo {author} {\bibfnamefont {P.}~\bibnamefont {Campinho}}, \bibinfo
  {author} {\bibfnamefont {R.}~\bibnamefont {Hauschild}}, \bibinfo {author}
  {\bibfnamefont {F.}~\bibnamefont {Oswald}}, \bibinfo {author} {\bibfnamefont
  {J.}~\bibnamefont {Roensch}}, \bibinfo {author} {\bibfnamefont {S.~W.}\
  \bibnamefont {Grill}},\ and\ \bibinfo {author} {\bibfnamefont {C.~P.}\
  \bibnamefont {Heisenberg}},\ }\bibfield  {title} {\bibinfo {title} {{Forces
  driving epithelial spreading in zebrafish gastrulation}},\ }\href
  {https://doi.org/10.1126/science.1224143} {\bibfield  {journal} {\bibinfo
  {journal} {Science}\ }\textbf {\bibinfo {volume} {338}},\ \bibinfo {pages}
  {257} (\bibinfo {year} {2012})}\BibitemShut {NoStop}%
\bibitem [{\citenamefont {Streichan}\ \emph {et~al.}(2018)\citenamefont
  {Streichan}, \citenamefont {Lefebvre}, \citenamefont {Noll}, \citenamefont
  {Wieschaus},\ and\ \citenamefont {Shraiman}}]{Streichan2018}%
  \BibitemOpen
  \bibfield  {author} {\bibinfo {author} {\bibfnamefont {S.~J.}\ \bibnamefont
  {Streichan}}, \bibinfo {author} {\bibfnamefont {M.}~\bibnamefont {Lefebvre}},
  \bibinfo {author} {\bibfnamefont {N.}~\bibnamefont {Noll}}, \bibinfo {author}
  {\bibfnamefont {E.~F.}\ \bibnamefont {Wieschaus}},\ and\ \bibinfo {author}
  {\bibfnamefont {B.~I.}\ \bibnamefont {Shraiman}},\ }\bibfield  {title}
  {\bibinfo {title} {Global morphogenetic flow is accurately predicted by the
  spatial distribution of myosin motors},\ }\href
  {https://doi.org/10.7554/eLife.27454} {\bibfield  {journal} {\bibinfo
  {journal} {eLife}\ }\textbf {\bibinfo {volume} {7}},\ \bibinfo {pages}
  {e27454} (\bibinfo {year} {2018})}\BibitemShut {NoStop}%
\bibitem [{\citenamefont {Stokkermans}\ \emph {et~al.}(2022)\citenamefont
  {Stokkermans}, \citenamefont {Chakrabarti}, \citenamefont {Subramanian},
  \citenamefont {Wang}, \citenamefont {Yin}, \citenamefont {Moghe},
  \citenamefont {Steenbergen}, \citenamefont {M{\"{o}}nke}, \citenamefont
  {Hiiragi}, \citenamefont {Prevedel}, \citenamefont {Mahadevan},\ and\
  \citenamefont {Ikmi}}]{Stokkermans2022}%
  \BibitemOpen
  \bibfield  {author} {\bibinfo {author} {\bibfnamefont {A.}~\bibnamefont
  {Stokkermans}}, \bibinfo {author} {\bibfnamefont {A.}~\bibnamefont
  {Chakrabarti}}, \bibinfo {author} {\bibfnamefont {K.}~\bibnamefont
  {Subramanian}}, \bibinfo {author} {\bibfnamefont {L.}~\bibnamefont {Wang}},
  \bibinfo {author} {\bibfnamefont {S.}~\bibnamefont {Yin}}, \bibinfo {author}
  {\bibfnamefont {P.}~\bibnamefont {Moghe}}, \bibinfo {author} {\bibfnamefont
  {P.}~\bibnamefont {Steenbergen}}, \bibinfo {author} {\bibfnamefont
  {G.}~\bibnamefont {M{\"{o}}nke}}, \bibinfo {author} {\bibfnamefont
  {T.}~\bibnamefont {Hiiragi}}, \bibinfo {author} {\bibfnamefont
  {R.}~\bibnamefont {Prevedel}}, \bibinfo {author} {\bibfnamefont
  {L.}~\bibnamefont {Mahadevan}},\ and\ \bibinfo {author} {\bibfnamefont
  {A.}~\bibnamefont {Ikmi}},\ }\bibfield  {title} {\bibinfo {title} {{Muscular
  hydraulics drive larva-polyp morphogenesis}},\ }\href
  {https://doi.org/10.1016/j.cub.2022.08.065} {\bibfield  {journal} {\bibinfo
  {journal} {Current Biology}\ }\textbf {\bibinfo {volume} {32}},\ \bibinfo
  {pages} {4707} (\bibinfo {year} {2022})}\BibitemShut {NoStop}%
\bibitem [{\citenamefont {Gehrels}\ \emph {et~al.}(2023)\citenamefont
  {Gehrels}, \citenamefont {Chakrabortty}, \citenamefont {Perrin},
  \citenamefont {Merkel},\ and\ \citenamefont {Lecuit}}]{Gehrels2023}%
  \BibitemOpen
  \bibfield  {author} {\bibinfo {author} {\bibfnamefont {E.~W.}\ \bibnamefont
  {Gehrels}}, \bibinfo {author} {\bibfnamefont {B.}~\bibnamefont
  {Chakrabortty}}, \bibinfo {author} {\bibfnamefont {M.-E.}\ \bibnamefont
  {Perrin}}, \bibinfo {author} {\bibfnamefont {M.}~\bibnamefont {Merkel}},\
  and\ \bibinfo {author} {\bibfnamefont {T.}~\bibnamefont {Lecuit}},\
  }\bibfield  {title} {\bibinfo {title} {Curvature gradient drives polarized
  tissue flow in the {{Drosophila}} embryo},\ }\href
  {https://doi.org/10.1073/pnas.2214205120} {\bibfield  {journal} {\bibinfo
  {journal} {Proceedings of the National Academy of Sciences}\ }\textbf
  {\bibinfo {volume} {120}},\ \bibinfo {pages} {e2214205120} (\bibinfo {year}
  {2023})}\BibitemShut {NoStop}%
\bibitem [{\citenamefont {Dye}\ \emph {et~al.}(2021)\citenamefont {Dye},
  \citenamefont {Popovic}, \citenamefont {Iyer}, \citenamefont {Fuhrmann},
  \citenamefont {Piscitello-Gómez}, \citenamefont {Eaton},\ and\ \citenamefont
  {Julicher}}]{Dye2021}%
  \BibitemOpen
  \bibfield  {author} {\bibinfo {author} {\bibfnamefont {N.~A.}\ \bibnamefont
  {Dye}}, \bibinfo {author} {\bibfnamefont {M.}~\bibnamefont {Popovic}},
  \bibinfo {author} {\bibfnamefont {K.~V.}\ \bibnamefont {Iyer}}, \bibinfo
  {author} {\bibfnamefont {J.~F.}\ \bibnamefont {Fuhrmann}}, \bibinfo {author}
  {\bibfnamefont {R.}~\bibnamefont {Piscitello-Gómez}}, \bibinfo {author}
  {\bibfnamefont {S.}~\bibnamefont {Eaton}},\ and\ \bibinfo {author}
  {\bibfnamefont {F.}~\bibnamefont {Julicher}},\ }\bibfield  {title} {\bibinfo
  {title} {Self-organized patterning of cell morphology via mechanosensitive
  feedback},\ }\bibfield  {journal} {\bibinfo  {journal} {eLife}\ }\textbf
  {\bibinfo {volume} {10}},\ \href {https://doi.org/10.7554/ELIFE.57964}
  {10.7554/ELIFE.57964} (\bibinfo {year} {2021})\BibitemShut {NoStop}%
\bibitem [{\citenamefont {Serra}\ \emph {et~al.}(2023)\citenamefont {Serra},
  \citenamefont {Nájera}, \citenamefont {Chuai}, \citenamefont {Plum},
  \citenamefont {Santhosh}, \citenamefont {Spandan}, \citenamefont {Weijer},\
  and\ \citenamefont {Mahadevan}}]{Serra2023}%
  \BibitemOpen
  \bibfield  {author} {\bibinfo {author} {\bibfnamefont {M.}~\bibnamefont
  {Serra}}, \bibinfo {author} {\bibfnamefont {G.~S.}\ \bibnamefont {Nájera}},
  \bibinfo {author} {\bibfnamefont {M.}~\bibnamefont {Chuai}}, \bibinfo
  {author} {\bibfnamefont {A.~M.}\ \bibnamefont {Plum}}, \bibinfo {author}
  {\bibfnamefont {S.}~\bibnamefont {Santhosh}}, \bibinfo {author}
  {\bibfnamefont {V.}~\bibnamefont {Spandan}}, \bibinfo {author} {\bibfnamefont
  {C.~J.}\ \bibnamefont {Weijer}},\ and\ \bibinfo {author} {\bibfnamefont
  {L.}~\bibnamefont {Mahadevan}},\ }\bibfield  {title} {\bibinfo {title} {A
  mechanochemical model recapitulates distinct vertebrate gastrulation modes},\
  }\href {https://doi.org/10.1126/SCIADV.ADH8152} {\bibfield  {journal}
  {\bibinfo  {journal} {Science advances}\ }\textbf {\bibinfo {volume} {9}},\
  \bibinfo {pages} {eadh8152} (\bibinfo {year} {2023})}\BibitemShut {NoStop}%
\bibitem [{\citenamefont {Gsell}\ \emph {et~al.}(2024)\citenamefont {Gsell},
  \citenamefont {Tlili}, \citenamefont {Merkel},\ and\ \citenamefont
  {Lenne}}]{Gsell2023}%
  \BibitemOpen
  \bibfield  {author} {\bibinfo {author} {\bibfnamefont {S.}~\bibnamefont
  {Gsell}}, \bibinfo {author} {\bibfnamefont {S.}~\bibnamefont {Tlili}},
  \bibinfo {author} {\bibfnamefont {M.}~\bibnamefont {Merkel}},\ and\ \bibinfo
  {author} {\bibfnamefont {P.-F.}\ \bibnamefont {Lenne}},\ }\bibfield  {title}
  {\bibinfo {title} {Marangoni-like tissue flows enhance symmetry breaking of
  embryonic organoids},\ }\bibfield  {journal} {\bibinfo  {journal} {bioRxiv}\
  }\href {https://doi.org/10.1101/2023.09.22.559003}
  {10.1101/2023.09.22.559003} (\bibinfo {year} {2024})\BibitemShut {NoStop}%
\bibitem [{\citenamefont {Barrett}\ \emph {et~al.}(2024)\citenamefont
  {Barrett}, \citenamefont {Anand}, \citenamefont {Thome}, \citenamefont
  {Lenne},\ and\ \citenamefont {Merkel}}]{Barrett2024}%
  \BibitemOpen
  \bibfield  {author} {\bibinfo {author} {\bibfnamefont {K.}~\bibnamefont
  {Barrett}}, \bibinfo {author} {\bibfnamefont {S.}~\bibnamefont {Anand}},
  \bibinfo {author} {\bibfnamefont {V.}~\bibnamefont {Thome}}, \bibinfo
  {author} {\bibfnamefont {P.-F.}\ \bibnamefont {Lenne}},\ and\ \bibinfo
  {author} {\bibfnamefont {M.}~\bibnamefont {Merkel}},\ }\bibfield  {title}
  {\bibinfo {title} {Epithelial-mesenchymal boundary guides cell shapes and
  axis elongation in embryonic explants},\ }\bibfield  {journal} {\bibinfo
  {journal} {bioRxiv}\ }\href {https://doi.org/10.1101/2024.08.20.608779}
  {10.1101/2024.08.20.608779} (\bibinfo {year} {2024})\BibitemShut {NoStop}%
\bibitem [{\citenamefont {Claussen}\ \emph {et~al.}(2024)\citenamefont
  {Claussen}, \citenamefont {Brauns},\ and\ \citenamefont
  {Shraiman}}]{Claussen2023}%
  \BibitemOpen
  \bibfield  {author} {\bibinfo {author} {\bibfnamefont {N.~H.}\ \bibnamefont
  {Claussen}}, \bibinfo {author} {\bibfnamefont {F.}~\bibnamefont {Brauns}},\
  and\ \bibinfo {author} {\bibfnamefont {B.~I.}\ \bibnamefont {Shraiman}},\
  }\bibfield  {title} {\bibinfo {title} {A geometric-tension-dynamics model of
  epithelial convergent extension},\ }\href
  {https://doi.org/10.1073/pnas.2321928121} {\bibfield  {journal} {\bibinfo
  {journal} {Proceedings of the National Academy of Sciences}\ }\textbf
  {\bibinfo {volume} {121}},\ \bibinfo {pages} {e2321928121} (\bibinfo {year}
  {2024})},\ \Eprint
  {https://arxiv.org/abs/https://www.pnas.org/doi/pdf/10.1073/pnas.2321928121}
  {https://www.pnas.org/doi/pdf/10.1073/pnas.2321928121} \BibitemShut {NoStop}%
\bibitem [{\citenamefont {Ioratim-Uba}\ \emph {et~al.}(2023)\citenamefont
  {Ioratim-Uba}, \citenamefont {Liverpool},\ and\ \citenamefont
  {Henkes}}]{Uba2023}%
  \BibitemOpen
  \bibfield  {author} {\bibinfo {author} {\bibfnamefont {A.}~\bibnamefont
  {Ioratim-Uba}}, \bibinfo {author} {\bibfnamefont {T.~B.}\ \bibnamefont
  {Liverpool}},\ and\ \bibinfo {author} {\bibfnamefont {S.}~\bibnamefont
  {Henkes}},\ }\bibfield  {title} {\bibinfo {title} {Mechanochemical active
  feedback generates convergence extension in epithelial tissue},\ }\href
  {https://doi.org/10.1103/PHYSREVLETT.131.238301/FIGURES/4/MEDIUM} {\bibfield
  {journal} {\bibinfo  {journal} {Physical Review Letters}\ }\textbf {\bibinfo
  {volume} {131}},\ \bibinfo {pages} {238301} (\bibinfo {year}
  {2023})}\BibitemShut {NoStop}%
\bibitem [{\citenamefont {Simha}\ and\ \citenamefont
  {Ramaswamy}(2002)}]{Simha2002}%
  \BibitemOpen
  \bibfield  {author} {\bibinfo {author} {\bibfnamefont {R.~A.}\ \bibnamefont
  {Simha}}\ and\ \bibinfo {author} {\bibfnamefont {S.}~\bibnamefont
  {Ramaswamy}},\ }\bibfield  {title} {\bibinfo {title} {{Hydrodynamic
  fluctuations and instabilities in ordered suspensions of self-propelled
  particles}},\ }\href {https://doi.org/10.1103/PhysRevLett.89.058101}
  {\bibfield  {journal} {\bibinfo  {journal} {Phys. Rev. Lett.}\ }\textbf
  {\bibinfo {volume} {89}},\ \bibinfo {pages} {058101} (\bibinfo {year}
  {2002})},\ \Eprint {https://arxiv.org/abs/0108301v2} {arXiv:0108301v2
  [arXiv:cond-mat]} \BibitemShut {NoStop}%
\bibitem [{\citenamefont {Voituriez}\ \emph {et~al.}(2005)\citenamefont
  {Voituriez}, \citenamefont {Joanny},\ and\ \citenamefont
  {Prost}}]{Voituriez2005}%
  \BibitemOpen
  \bibfield  {author} {\bibinfo {author} {\bibfnamefont {R.}~\bibnamefont
  {Voituriez}}, \bibinfo {author} {\bibfnamefont {J.~F.}\ \bibnamefont
  {Joanny}},\ and\ \bibinfo {author} {\bibfnamefont {J.}~\bibnamefont
  {Prost}},\ }\bibfield  {title} {\bibinfo {title} {{Spontaneous flow
  transition in active polar gels}},\ }\href
  {https://doi.org/10.1209/epl/i2004-10501-2} {\bibfield  {journal} {\bibinfo
  {journal} {Europhys. Lett.}\ }\textbf {\bibinfo {volume} {70}},\ \bibinfo
  {pages} {404} (\bibinfo {year} {2005})},\ \Eprint
  {https://arxiv.org/abs/0503022} {arXiv:0503022 [q-bio]} \BibitemShut
  {NoStop}%
\bibitem [{\citenamefont {Marchetti}\ \emph {et~al.}(2013)\citenamefont
  {Marchetti}, \citenamefont {Joanny}, \citenamefont {Ramaswamy}, \citenamefont
  {Liverpool}, \citenamefont {Prost}, \citenamefont {Rao},\ and\ \citenamefont
  {Simha}}]{Marchetti2013}%
  \BibitemOpen
  \bibfield  {author} {\bibinfo {author} {\bibfnamefont {M.~C.}\ \bibnamefont
  {Marchetti}}, \bibinfo {author} {\bibfnamefont {J.~F.}\ \bibnamefont
  {Joanny}}, \bibinfo {author} {\bibfnamefont {S.}~\bibnamefont {Ramaswamy}},
  \bibinfo {author} {\bibfnamefont {T.~B.}\ \bibnamefont {Liverpool}}, \bibinfo
  {author} {\bibfnamefont {J.}~\bibnamefont {Prost}}, \bibinfo {author}
  {\bibfnamefont {M.}~\bibnamefont {Rao}},\ and\ \bibinfo {author}
  {\bibfnamefont {R.~A.}\ \bibnamefont {Simha}},\ }\bibfield  {title} {\bibinfo
  {title} {Hydrodynamics of soft active matter},\ }\href
  {https://doi.org/10.1103/RevModPhys.85.1143} {\bibfield  {journal} {\bibinfo
  {journal} {Reviews of Modern Physics}\ }\textbf {\bibinfo {volume} {85}},\
  \bibinfo {pages} {1143} (\bibinfo {year} {2013})},\ \Eprint
  {https://arxiv.org/abs/1207.2929v1} {arXiv:1207.2929v1} \BibitemShut
  {NoStop}%
\bibitem [{\citenamefont {von~der Hardt}\ \emph {et~al.}(2007)\citenamefont
  {von~der Hardt}, \citenamefont {Bakkers}, \citenamefont {Inbal},
  \citenamefont {Carvalho}, \citenamefont {Solnica-Krezel}, \citenamefont
  {Heisenberg},\ and\ \citenamefont {Hammerschmidt}}]{VonderHardt2007}%
  \BibitemOpen
  \bibfield  {author} {\bibinfo {author} {\bibfnamefont {S.}~\bibnamefont
  {von~der Hardt}}, \bibinfo {author} {\bibfnamefont {J.}~\bibnamefont
  {Bakkers}}, \bibinfo {author} {\bibfnamefont {A.}~\bibnamefont {Inbal}},
  \bibinfo {author} {\bibfnamefont {L.}~\bibnamefont {Carvalho}}, \bibinfo
  {author} {\bibfnamefont {L.}~\bibnamefont {Solnica-Krezel}}, \bibinfo
  {author} {\bibfnamefont {C.~P.}\ \bibnamefont {Heisenberg}},\ and\ \bibinfo
  {author} {\bibfnamefont {M.}~\bibnamefont {Hammerschmidt}},\ }\bibfield
  {title} {\bibinfo {title} {{The Bmp Gradient of the Zebrafish Gastrula Guides
  Migrating Lateral Cells by Regulating Cell-Cell Adhesion}},\ }\href
  {https://doi.org/10.1016/j.cub.2007.02.013} {\bibfield  {journal} {\bibinfo
  {journal} {Current Biology}\ }\textbf {\bibinfo {volume} {17}},\ \bibinfo
  {pages} {475} (\bibinfo {year} {2007})}\BibitemShut {NoStop}%
\bibitem [{\citenamefont {Rogers}\ and\ \citenamefont
  {Schier}()}]{Rogers2011a}%
  \BibitemOpen
  \bibfield  {author} {\bibinfo {author} {\bibfnamefont {K.~W.}\ \bibnamefont
  {Rogers}}\ and\ \bibinfo {author} {\bibfnamefont {A.~F.}\ \bibnamefont
  {Schier}},\ }\bibfield  {title} {\bibinfo {title} {Morphogen {{Gradients}}:
  {{From Generation}} to {{Interpretation}}},\ }\href
  {https://doi.org/10.1146/annurev-cellbio-092910-154148} {\bibfield  {journal}
  {\bibinfo  {journal} {Annual Review of Cell and Developmental Biology}\
  }\textbf {\bibinfo {volume} {27}},\ \bibinfo {pages} {377}}\BibitemShut
  {NoStop}%
\bibitem [{\citenamefont {M{\"{u}}ller}\ \emph {et~al.}(2013)\citenamefont
  {M{\"{u}}ller}, \citenamefont {Rogers}, \citenamefont {Yu}, \citenamefont
  {Brand},\ and\ \citenamefont {Schier}}]{Muller2013}%
  \BibitemOpen
  \bibfield  {author} {\bibinfo {author} {\bibfnamefont {P.}~\bibnamefont
  {M{\"{u}}ller}}, \bibinfo {author} {\bibfnamefont {K.~W.}\ \bibnamefont
  {Rogers}}, \bibinfo {author} {\bibfnamefont {S.~R.}\ \bibnamefont {Yu}},
  \bibinfo {author} {\bibfnamefont {M.}~\bibnamefont {Brand}},\ and\ \bibinfo
  {author} {\bibfnamefont {A.~F.}\ \bibnamefont {Schier}},\ }\bibfield  {title}
  {\bibinfo {title} {{Morphogen transport}},\ }\href
  {https://doi.org/10.1242/dev.083519} {\bibfield  {journal} {\bibinfo
  {journal} {Development (Cambridge)}\ }\textbf {\bibinfo {volume} {140}},\
  \bibinfo {pages} {1621} (\bibinfo {year} {2013})}\BibitemShut {NoStop}%
\bibitem [{\citenamefont {Tka{\v c}ik}\ \emph {et~al.}(2015)\citenamefont
  {Tka{\v c}ik}, \citenamefont {Dubuis}, \citenamefont {Petkova},\ and\
  \citenamefont {Gregor}}]{Tkacik2014a}%
  \BibitemOpen
  \bibfield  {author} {\bibinfo {author} {\bibfnamefont {G.}~\bibnamefont
  {Tka{\v c}ik}}, \bibinfo {author} {\bibfnamefont {J.~O.}\ \bibnamefont
  {Dubuis}}, \bibinfo {author} {\bibfnamefont {M.~D.}\ \bibnamefont
  {Petkova}},\ and\ \bibinfo {author} {\bibfnamefont {T.}~\bibnamefont
  {Gregor}},\ }\bibfield  {title} {\bibinfo {title} {Positional
  {{Information}}, {{Positional Error}}, and {{Readout Precision}} in
  {{Morphogenesis}}: {{A Mathematical Framework}}},\ }\href
  {https://doi.org/10.1534/genetics.114.171850} {\bibfield  {journal} {\bibinfo
   {journal} {Genetics}\ }\textbf {\bibinfo {volume} {199}},\ \bibinfo {pages}
  {39} (\bibinfo {year} {2015})},\ \Eprint {https://arxiv.org/abs/1404.5599}
  {arXiv:1404.5599} \BibitemShut {NoStop}%
\bibitem [{\citenamefont {Mosby}\ \emph {et~al.}()\citenamefont {Mosby},
  \citenamefont {Bowen},\ and\ \citenamefont {Hadjivasiliou}}]{Mosby2024}%
  \BibitemOpen
  \bibfield  {author} {\bibinfo {author} {\bibfnamefont {L.~S.}\ \bibnamefont
  {Mosby}}, \bibinfo {author} {\bibfnamefont {A.~E.}\ \bibnamefont {Bowen}},\
  and\ \bibinfo {author} {\bibfnamefont {Z.}~\bibnamefont {Hadjivasiliou}},\
  }\bibfield  {title} {\bibinfo {title} {Morphogens in the evolution of size,
  shape and patterning},\ }\href {https://doi.org/10.1242/dev.202412}
  {\bibfield  {journal} {\bibinfo  {journal} {Development}\ }\textbf {\bibinfo
  {volume} {151}},\ \bibinfo {pages} {dev202412}}\BibitemShut {NoStop}%
\bibitem [{\citenamefont {Ninomiya}\ \emph {et~al.}(2004)\citenamefont
  {Ninomiya}, \citenamefont {Elinson},\ and\ \citenamefont
  {Winklbauer}}]{Ninomiya2004}%
  \BibitemOpen
  \bibfield  {author} {\bibinfo {author} {\bibfnamefont {H.}~\bibnamefont
  {Ninomiya}}, \bibinfo {author} {\bibfnamefont {R.~P.}\ \bibnamefont
  {Elinson}},\ and\ \bibinfo {author} {\bibfnamefont {R.}~\bibnamefont
  {Winklbauer}},\ }\bibfield  {title} {\bibinfo {title} {{Antero-posterior
  tissue polarity links mesoderm convergent extension to axial patterning}},\
  }\href {https://doi.org/10.1038/nature02620} {\bibfield  {journal} {\bibinfo
  {journal} {Nature}\ }\textbf {\bibinfo {volume} {430}},\ \bibinfo {pages}
  {364} (\bibinfo {year} {2004})}\BibitemShut {NoStop}%
\bibitem [{\citenamefont {Lavalou}\ \emph {et~al.}(2021)\citenamefont
  {Lavalou}, \citenamefont {Mao}, \citenamefont {Harmansa}, \citenamefont
  {Kerridge}, \citenamefont {Lellouch}, \citenamefont {Philippe}, \citenamefont
  {Audebert}, \citenamefont {Camoin},\ and\ \citenamefont
  {Lecuit}}]{Lavalou2021}%
  \BibitemOpen
  \bibfield  {author} {\bibinfo {author} {\bibfnamefont {J.}~\bibnamefont
  {Lavalou}}, \bibinfo {author} {\bibfnamefont {Q.}~\bibnamefont {Mao}},
  \bibinfo {author} {\bibfnamefont {S.}~\bibnamefont {Harmansa}}, \bibinfo
  {author} {\bibfnamefont {S.}~\bibnamefont {Kerridge}}, \bibinfo {author}
  {\bibfnamefont {A.~C.}\ \bibnamefont {Lellouch}}, \bibinfo {author}
  {\bibfnamefont {J.~M.}\ \bibnamefont {Philippe}}, \bibinfo {author}
  {\bibfnamefont {S.}~\bibnamefont {Audebert}}, \bibinfo {author}
  {\bibfnamefont {L.}~\bibnamefont {Camoin}},\ and\ \bibinfo {author}
  {\bibfnamefont {T.}~\bibnamefont {Lecuit}},\ }\bibfield  {title} {\bibinfo
  {title} {{Formation of polarized contractile interfaces by self-organized
  Toll-8/Cirl GPCR asymmetry}},\ }\href
  {https://doi.org/10.1016/j.devcel.2021.03.030} {\bibfield  {journal}
  {\bibinfo  {journal} {Dev. Cell}\ }\textbf {\bibinfo {volume} {56}},\
  \bibinfo {pages} {1574} (\bibinfo {year} {2021})}\BibitemShut {NoStop}%
\bibitem [{\citenamefont {Wang}\ \emph {et~al.}(2023)\citenamefont {Wang},
  \citenamefont {Marchetti},\ and\ \citenamefont {Brauns}}]{Wang2023}%
  \BibitemOpen
  \bibfield  {author} {\bibinfo {author} {\bibfnamefont {Z.}~\bibnamefont
  {Wang}}, \bibinfo {author} {\bibfnamefont {M.~C.}\ \bibnamefont
  {Marchetti}},\ and\ \bibinfo {author} {\bibfnamefont {F.}~\bibnamefont
  {Brauns}},\ }\bibfield  {title} {\bibinfo {title} {{Patterning of
  morphogenetic anisotropy fields}},\ }\href
  {https://doi.org/10.1073/pnas.2220167120} {\bibfield  {journal} {\bibinfo
  {journal} {Proceedings of the National Academy of Sciences of the United
  States of America}\ }\textbf {\bibinfo {volume} {120}},\ \bibinfo {pages}
  {e2220167120} (\bibinfo {year} {2023})},\ \Eprint
  {https://arxiv.org/abs/2212.12215} {arXiv:2212.12215} \BibitemShut {NoStop}%
\bibitem [{\citenamefont {Ibrahimi}\ and\ \citenamefont
  {Merkel}(2023)}]{Ibrahimi2023}%
  \BibitemOpen
  \bibfield  {author} {\bibinfo {author} {\bibfnamefont {M.}~\bibnamefont
  {Ibrahimi}}\ and\ \bibinfo {author} {\bibfnamefont {M.}~\bibnamefont
  {Merkel}},\ }\bibfield  {title} {\bibinfo {title} {{Deforming polar active
  matter in a scalar field gradient}},\ }\href
  {https://doi.org/10.1088/1367-2630/ACB2E5} {\bibfield  {journal} {\bibinfo
  {journal} {New Journal of Physics}\ }\textbf {\bibinfo {volume} {25}},\
  \bibinfo {pages} {013022} (\bibinfo {year} {2023})},\ \Eprint
  {https://arxiv.org/abs/2206.12850} {arXiv:2206.12850} \BibitemShut {NoStop}%
\bibitem [{\citenamefont {Lefebvre}\ \emph {et~al.}(2023)\citenamefont
  {Lefebvre}, \citenamefont {Claussen}, \citenamefont {Mitchell}, \citenamefont
  {Gustafson},\ and\ \citenamefont {Streichan}}]{Lefebvre2022}%
  \BibitemOpen
  \bibfield  {author} {\bibinfo {author} {\bibfnamefont {M.~F.}\ \bibnamefont
  {Lefebvre}}, \bibinfo {author} {\bibfnamefont {N.~H.}\ \bibnamefont
  {Claussen}}, \bibinfo {author} {\bibfnamefont {N.~P.}\ \bibnamefont
  {Mitchell}}, \bibinfo {author} {\bibfnamefont {H.~J.}\ \bibnamefont
  {Gustafson}},\ and\ \bibinfo {author} {\bibfnamefont {S.~J.}\ \bibnamefont
  {Streichan}},\ }\bibfield  {title} {\bibinfo {title} {{Geometric control of
  Myosin-II orientation during axis elongation}},\ }\href
  {https://doi.org/10.7554/eLife.78787} {\bibfield  {journal} {\bibinfo
  {journal} {eLife}\ }\textbf {\bibinfo {volume} {12}},\ \bibinfo {pages}
  {2022.01.12.476069} (\bibinfo {year} {2023})}\BibitemShut {NoStop}%
\bibitem [{\citenamefont {Plum}\ and\ \citenamefont {Serra}(2025)}]{Plum2025b}%
  \BibitemOpen
  \bibfield  {author} {\bibinfo {author} {\bibfnamefont {A.~M.}\ \bibnamefont
  {Plum}}\ and\ \bibinfo {author} {\bibfnamefont {M.}~\bibnamefont {Serra}},\
  }\href {https://doi.org/10.1101/2025.01.04.631293} {\bibinfo {title}
  {Morphogen {{Patterning}} in {{Dynamic Tissues}}}} (\bibinfo {year}
  {2025})\BibitemShut {NoStop}%
\bibitem [{\citenamefont {Wartlick}\ \emph {et~al.}(2011)\citenamefont
  {Wartlick}, \citenamefont {Mumcu}, \citenamefont {Kicheva}, \citenamefont
  {Bittig}, \citenamefont {Seum}, \citenamefont {J{\"{u}}licher},\ and\
  \citenamefont {Gonz{\'{a}}lez-Gait{\'{a}}n}}]{Wartlick2011}%
  \BibitemOpen
  \bibfield  {author} {\bibinfo {author} {\bibfnamefont {O.}~\bibnamefont
  {Wartlick}}, \bibinfo {author} {\bibfnamefont {P.}~\bibnamefont {Mumcu}},
  \bibinfo {author} {\bibfnamefont {A.}~\bibnamefont {Kicheva}}, \bibinfo
  {author} {\bibfnamefont {T.}~\bibnamefont {Bittig}}, \bibinfo {author}
  {\bibfnamefont {C.}~\bibnamefont {Seum}}, \bibinfo {author} {\bibfnamefont
  {F.}~\bibnamefont {J{\"{u}}licher}},\ and\ \bibinfo {author} {\bibfnamefont
  {M.}~\bibnamefont {Gonz{\'{a}}lez-Gait{\'{a}}n}},\ }\bibfield  {title}
  {\bibinfo {title} {{Dynamics of Dpp signaling and proliferation control}},\
  }\href {https://doi.org/10.1126/science.1200037} {\bibfield  {journal}
  {\bibinfo  {journal} {Science}\ }\textbf {\bibinfo {volume} {331}},\ \bibinfo
  {pages} {1154} (\bibinfo {year} {2011})}\BibitemShut {NoStop}%
\bibitem [{\citenamefont {Mateus}\ \emph {et~al.}(2020)\citenamefont {Mateus},
  \citenamefont {Holtzer}, \citenamefont {Seum}, \citenamefont {Hadjivasiliou},
  \citenamefont {Dubois}, \citenamefont {J{\"u}licher},\ and\ \citenamefont
  {{Gonzalez-Gaitan}}}]{Mateus2020a}%
  \BibitemOpen
  \bibfield  {author} {\bibinfo {author} {\bibfnamefont {R.}~\bibnamefont
  {Mateus}}, \bibinfo {author} {\bibfnamefont {L.}~\bibnamefont {Holtzer}},
  \bibinfo {author} {\bibfnamefont {C.}~\bibnamefont {Seum}}, \bibinfo {author}
  {\bibfnamefont {Z.}~\bibnamefont {Hadjivasiliou}}, \bibinfo {author}
  {\bibfnamefont {M.}~\bibnamefont {Dubois}}, \bibinfo {author} {\bibfnamefont
  {F.}~\bibnamefont {J{\"u}licher}},\ and\ \bibinfo {author} {\bibfnamefont
  {M.}~\bibnamefont {{Gonzalez-Gaitan}}},\ }\bibfield  {title} {\bibinfo
  {title} {{{BMP Signaling Gradient Scaling}} in the {{Zebrafish Pectoral
  Fin}}},\ }\href {https://doi.org/10.1016/j.celrep.2020.03.024} {\bibfield
  {journal} {\bibinfo  {journal} {Cell Reports}\ }\textbf {\bibinfo {volume}
  {30}},\ \bibinfo {pages} {4292} (\bibinfo {year} {2020})}\BibitemShut
  {NoStop}%
\bibitem [{\citenamefont {Romanova-Michaelides}\ \emph
  {et~al.}(2022)\citenamefont {Romanova-Michaelides}, \citenamefont
  {Hadjivasiliou}, \citenamefont {Aguilar-Hidalgo}, \citenamefont
  {Basagiannis}, \citenamefont {Seum}, \citenamefont {Dubois}, \citenamefont
  {J{\"{u}}licher},\ and\ \citenamefont
  {Gonzalez-Gaitan}}]{Romanova-Michaelides2022}%
  \BibitemOpen
  \bibfield  {author} {\bibinfo {author} {\bibfnamefont {M.}~\bibnamefont
  {Romanova-Michaelides}}, \bibinfo {author} {\bibfnamefont {Z.}~\bibnamefont
  {Hadjivasiliou}}, \bibinfo {author} {\bibfnamefont {D.}~\bibnamefont
  {Aguilar-Hidalgo}}, \bibinfo {author} {\bibfnamefont {D.}~\bibnamefont
  {Basagiannis}}, \bibinfo {author} {\bibfnamefont {C.}~\bibnamefont {Seum}},
  \bibinfo {author} {\bibfnamefont {M.}~\bibnamefont {Dubois}}, \bibinfo
  {author} {\bibfnamefont {F.}~\bibnamefont {J{\"{u}}licher}},\ and\ \bibinfo
  {author} {\bibfnamefont {M.}~\bibnamefont {Gonzalez-Gaitan}},\ }\bibfield
  {title} {\bibinfo {title} {{Morphogen gradient scaling by recycling of
  intracellular Dpp}},\ }\href {https://doi.org/10.1038/s41586-021-04346-w}
  {\bibfield  {journal} {\bibinfo  {journal} {Nature}\ }\textbf {\bibinfo
  {volume} {602}},\ \bibinfo {pages} {287} (\bibinfo {year}
  {2022})}\BibitemShut {NoStop}%
\bibitem [{\citenamefont {Wartlick}\ \emph {et~al.}(2009)\citenamefont
  {Wartlick}, \citenamefont {Kicheva},\ and\ \citenamefont
  {Gonz{\'{a}}lez-Gait{\'{a}}n}}]{Wartlick2009}%
  \BibitemOpen
  \bibfield  {author} {\bibinfo {author} {\bibfnamefont {O.}~\bibnamefont
  {Wartlick}}, \bibinfo {author} {\bibfnamefont {A.}~\bibnamefont {Kicheva}},\
  and\ \bibinfo {author} {\bibfnamefont {M.}~\bibnamefont
  {Gonz{\'{a}}lez-Gait{\'{a}}n}},\ }\bibfield  {title} {\bibinfo {title}
  {{Morphogen gradient formation.}},\ }\bibfield  {journal} {\bibinfo
  {journal} {Cold Spring Harbor perspectives in biology}\ }\textbf {\bibinfo
  {volume} {1}},\ \href {https://doi.org/10.1101/cshperspect.a001255}
  {10.1101/cshperspect.a001255} (\bibinfo {year} {2009})\BibitemShut {NoStop}%
\bibitem [{\citenamefont {Wang}\ \emph {et~al.}(2016)\citenamefont {Wang},
  \citenamefont {Wang}, \citenamefont {Wohland},\ and\ \citenamefont
  {Sampath}}]{Wang2016}%
  \BibitemOpen
  \bibfield  {author} {\bibinfo {author} {\bibfnamefont {Y.}~\bibnamefont
  {Wang}}, \bibinfo {author} {\bibfnamefont {X.}~\bibnamefont {Wang}}, \bibinfo
  {author} {\bibfnamefont {T.}~\bibnamefont {Wohland}},\ and\ \bibinfo {author}
  {\bibfnamefont {K.}~\bibnamefont {Sampath}},\ }\bibfield  {title} {\bibinfo
  {title} {{Extracellular interactions and ligand degradation shape the nodal
  morphogen gradient}},\ }\bibfield  {journal} {\bibinfo  {journal} {eLife}\
  }\textbf {\bibinfo {volume} {5}},\ \href
  {https://doi.org/10.7554/eLife.13879} {10.7554/eLife.13879} (\bibinfo {year}
  {2016})\BibitemShut {NoStop}%
\bibitem [{\citenamefont {Huang}\ \emph {et~al.}(2017)\citenamefont {Huang},
  \citenamefont {Amourda}, \citenamefont {Zhang}, \citenamefont {Tolwinski},\
  and\ \citenamefont {Saunders}}]{Huang2017}%
  \BibitemOpen
  \bibfield  {author} {\bibinfo {author} {\bibfnamefont {A.}~\bibnamefont
  {Huang}}, \bibinfo {author} {\bibfnamefont {C.}~\bibnamefont {Amourda}},
  \bibinfo {author} {\bibfnamefont {S.}~\bibnamefont {Zhang}}, \bibinfo
  {author} {\bibfnamefont {N.~S.}\ \bibnamefont {Tolwinski}},\ and\ \bibinfo
  {author} {\bibfnamefont {T.~E.}\ \bibnamefont {Saunders}},\ }\bibfield
  {title} {\bibinfo {title} {Decoding temporal interpretation of the morphogen
  bicoid in the early drosophila embryo},\ }\href
  {https://doi.org/10.7554/eLife.26258} {\bibfield  {journal} {\bibinfo
  {journal} {eLife}\ }\textbf {\bibinfo {volume} {6}},\ \bibinfo {pages} {1}
  (\bibinfo {year} {2017})}\BibitemShut {NoStop}%
\bibitem [{\citenamefont {Par{\'{e}}}\ \emph {et~al.}(2014)\citenamefont
  {Par{\'{e}}}, \citenamefont {Vichas}, \citenamefont {Fincher}, \citenamefont
  {Mirman}, \citenamefont {Farrell}, \citenamefont {Mainieri},\ and\
  \citenamefont {Zallen}}]{Pare2014}%
  \BibitemOpen
  \bibfield  {author} {\bibinfo {author} {\bibfnamefont {A.~C.}\ \bibnamefont
  {Par{\'{e}}}}, \bibinfo {author} {\bibfnamefont {A.}~\bibnamefont {Vichas}},
  \bibinfo {author} {\bibfnamefont {C.~T.}\ \bibnamefont {Fincher}}, \bibinfo
  {author} {\bibfnamefont {Z.}~\bibnamefont {Mirman}}, \bibinfo {author}
  {\bibfnamefont {D.~L.}\ \bibnamefont {Farrell}}, \bibinfo {author}
  {\bibfnamefont {A.}~\bibnamefont {Mainieri}},\ and\ \bibinfo {author}
  {\bibfnamefont {J.~A.}\ \bibnamefont {Zallen}},\ }\bibfield  {title}
  {\bibinfo {title} {{A positional Toll receptor code directs convergent
  extension in Drosophila}},\ }\href {https://doi.org/10.1038/nature13953}
  {\bibfield  {journal} {\bibinfo  {journal} {Nature}\ }\textbf {\bibinfo
  {volume} {515}},\ \bibinfo {pages} {523} (\bibinfo {year}
  {2014})}\BibitemShut {NoStop}%
\bibitem [{\citenamefont {Benton}\ \emph {et~al.}(2016)\citenamefont {Benton},
  \citenamefont {Pechmann}, \citenamefont {Frey}, \citenamefont {Stappert},
  \citenamefont {Conrads}, \citenamefont {Chen}, \citenamefont {Stamataki},
  \citenamefont {Pavlopoulos},\ and\ \citenamefont {Roth}}]{Benton2016}%
  \BibitemOpen
  \bibfield  {author} {\bibinfo {author} {\bibfnamefont {M.~A.}\ \bibnamefont
  {Benton}}, \bibinfo {author} {\bibfnamefont {M.}~\bibnamefont {Pechmann}},
  \bibinfo {author} {\bibfnamefont {N.}~\bibnamefont {Frey}}, \bibinfo {author}
  {\bibfnamefont {D.}~\bibnamefont {Stappert}}, \bibinfo {author}
  {\bibfnamefont {K.~H.}\ \bibnamefont {Conrads}}, \bibinfo {author}
  {\bibfnamefont {Y.~T.}\ \bibnamefont {Chen}}, \bibinfo {author}
  {\bibfnamefont {E.}~\bibnamefont {Stamataki}}, \bibinfo {author}
  {\bibfnamefont {A.}~\bibnamefont {Pavlopoulos}},\ and\ \bibinfo {author}
  {\bibfnamefont {S.}~\bibnamefont {Roth}},\ }\bibfield  {title} {\bibinfo
  {title} {{Toll Genes Have an Ancestral Role in Axis Elongation}},\ }\href
  {https://doi.org/10.1016/j.cub.2016.04.055} {\bibfield  {journal} {\bibinfo
  {journal} {Current Biology}\ }\textbf {\bibinfo {volume} {26}},\ \bibinfo
  {pages} {1609} (\bibinfo {year} {2016})}\BibitemShut {NoStop}%
\bibitem [{\citenamefont {Jia}\ \emph {et~al.}(2016)\citenamefont {Jia},
  \citenamefont {Xu}, \citenamefont {Xie}, \citenamefont {Mio},\ and\
  \citenamefont {Deng}}]{Jia2016}%
  \BibitemOpen
  \bibfield  {author} {\bibinfo {author} {\bibfnamefont {D.}~\bibnamefont
  {Jia}}, \bibinfo {author} {\bibfnamefont {Q.}~\bibnamefont {Xu}}, \bibinfo
  {author} {\bibfnamefont {Q.}~\bibnamefont {Xie}}, \bibinfo {author}
  {\bibfnamefont {W.}~\bibnamefont {Mio}},\ and\ \bibinfo {author}
  {\bibfnamefont {W.~M.}\ \bibnamefont {Deng}},\ }\bibfield  {title} {\bibinfo
  {title} {{Automatic stage identification of Drosophila egg chamber based on
  DAPI images}},\ }\href {https://doi.org/10.1038/srep18850} {\bibfield
  {journal} {\bibinfo  {journal} {Scientific Reports}\ }\textbf {\bibinfo
  {volume} {6}},\ \bibinfo {pages} {1} (\bibinfo {year} {2016})}\BibitemShut
  {NoStop}%
\bibitem [{\citenamefont {Al{\'{e}}got}\ \emph {et~al.}(2018)\citenamefont
  {Al{\'{e}}got}, \citenamefont {Pouchin}, \citenamefont {Bardot},\ and\
  \citenamefont {Mirouse}}]{Alegot2018}%
  \BibitemOpen
  \bibfield  {author} {\bibinfo {author} {\bibfnamefont {H.}~\bibnamefont
  {Al{\'{e}}got}}, \bibinfo {author} {\bibfnamefont {P.}~\bibnamefont
  {Pouchin}}, \bibinfo {author} {\bibfnamefont {O.}~\bibnamefont {Bardot}},\
  and\ \bibinfo {author} {\bibfnamefont {V.}~\bibnamefont {Mirouse}},\
  }\bibfield  {title} {\bibinfo {title} {{Jak-stat pathway induces Drosophila
  follicle elongation by a gradient of apical contractility}},\ }\bibfield
  {journal} {\bibinfo  {journal} {eLife}\ }\textbf {\bibinfo {volume} {7}},\
  \href {https://doi.org/10.7554/eLife.32943} {10.7554/eLife.32943} (\bibinfo
  {year} {2018})\BibitemShut {NoStop}%
\bibitem [{\citenamefont {Iwaki}\ \emph {et~al.}(2001)\citenamefont {Iwaki},
  \citenamefont {Johansen}, \citenamefont {Singer},\ and\ \citenamefont
  {Lengyel}}]{Iwaki2001}%
  \BibitemOpen
  \bibfield  {author} {\bibinfo {author} {\bibfnamefont {D.~D.}\ \bibnamefont
  {Iwaki}}, \bibinfo {author} {\bibfnamefont {K.~A.}\ \bibnamefont {Johansen}},
  \bibinfo {author} {\bibfnamefont {J.~B.}\ \bibnamefont {Singer}},\ and\
  \bibinfo {author} {\bibfnamefont {J.~A.}\ \bibnamefont {Lengyel}},\
  }\bibfield  {title} {\bibinfo {title} {{Drumstick, bowl, and lines are
  required for patterning and cell rearrangement in the Drosophila embryonic
  hindgut}},\ }\href {https://doi.org/10.1006/dbio.2001.0483} {\bibfield
  {journal} {\bibinfo  {journal} {Developmental Biology}\ }\textbf {\bibinfo
  {volume} {240}},\ \bibinfo {pages} {611} (\bibinfo {year}
  {2001})}\BibitemShut {NoStop}%
\bibitem [{\citenamefont {Williams}\ and\ \citenamefont
  {Solnica-Krezel}(2020)}]{Williams2020}%
  \BibitemOpen
  \bibfield  {author} {\bibinfo {author} {\bibfnamefont {M.~L.}\ \bibnamefont
  {Williams}}\ and\ \bibinfo {author} {\bibfnamefont {L.}~\bibnamefont
  {Solnica-Krezel}},\ }\bibfield  {title} {\bibinfo {title} {{Nodal and planar
  cell polarity signaling cooperate to regulate zebrafish convergence and
  extension gastrulation movements}},\ }\bibfield  {journal} {\bibinfo
  {journal} {eLife}\ }\textbf {\bibinfo {volume} {9}},\ \href
  {https://doi.org/10.7554/eLife.54445} {10.7554/eLife.54445} (\bibinfo {year}
  {2020})\BibitemShut {NoStop}%
\bibitem [{\citenamefont {Steventon}\ \emph {et~al.}(2016)\citenamefont
  {Steventon}, \citenamefont {Duarte}, \citenamefont {Lagadec}, \citenamefont
  {Mazan}, \citenamefont {Nicolas},\ and\ \citenamefont
  {Hirsinger}}]{Steventon2016}%
  \BibitemOpen
  \bibfield  {author} {\bibinfo {author} {\bibfnamefont {B.}~\bibnamefont
  {Steventon}}, \bibinfo {author} {\bibfnamefont {F.}~\bibnamefont {Duarte}},
  \bibinfo {author} {\bibfnamefont {R.}~\bibnamefont {Lagadec}}, \bibinfo
  {author} {\bibfnamefont {S.}~\bibnamefont {Mazan}}, \bibinfo {author}
  {\bibfnamefont {J.~F.}\ \bibnamefont {Nicolas}},\ and\ \bibinfo {author}
  {\bibfnamefont {E.}~\bibnamefont {Hirsinger}},\ }\bibfield  {title} {\bibinfo
  {title} {Species-specific contribution of volumetric growth and tissue
  convergence to posterior body elongation in vertebrates},\ }\href
  {https://doi.org/10.1242/dev.126375} {\bibfield  {journal} {\bibinfo
  {journal} {Development (Cambridge)}\ }\textbf {\bibinfo {volume} {143}},\
  \bibinfo {pages} {1732} (\bibinfo {year} {2016})}\BibitemShut {NoStop}%
\bibitem [{\citenamefont {Kicheva}\ \emph {et~al.}(2007)\citenamefont
  {Kicheva}, \citenamefont {Pantazis}, \citenamefont {Bollenbach},
  \citenamefont {Kalaidzidis}, \citenamefont {Bittig}, \citenamefont
  {J{\"{u}}licher},\ and\ \citenamefont
  {Gonz{\'{a}}lez-Gait{\'{a}}n}}]{Kicheva2007}%
  \BibitemOpen
  \bibfield  {author} {\bibinfo {author} {\bibfnamefont {A.}~\bibnamefont
  {Kicheva}}, \bibinfo {author} {\bibfnamefont {P.}~\bibnamefont {Pantazis}},
  \bibinfo {author} {\bibfnamefont {T.}~\bibnamefont {Bollenbach}}, \bibinfo
  {author} {\bibfnamefont {Y.}~\bibnamefont {Kalaidzidis}}, \bibinfo {author}
  {\bibfnamefont {T.}~\bibnamefont {Bittig}}, \bibinfo {author} {\bibfnamefont
  {F.}~\bibnamefont {J{\"{u}}licher}},\ and\ \bibinfo {author} {\bibfnamefont
  {M.}~\bibnamefont {Gonz{\'{a}}lez-Gait{\'{a}}n}},\ }\bibfield  {title}
  {\bibinfo {title} {{Kinetics of morphogen gradient formation}},\ }\href
  {https://doi.org/10.1126/science.1135774} {\bibfield  {journal} {\bibinfo
  {journal} {Science}\ }\textbf {\bibinfo {volume} {315}},\ \bibinfo {pages}
  {521} (\bibinfo {year} {2007})}\BibitemShut {NoStop}%
\bibitem [{\citenamefont {Tiribocchi}\ \emph {et~al.}(2015)\citenamefont
  {Tiribocchi}, \citenamefont {Wittkowski}, \citenamefont {Marenduzzo},\ and\
  \citenamefont {Cates}}]{Tiribocchi2015a}%
  \BibitemOpen
  \bibfield  {author} {\bibinfo {author} {\bibfnamefont {A.}~\bibnamefont
  {Tiribocchi}}, \bibinfo {author} {\bibfnamefont {R.}~\bibnamefont
  {Wittkowski}}, \bibinfo {author} {\bibfnamefont {D.}~\bibnamefont
  {Marenduzzo}},\ and\ \bibinfo {author} {\bibfnamefont {M.~E.}\ \bibnamefont
  {Cates}},\ }\bibfield  {title} {\bibinfo {title} {{Active Model H: Scalar
  Active Matter in a Momentum-Conserving Fluid}},\ }\href
  {https://doi.org/10.1103/PhysRevLett.115.188302} {\bibfield  {journal}
  {\bibinfo  {journal} {Phys. Rev. Lett.}\ }\textbf {\bibinfo {volume} {115}},\
  \bibinfo {pages} {1} (\bibinfo {year} {2015})},\ \Eprint
  {https://arxiv.org/abs/1504.07447v1} {arXiv:1504.07447v1} \BibitemShut
  {NoStop}%
\bibitem [{\citenamefont {Kirkpatrick}\ and\ \citenamefont
  {Bhattacherjee}(2019)}]{Kirkpatrick2019a}%
  \BibitemOpen
  \bibfield  {author} {\bibinfo {author} {\bibfnamefont {T.~R.}\ \bibnamefont
  {Kirkpatrick}}\ and\ \bibinfo {author} {\bibfnamefont {J.~K.}\ \bibnamefont
  {Bhattacherjee}},\ }\bibfield  {title} {\bibinfo {title} {{Driven active
  matter: Fluctuations and a hydrodynamic instability}},\ }\href
  {https://doi.org/10.1103/PhysRevFluids.4.024306} {\bibfield  {journal}
  {\bibinfo  {journal} {Phys. Rev. Fluids}\ }\textbf {\bibinfo {volume} {4}},\
  \bibinfo {pages} {1} (\bibinfo {year} {2019})}\BibitemShut {NoStop}%
\bibitem [{\citenamefont {Gao}(2012)}]{Gao2012}%
  \BibitemOpen
  \bibfield  {author} {\bibinfo {author} {\bibfnamefont {B.}~\bibnamefont
  {Gao}},\ }\bibfield  {title} {\bibinfo {title} {{Wnt Regulation of Planar
  Cell Polarity (PCP)}},\ }\href
  {https://doi.org/10.1016/B978-0-12-394592-1.00008-9} {\bibfield  {journal}
  {\bibinfo  {journal} {Current Topics in Developmental Biology}\ }\textbf
  {\bibinfo {volume} {101}},\ \bibinfo {pages} {263} (\bibinfo {year}
  {2012})}\BibitemShut {NoStop}%
\bibitem [{\citenamefont {Merkel}\ \emph {et~al.}(2014)\citenamefont {Merkel},
  \citenamefont {Sagner}, \citenamefont {Gruber}, \citenamefont {Etournay},
  \citenamefont {Blasse}, \citenamefont {Myers}, \citenamefont {Eaton},\ and\
  \citenamefont {J{\"u}licher}}]{Merkel2014}%
  \BibitemOpen
  \bibfield  {author} {\bibinfo {author} {\bibfnamefont {M.}~\bibnamefont
  {Merkel}}, \bibinfo {author} {\bibfnamefont {A.}~\bibnamefont {Sagner}},
  \bibinfo {author} {\bibfnamefont {F.~S.}\ \bibnamefont {Gruber}}, \bibinfo
  {author} {\bibfnamefont {R.}~\bibnamefont {Etournay}}, \bibinfo {author}
  {\bibfnamefont {C.}~\bibnamefont {Blasse}}, \bibinfo {author} {\bibfnamefont
  {E.}~\bibnamefont {Myers}}, \bibinfo {author} {\bibfnamefont
  {S.}~\bibnamefont {Eaton}},\ and\ \bibinfo {author} {\bibfnamefont
  {F.}~\bibnamefont {J{\"u}licher}},\ }\bibfield  {title} {\bibinfo {title}
  {The balance of prickle/spiny-legs isoforms controls the amount of coupling
  between core and fat {{PCP}} systems},\ }\href
  {https://doi.org/10.1016/j.cub.2014.08.005} {\bibfield  {journal} {\bibinfo
  {journal} {Current Biology}\ }\textbf {\bibinfo {volume} {24}},\ \bibinfo
  {pages} {2111} (\bibinfo {year} {2014})}\BibitemShut {NoStop}%
\bibitem [{Note1()}]{Note1}%
  \BibitemOpen
  \bibinfo {note} {See Supplemental Material at \protect \url
  {https://arxiv.org/abs/2412.15774} for movies showing the
  simulations.}\BibitemShut {Stop}%
\bibitem [{\citenamefont {Lienkamp}\ \emph {et~al.}(2012)\citenamefont
  {Lienkamp}, \citenamefont {Liu}, \citenamefont {Karner}, \citenamefont
  {Carroll}, \citenamefont {Ronneberger}, \citenamefont {Wallingford},\ and\
  \citenamefont {Walz}}]{Lienkamp2012}%
  \BibitemOpen
  \bibfield  {author} {\bibinfo {author} {\bibfnamefont {S.~S.}\ \bibnamefont
  {Lienkamp}}, \bibinfo {author} {\bibfnamefont {K.}~\bibnamefont {Liu}},
  \bibinfo {author} {\bibfnamefont {C.~M.}\ \bibnamefont {Karner}}, \bibinfo
  {author} {\bibfnamefont {T.~J.}\ \bibnamefont {Carroll}}, \bibinfo {author}
  {\bibfnamefont {O.}~\bibnamefont {Ronneberger}}, \bibinfo {author}
  {\bibfnamefont {J.~B.}\ \bibnamefont {Wallingford}},\ and\ \bibinfo {author}
  {\bibfnamefont {G.}~\bibnamefont {Walz}},\ }\bibfield  {title} {\bibinfo
  {title} {Vertebrate kidney tubules elongate using a planar cell
  polarity-dependent, rosette-based mechanism of convergent extension},\ }\href
  {https://doi.org/10.1038/ng.2452} {\bibfield  {journal} {\bibinfo  {journal}
  {Nature Genetics}\ }\textbf {\bibinfo {volume} {44}},\ \bibinfo {pages}
  {1382} (\bibinfo {year} {2012})}\BibitemShut {NoStop}%
\bibitem [{\citenamefont {Ossipova}\ \emph {et~al.}(2015)\citenamefont
  {Ossipova}, \citenamefont {Kim},\ and\ \citenamefont {Soko}}]{Ossipova2015}%
  \BibitemOpen
  \bibfield  {author} {\bibinfo {author} {\bibfnamefont {O.}~\bibnamefont
  {Ossipova}}, \bibinfo {author} {\bibfnamefont {K.}~\bibnamefont {Kim}},\ and\
  \bibinfo {author} {\bibfnamefont {S.~Y.}\ \bibnamefont {Soko}},\ }\bibfield
  {title} {\bibinfo {title} {Planar polarization of vangl2 in the vertebrate
  neural plate is controlled by wnt and myosin ii signaling},\ }\href
  {https://doi.org/10.1242/bio.201511676} {\bibfield  {journal} {\bibinfo
  {journal} {Biology Open}\ }\textbf {\bibinfo {volume} {4}},\ \bibinfo {pages}
  {722} (\bibinfo {year} {2015})}\BibitemShut {NoStop}%
\bibitem [{\citenamefont {Ibrahimi}(2022)}]{Ibrahimi2022}%
  \BibitemOpen
  \bibfield  {author} {\bibinfo {author} {\bibfnamefont {M.}~\bibnamefont
  {Ibrahimi}},\ }\emph {\bibinfo {title} {Robustness of active anisotropic
  deformation in developing biological tissues}},\ \href
  {http://www.theses.fr/2022AIXM0523} {Ph.D. thesis},\ \bibinfo  {school}
  {Aix-Marseille Université} (\bibinfo {year} {2022}),\ \bibinfo {note}
  {2022AIXM0523}\BibitemShut {NoStop}%
\bibitem [{\citenamefont {Lefebvre}\ \emph {et~al.}(2024)\citenamefont
  {Lefebvre}, \citenamefont {Colen}, \citenamefont {Claussen}, \citenamefont
  {Brauns}, \citenamefont {Raich}, \citenamefont {Mitchell}, \citenamefont
  {Fruchart}, \citenamefont {Vitelli},\ and\ \citenamefont
  {Streichan}}]{Lefebvre2024}%
  \BibitemOpen
  \bibfield  {author} {\bibinfo {author} {\bibfnamefont {M.}~\bibnamefont
  {Lefebvre}}, \bibinfo {author} {\bibfnamefont {J.}~\bibnamefont {Colen}},
  \bibinfo {author} {\bibfnamefont {N.}~\bibnamefont {Claussen}}, \bibinfo
  {author} {\bibfnamefont {F.}~\bibnamefont {Brauns}}, \bibinfo {author}
  {\bibfnamefont {M.}~\bibnamefont {Raich}}, \bibinfo {author} {\bibfnamefont
  {N.}~\bibnamefont {Mitchell}}, \bibinfo {author} {\bibfnamefont
  {M.}~\bibnamefont {Fruchart}}, \bibinfo {author} {\bibfnamefont
  {V.}~\bibnamefont {Vitelli}},\ and\ \bibinfo {author} {\bibfnamefont {S.~J.}\
  \bibnamefont {Streichan}},\ }\href@noop {} {\bibinfo {title} {Learning a
  conserved mechanism for early neuroectoderm morphogenesis}},\ \bibinfo
  {howpublished} {https://arxiv.org/abs/2405.18382v1} (\bibinfo {year}
  {2024})\BibitemShut {NoStop}%
\bibitem [{\citenamefont {Roberts}\ and\ \citenamefont
  {Bowman}(2011)}]{Roberts2011}%
  \BibitemOpen
  \bibfield  {author} {\bibinfo {author} {\bibfnamefont {M.}~\bibnamefont
  {Roberts}}\ and\ \bibinfo {author} {\bibfnamefont {J.~C.}\ \bibnamefont
  {Bowman}},\ }\bibfield  {title} {\bibinfo {title} {Dealiased convolutions for
  pseudospectral simulations},\ }\href
  {https://doi.org/10.1088/1742-6596/318/7/072037} {\bibfield  {journal}
  {\bibinfo  {journal} {Journal of Physics: Conference Series}\ }\textbf
  {\bibinfo {volume} {318}},\ \bibinfo {pages} {072037} (\bibinfo {year}
  {2011})}\BibitemShut {NoStop}%
\bibitem [{\citenamefont {Bowman}\ and\ \citenamefont
  {Roberts}(2011)}]{Bowman2011}%
  \BibitemOpen
  \bibfield  {author} {\bibinfo {author} {\bibfnamefont {J.~C.}\ \bibnamefont
  {Bowman}}\ and\ \bibinfo {author} {\bibfnamefont {M.}~\bibnamefont
  {Roberts}},\ }\bibfield  {title} {\bibinfo {title} {Efficient dealiased
  convolutions without padding},\ }\href {https://doi.org/10.1137/100787933}
  {\bibfield  {journal} {\bibinfo  {journal} {SIAM Journal on Scientific
  Computing}\ }\textbf {\bibinfo {volume} {33}},\ \bibinfo {pages} {386}
  (\bibinfo {year} {2011})},\ \Eprint
  {https://arxiv.org/abs/https://doi.org/10.1137/100787933}
  {https://doi.org/10.1137/100787933} \BibitemShut {NoStop}%
\end{thebibliography}%
\end{document}